\pdfoutput=1
\documentclass[
    aip,
    reprint,
    final,
    preprintnumbers,
    longbibliography,
    nobibnotes,
    floatfix,
    nofootinbib,
    showkeys
]{revtex4-1}

\usepackage{mathtools}
\usepackage{BOONDOX-calo}   
\usepackage[export]{adjustbox}
\usepackage{algpseudocode}
\usepackage{algorithm}
\usepackage{alphalph}
\usepackage{amsmath}
\usepackage{amssymb}
\usepackage{amsthm}
\usepackage{amsbsy}
\usepackage{array}
\usepackage{appendix}
\usepackage{calc}
\usepackage{changepage}
\usepackage{dcolumn}        
\usepackage{enumitem}
\usepackage{epsfig}
\usepackage{esvect}
\usepackage[pscoord]{eso-pic}
\usepackage{etoolbox}
\usepackage{fancybox}
\usepackage{fancyhdr}
\usepackage{float}
\usepackage[T1]{fontenc}
\usepackage[hang,bottom]{footmisc}
\usepackage[nomessages]{fp}
\usepackage[
bindingoffset=0in,
top    = 0.8in,
bottom = 0.8in,
left   = 0.75in,
right  = 0.75in]{geometry}
\usepackage{graphicx}
\usepackage{grffile}    
\usepackage[hidelinks]{hyperref}
\usepackage{ifthen}
\usepackage[utf8]{inputenc}
\usepackage{lastpage}
\usepackage{letltxmacro}
\usepackage{mathptmx}
\usepackage{marvosym}
\usepackage{multirow}
\usepackage{natbib}
\usepackage{parskip}
\usepackage{pgffor}
\usepackage{pgfplots}
\usepackage{pifont}
\usepackage{ragged2e}
\usepackage{scrextend}
\usepackage{setspace}
\usepackage{siunitx}
\usepackage{soul}
\usepackage[caption=false]{subfig}
\usepackage{tabularx}
\usepackage{titlesec}
\usepackage{textcomp}
\usepackage{tikz}
\usepackage{txfonts}
\usepackage[normalem]{ulem}
\usepackage{url}
\usepackage{varwidth}
\usepackage{verbatim}
\usepackage{wasysym}
\usepackage{xcolor}
\usepackage{xstring}

\newsavebox\CBox
\newcommand\hcancel[2][0.3pt]{%
    \ifmmode\sbox\CBox{$#2$}\else\sbox\CBox{#2}\fi%
    \makebox[0pt][l]{\usebox\CBox}%
    \rule[0.3\ht\CBox-#1/2]{\wd\CBox}{#1}}

\newcommand\bmm[1]{\boldsymbol{#1}}

\definecolor{green}{rgb}{0,0.79,0}

\hypersetup{
    linktoc     = all,
    colorlinks  = true,
    allcolors = [rgb]{0,0,0.75}, 
}
\hypersetup{final}

\usetikzlibrary{arrows,chains,matrix,positioning,scopes}

\makeatletter
\def\resetMathstrut@{%
    \setbox\z@\hbox{%
        \mathchardef\@tempa\mathcode`\\relax
        \def\@tempb##1"##2##3{\the\textfont"##3\char"}%
        \expandafter\@tempb\meaning\@tempa \relax
    }%
    \ht\Mathstrutbox@1.2\ht\z@ \dp\Mathstrutbox@1.2\dp\z@
}
\makeatother

\catcode`@=11
\def\caseswithdelim#1#2{\left#1\,\vcenter{\normalbaselines\m@th
        \ialign{\strut$##\hfil$&\quad##\hfil\crcr#2\crcr}}\right.}
\catcode`@=12

\def\breakloop{\fi\iffalse}

\titlespacing{\section}{0pt}{12pt}{5pt}
\titlespacing{\subsection}{0pt}{12pt}{5pt}
\titlespacing{\subsubsection}{0pt}{12pt}{5pt}

\newcommand{\fsize}{}
\newcommand{\fsizeiso}{}
\newcommand{\fdir}{}
\newcommand{\ilflag}{0}
\newcommand{\fvardir}{}
\newcommand{\fvar}{}
\newcommand{\fvartext}{}
\newcommand{\fvarlatex}{}
\newcommand{\ftype}{}
\newcommand{\filetype}{}
\newcommand{\figtype}{}
\newcommand{\fres}{}

\newcommand{\slice}{}
\newcommand{\slicetype}{}
\newcommand{\ftime}{}
\newcommand{\dt}{}

\newcommand{\myresult}{}
\newcommand{\forloop}[5][1]%
{%
    \setcounter{#2}{#3}%
    \ifthenelse{#4}%
    {%
        #5%
        \addtocounter{#2}{#1}%
        \forloop[#1]{#2}{\value{#2}}{#4}{#5}%
    }%
    {%
    }%
}%

\newcounter{il}
\newcounter{jl}

\newcommand*\overbar[1]{%
    \vbox{%
        \hrule height 0.9pt%
        \kern0.35ex%
        \hbox{%
            \kern-0.1em%
            \ifmmode#1\else\ensuremath{#1}\fi%
            \kern0.0em%
        }
    }
}

\renewcommand{\algorithmiccomment}[1]{\bgroup\hfill!#1\egroup}

\renewcommand{\arraystretch}{1.1}

\definecolor{bluesteel}{rgb}{0.392, 0.392, 0.498}
\definecolor{lightgray}{gray}{1.0}
\renewcommand{\figtype}{pdf}

\begin{document}


\title{The effect of entrance flow development on vortex formation and wall shear stress in a curved artery model}

\author{Christopher Cox}
\altaffiliation[Also at ]{Lawrence Livermore National Laboratory, Livermore, CA, USA}
\email[\\email:~]{ccox@gwmail.gwu.edu}
\author{Michael W. Plesniak}
\email[email:~]{plesniak@gwu.edu}
\noaffiliation
\affiliation{
    Department of Mechanical and Aerospace Engineering,
    The George Washington University,\\
    Washington, DC 20052, USA
}

\begin{abstract}
    We numerically investigate the effect of entrance condition on the spatial and temporal evolution of multiple three-dimensional vortex pairs and wall shear stress distribution in a curved artery model. We perform this study using a Newtonian blood-analog fluid subjected to a pulsatile flow with two inflow conditions. The first flow condition is fully developed while the second condition is undeveloped (i.e. uniform). We discuss the connection along the axial direction between regions of organized vorticity observed at various cross-sections of the model and compare results between the different entrance conditions. We model a human artery with a simple, rigid $180^\circ$ curved pipe with circular cross-section and constant curvature, neglecting effects of taper, torsion and elasticity. Numerical results are computed from a discontinuous high-order spectral element flow solver. The flow rate used in this study is physiological. We observe differences in secondary flow patterns, especially during the deceleration phase of the physiological waveform where multiple vortical structures of both Dean-type and Lyne-type coexist. We highlight the effect of the entrance condition on the formation of these structures and subsequent appearance of abnormal inner wall shear stresses---a potentially significant correlation since cardiovascular disease is known to progress along the inner wall of curved arteries under varying degrees of flow development.
    
\end{abstract}


\maketitle

\section{INTRODUCTION}
\label{s:curved_pipe_introduction}

\begin{figure*}[t]
    \captionsetup[subfigure]{labelformat=parens}
    \renewcommand{\fsize}{44mm}
    \centering\setcounter{subfigure}{0}
    \subfloat[]{
        \includegraphics[height=\fsize,keepaspectratio]
        {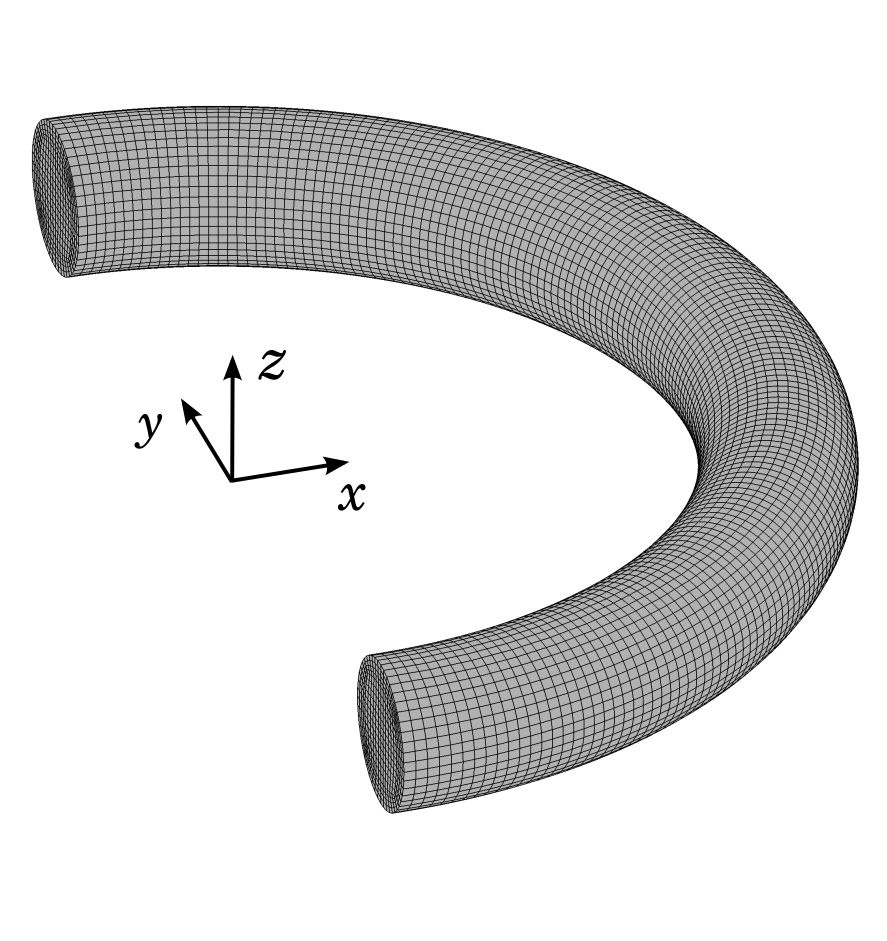}
        \label{f:pipe_mesh}
    }
    \subfloat[]{
        \includegraphics[height=\fsize,keepaspectratio]
        {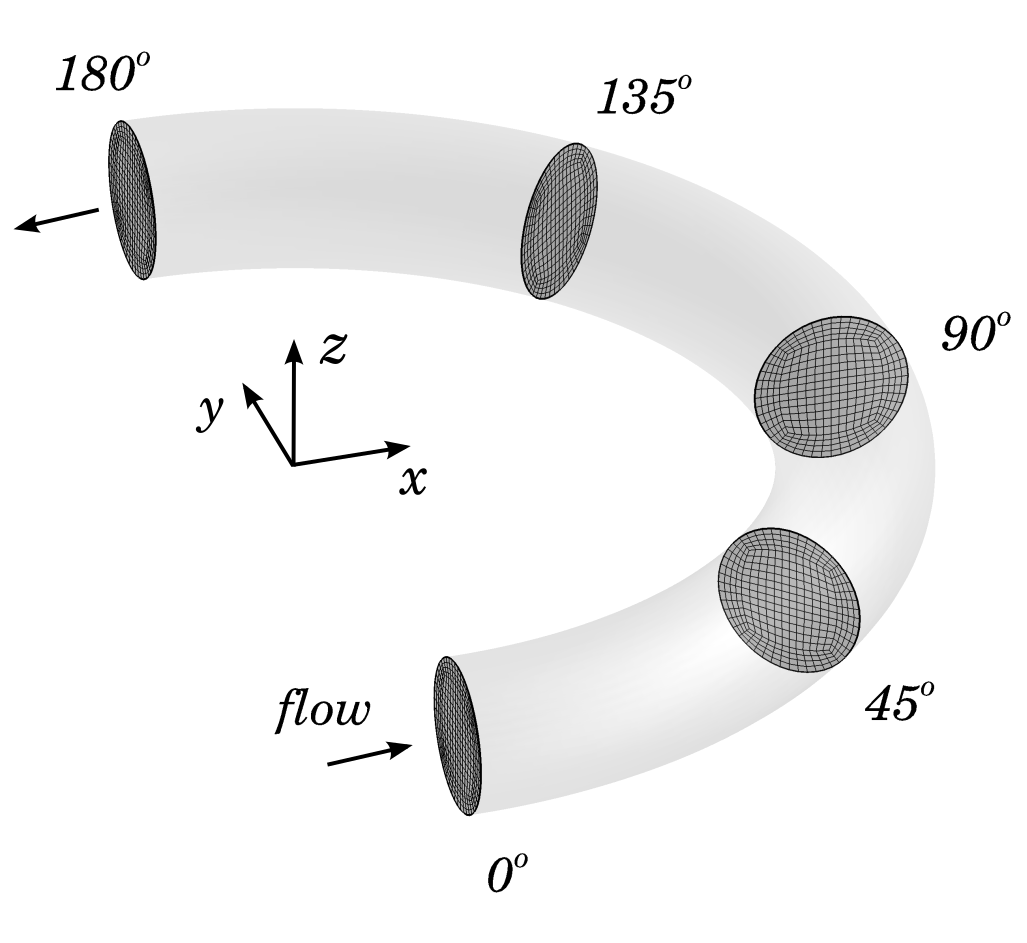}
        \label{f:pipe_cross_section}
    }
    \subfloat[]{
        \includegraphics[height=\fsize,keepaspectratio]
        {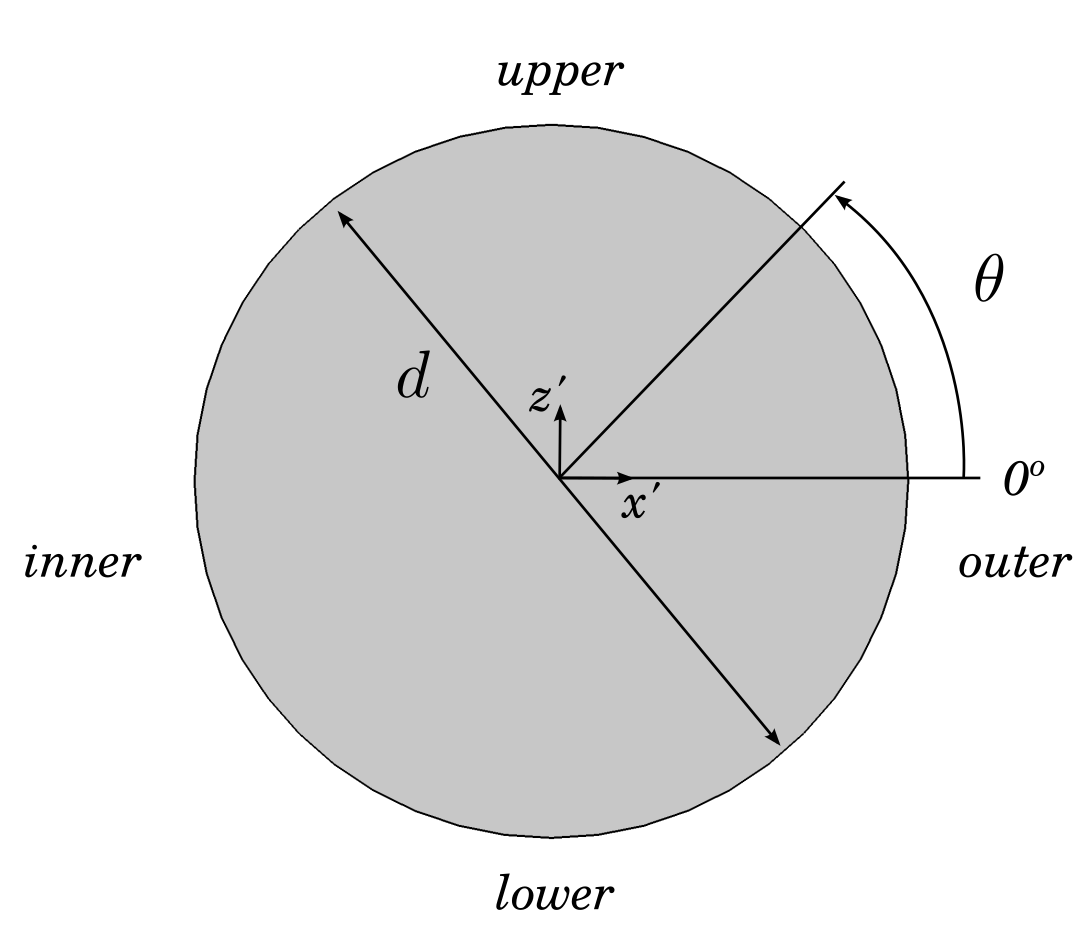}
        \label{f:theta_planar_view}
    }
    \subfloat[]{
        \includegraphics[height=\fsize,keepaspectratio]
        {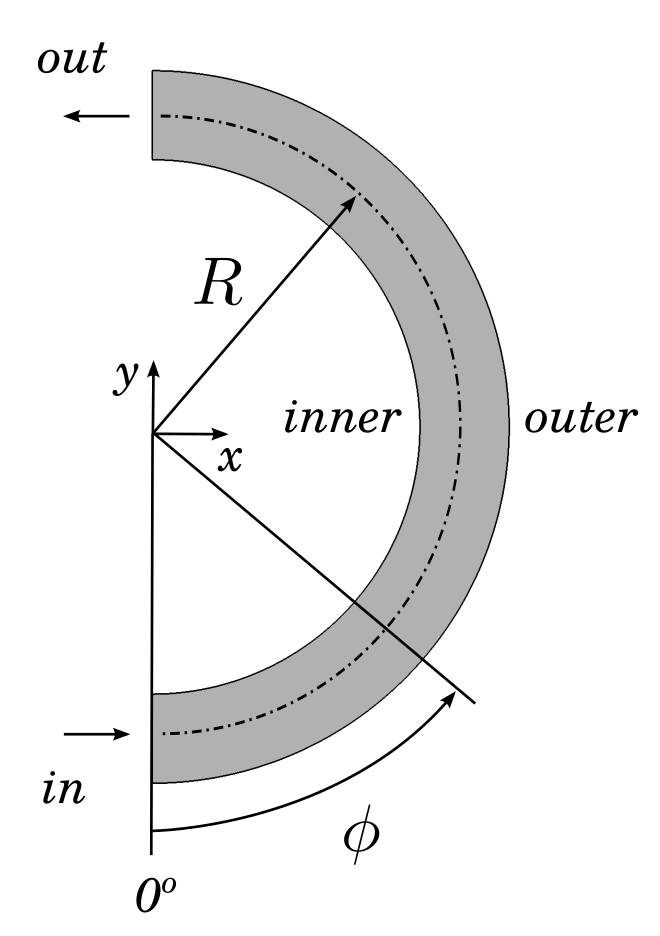}
        \label{f:phi_top_view}
    }
    
    \caption{Curved pipe geometry and orientation. (a) curvature mesh, (b) curved pipe with flow direction and various cross-sections, (c) $x'z'$ cross-sectional plane viewed from upstream, (d) $xy$ plane of symmetry.}
    \label{f:curved_pipe_mesh_geometry_N4}
\end{figure*}

Cardiovascular flows are pulsatile, incompressible flows that exist in complex geometries with compliant walls. Together, these factors produce a vortex rich environment. The physiologically and pathologically relevant forces at play in the cardiovascular system are wall shear stress, which acts on the endothelium, and circumferential strain, which acts on endothelial and smooth muscle cells. These forces are important because atherosclerotic regions are strongly correlated with curvature and branching in the human vasculature, where there exists oscillatory shear stress with a low time-averaged value and spatial/temporal shear gradients.~\cite{davies:2008,glagov-zarins-giddens-ku:1988} Ultimately, vortices in the flow can affect the progression of cardiovascular disease by altering wall shear stresses.

Multidirectionality of the flow may also play an important role in the prevalence of atherosclerotic disease.~\cite{mantha:2006,chakraborty:2012,peiffer-sherwin-weinberg:2013} Relevant haemodynamic metrics used to assess the local variation in blood flow characteristics as it relates to atherosclerotic lesions are low {\it time-averaged wall shear stress}, {\it oscillatory shear index}~\cite{he-ku:1996} and {\it relative residence time};~\cite{himburg:2004} although, evidence for the low temporal mean and oscillatory shear stress concept is less robust than previously assumed by the community.~\cite{peiffer-sherwin-weinberg:2013b} In light of this evidence, another metric---{\it transverse wall shear stress}~\cite{peiffer-sherwin-weinberg:2013}---was designed to account for multidirectionality of the wall shear stress vector.

Human arteries often follow a path that can curve, twist, taper, bifurcate, and vary in cross-sectional shape. To locate regions where atherosclerosis progresses, many researchers perform numerical simulations using patient-specific data and geometry to study fluid flow. Other researchers adopt a more idealized approach in order to grasp the underlying flow physics without the added complexity of variable geometry. In this case, a simplified model, such as a curved pipe, is needed. We adopt the simplified approach in this work, modeling a human artery with a rigid $180^\circ$ curved pipe with circular cross-section and constant curvature. This assumption is thought to be valid since healthy arteries exhibit a more circular cross-section, whereby segments of curved arteries can be modeled as curved pipes, neglecting effects of taper, torsion and elasticity.

The body of research on flows through curved pipes in the 20th century is vast, beginning in 1902 with Williams et al.~\cite{williams:1902} who first noticed the velocity profile shift towards the outer wall. Eustice~\cite{eustice:1910,eustice:1911} showed the existence of secondary flow\footnote{In curved pipes, the primary flow is defined to be parallel to the pipe axis whereas the secondary flow is superimposed on and perpendicular to the primary flow.} through a coiled pipe. Steady fluid motion through a curved pipe consists of a pair of counter-rotating helical vortices that are symmetric with respect to the plane of symmetry separating the upper and lower halves of the pipe. The development of secondary flow creates vortical structures that affect wall shear stress patterns, which may play a significant role in the progression of cardiovascular disease.~\cite{cox-najjari-plesniak:2019}

The mechanism of secondary flow in a curved pipe under fully developed steady flow conditions was originally put forth by Dean,~\cite{dean:1927,dean:1928} who found that fluid motion is dependent upon a parameter $\kappa$ referred to thereafter as the {\it Dean number} and defined as~\cite{berger-talbot-yao:1983}
\begin{align}
    \kappa = Re \sqrt{\delta}.
    \label{e:dean_number_bty}
\end{align}
The curvature ratio $\delta = r/R$ is a measure of the geometric effect and extent to which centrifugal forcing varies in the cross-section. In this form of the Dean number, the Reynolds number $Re=\overbar{u}d/\nu$ is defined based on bulk velocity $\overbar{u}$ and pipe diameter $d$, where $\nu$ is the kinematic viscosity of the fluid. Although there exist other definitions of the Dean number (e.g. see Berger et al.~\cite{berger-talbot-yao:1983}), versions of the Dean number based on bulk velocity are natural for experimentalists because this quantity can readily be measured, providing a convenient characterization of the flow. A physical interpretation of the Dean number is the ratio of centrifugal and convective inertial forces to viscous forces. Since secondary flow is induced by centrifugal forces and their interaction with viscous forces, $\kappa$ is thought of as a measure of the degree of secondary flow. In addition to the Dean number, the degree to which the flow is developed at the entrance to the curvature affects secondary flow patterns. We discuss this fact in the current work as it relates to the production of vortical structures and abnormal wall shear stresses.

In addition ot the Dean number, fully developed pulsatile fluid motion in a tube is also dependent upon a parameter $\alpha$ commonly referred to as the {\it Womersley number}~\cite{womersley:1955} and written as
\begin{align}
    \alpha = r \sqrt{\frac{\omega}{\nu}},
    \label{e:womersley_number}
\end{align}
where $\omega=2 \pi f$ is the angular frequency of pulsation. The Womersley number, which can be re-written as $\alpha=(d/2) \sqrt{2 \pi/(\nu T)}$, can be interpreted physically~\cite{doorly-sherwin:2009} as the ratio of pipe diameter to growth of the laminar boundary layer during the pulsatile waveform period $T$. For an extensive review of the 20th century literature on flows in curved pipes, see Berger et al.~\cite{berger-talbot-yao:1983}
    
Fully developed flows driven by a pressure gradient that is sinusoidally varying in time about a non-zero mean was investigated by Smith.~\cite{smith:1975} The analysis revealed a number of pulsatile motions and the manner in which secondary motion can change its direction from inward to outward ``centrifuging'' at high frequencies.
Sudo et al.~\cite{sudo-sumida-yamane:1992} performed experimental and numerical studies of secondary flow induced by fully developed oscillatory flow in a curved pipe for a range of Dean and Womersley numbers. Their results suggested that secondary flows can be classified into five circulation patterns. Boiron et al.~\cite{boiron-deplano-pelissier:2007} also conducted experimental and numerical studies in a $180^\circ$ curved pipe, focusing on the starting effect of the fluid on secondary flow for a range of Dean and Womersley numbers. In addition, they studied the balance of centrifugal force and radial pressure gradient in the symmetry plane as it related to the appearance of Lyne vortices. Further experimental and numerical work was done by Timite et al.,~\cite{timite-castelain-peerhossaini:2010} who studied developing pulsatile flow in a $90^\circ$ curved pipe. Their results demonstrated that for increased Womersley number, the intensity of secondary flow decreased during the acceleration phase of the pulsatile waveform and increased during the deceleration phase due to the effect of reverse flow. A detailed investigation of the mechanisms by which curvature and torsion can affect blood flow in the human vasculature was performed by Alastruey et al.,~\cite{alastruey:2012} where they simulated steady-state flow in single bends, helices, double bends and a rabbit thoracic aorta. It was shown that the curvature-dependent Coriolis force directly links motions in the radius of curvature direction to axial accelerations, whereby it enhances/slows the axial bulk flow if secondary motions are toward/away from the center of curvature such that as the flow develops in the curve the Coriolis force helps shape the velocity profile. More recently, van Wyk et al.~\cite{vanwyk:2015} conducted numerical simulations of flow through a rigid $180^\circ$ curved pipe and compared results generated using Newtonian and non-Newtonian blood-analog fluids. They used a physiological waveform from a human carotid artery similar to the one used in this work.

Krishna et al.~\cite{krishna-arakeri:2017} performed both numerical and experimental flow visualization studies of pulsatile flow in a square cross-sectional tube with a high curvature ratio ($\approx 0.3$) over a range of Dean and Womersley numbers that included values observed in a human ascending aorta. They showed that the radially outward moving wall jet reversed direction and caused flow separation near the inner wall during the deceleration phase---this flow separation they attributed to secondary flow. A study by Najjari et al.~\cite{najjari-cox-plesniak:2019} on the formation and interaction of multiple vortical structures in a curved pipe under transient and oscillatory flows revealed that vorticity associated with the Lyne-type vortex is transported by convection from the inner wall region---just downstream of the entrance to the curve---during its formation phase. Their analysis suggested that the morphology of the Lyne-type vortex is similar to that of a hairpin vortex.

Ku~\cite{ku:1997} highlighted the fact that flow through an artery may not be fully developed as is the case when blood emanates from the heart, which functions as a large pressure reservoir. Near the heart, flow entering curved or branched arteries is not fully developed---the velocity profiles are flattened near the center, signifying a developing boundary near the wall and an inviscid core in the center. This concept of blood flow development upstream to a curved artery motivated the current study.

Despite the vast amount of research in the 20th century on flow through curved conduits, the appearance of vortical structures under pulsatile developing flow and their role in vascular flows still warrants further investigation. Recent research by Cox et al.~\cite{cox-najjari-plesniak:2019} characterized the spatial and temporal evolution of three-dimensional vortical structures under pulsatile flow through a curved artery model, specifically capturing both Dean-type and Lyne-type vortices throughout the deceleration phase of the physiologically relevant flow rate. From these results, the authors discussed the connection along the axial direction between vortical structures observed at various cross sections of the model, supplementing previously limited two-dimensional analysis. Identifying and understanding the formation of these structures allowed for correlations to be made between vortices produced under physiological conditions and wall shear stress distributions. The work presented herein follows up that research to study the effect of flow development at the entrance to a curved artery on the formation of vortical structures and subsequent wall shear stress patterns. We perform numerical simulations of a Newtonian blood-analog fluid using a physiological pulsatile waveform and two inflow conditions to a rigid $180^\circ$ curved pipe with circular cross section and constant curvature, neglecting effects of taper, torsion, and elasticity. The first pulsatile entrance flow condition is one that is fully developed while the second condition is undeveloped (i.e. uniform). The pulsatile flow rates under both scenarios are equivalent, and the flow develops in the curve without any symmetry condition imposed between the upper and lower halves of the pipe. The curved geometry, physiological waveform, and Womersley number are similar to those used in previous studies.~\cite{vanwyk:2015,najjari-plesniak:2016}

We present results of velocity profiles and surfaces along with secondary velocities, vorticity and instantaneous wall shear stress under both entrance conditions. We identify vortical structures using an appropriate vortex identification method and characterize their evolution throughout the deceleration phase of the physiologically relevant flow rate, capturing both Dean-type and Lyne-type vortices. From these results we can then discuss the influence of curvature entrance flow development on vortex formation and how the inflow condition ultimately affects wall shear stress distributions and potentially impacts the progression of cardiovascular disease.

\begin{table}[t]
    \caption{Pulsatile waveform parameters.\hfill}
    \label{t:waveform_stats}
    {\footnotesize
        \begin{ruledtabular}
            \begin{tabular}{lll}
                Parameter       & Expression & Value \\ \hline
                $Q_{min}$       & -- & 1.0~mL/s   \\
                $Q_{mean}$      & -- & 13.4~mL/s  \\
                $Q_{max}$       & -- & 53.2~mL/s  \\ \hline
                $\overbar{u}_{min}$ & $4 Q_{min}  / \pi d^2$ & 7.8~mm/s    \\
                $\overbar{u}_{mean}$& $4 Q_{mean} / \pi d^2$ & 105.5~mm/s  \\
                $\overbar{u}_{max}$ & $4 Q_{max}  / \pi d^2$ & 419.6~mm/s  \\
                $u_{red}$       & $\overbar{u}_{mean} T / d$ & 33.2        \\
                $\overbar{u}_{peak-to-mean}$ & $\overbar{u}_{max} / \overbar{u}_{mean}$ & 4.0  \\ \hline
                $Re_{min}$      & $\overbar{u}_{min} d  / \nu$ & 28       \\
                $Re_{mean}$     & $\overbar{u}_{mean} d / \nu$ & 377      \\
                $Re_{max}$      & $\overbar{u}_{max} d  / \nu$ & 1\,501   \\ \hline
                $\kappa_{min}$  & $Re_{min}\sqrt{\delta}$      & 10       \\
                $\kappa_{mean}$ & $Re_{mean}\sqrt{\delta}$     & 143      \\
                $\kappa_{max}$  & $Re_{max}\sqrt{\delta}$      & 567
            \end{tabular}
        \end{ruledtabular}
    }
\end{table}

\section{NUMERICAL METHOD}
\label{s:numerics}

\subsection{Simulation setup and physiological flow rate}
\label{s:cfd_geometry}

A model of the curved section of pipe that we use for our numerical simulations and presentation of results is depicted in Fig.~\ref{f:curved_pipe_mesh_geometry_N4}. The origin of the Cartesian coordinate system $(x,y,z)=(0,0,0)$ is defined in Fig.~\ref{f:curved_pipe_mesh_geometry_N4}\subref{f:pipe_mesh} along with the various cross-sections of interest in Fig.~\ref{f:curved_pipe_mesh_geometry_N4}\subref{f:pipe_cross_section}, the diameter $d$ in Fig.~\ref{f:curved_pipe_mesh_geometry_N4}\subref{f:theta_planar_view}, and the radius of curvature $R$, flow direction, and location of the inner, outer, upper and lower walls in Fig.~\ref{f:curved_pipe_mesh_geometry_N4}\subref{f:phi_top_view}. For these simulations, the curvature ratio $r/R=1/7$ matches that used in previous work reported in the literature including Talbot and Gong,~\cite{talbot-gong:1983} Soh and Berger,~\cite{soh-berger:1984} Hamakiotes and Berger,~\cite{hamakiotes-berger:1988} van Wyk et al.,~\cite{vanwyk:2015}, Plesniak and Bulusu,~\cite{plesniak-bulusu:2016} and Najjari and Plesniak.~\cite{najjari-plesniak:2016}  Flow through large curvature ratios in human arteries, such as that seen in the aortic arch ($r/R=0.4$), can be dominated by developing secondary flow and experience significant flow separation at large flow rates (i.e. large Dean numbers). A smaller curvature ratio like the one used in the current work is less susceptible to flow separation for this range of Reynolds numbers and facilitates meaningful interpretation of the entrance effect on vortical structures and wall shear stress distributions that appear during the deceleration phase of the physiological waveform.

\begin{figure}[t]
    \centering\setcounter{subfigure}{0}
    \includegraphics[width=0.48\textwidth,keepaspectratio]
    {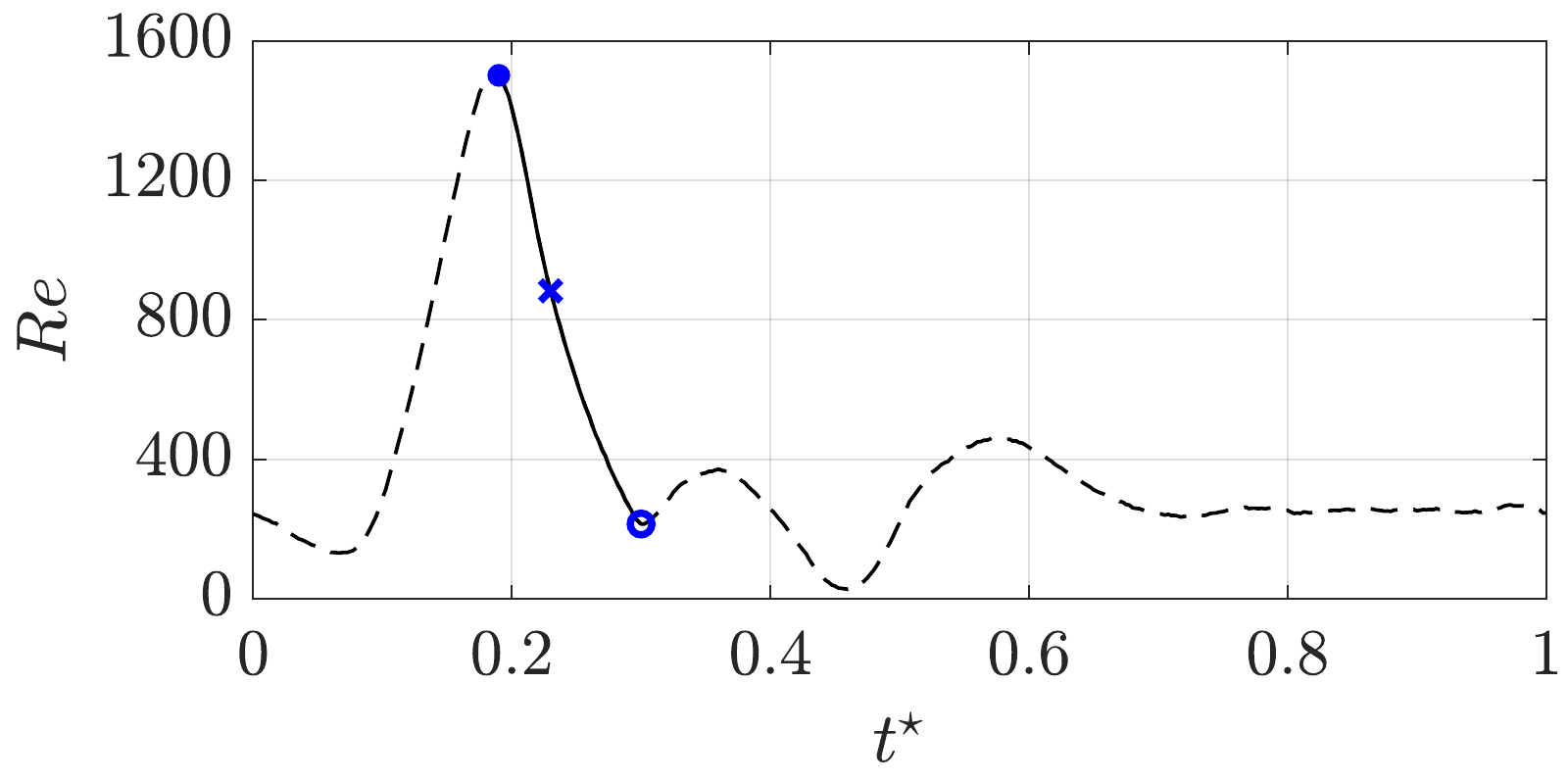}
    
    \caption{Physiological (pulsatile) waveform in terms of Reynolds number $Re$ as a function of nondimensional pulse period $t^\star=t/T$ (dash), displaying rapid deceleration (solid) during the interval $0.19 < t^\star < 0.30$. $Re|_{t^\star=0.19}=1\,501$ (blue dot), $Re|_{t^\star=0.23}=883$ (blue cross), $Re|_{t^\star=0.30}=215$ (blue circle).}
    \label{f:waveform_Re}
\end{figure}

The mesh contains a straight section of length $5d$ upstream to the curvature entrance and another straight section of length $15d$ downstream of the curvature exit. To help describe results within the curve in the following sections, we define $\theta=[0^\circ,180^\circ]$ to represent the poloidal angle measured along the azimuthal direction from the outer wall and $\phi=[0^\circ,180^\circ]$ to represent the toroidal angle measured along the streamwise/axial direction from the inlet (see Fig.~\ref{f:curved_pipe_mesh_geometry_N4}\subref{f:theta_planar_view}\subref{f:phi_top_view}). At the given radius of curvature---measured from the domain origin to the axis of the curved pipe---the distance from the entrance at $\phi=0^\circ$ to the exit at $\phi=180^\circ$ is $7 \pi d/2$. 

The pulsatile inflow waveform used in these numerical simulations and provided in Fig.~\ref{f:waveform_Re} is that reported by Holdsworth et al.~\cite{holdsworth:1999} measured in the carotid artery of healthy human subjects. It is prototypical of pulsatile waveforms found in the human vasculature. We note that systemic compliance is inherent in the waveform and, therefore, taken into account. This waveform exhibits characteristic flow rate acceleration and deceleration, as a result of systolic and diastolic phases, and has been used in previous studies~\cite{vanwyk:2015,plesniak-bulusu:2016,najjari-plesniak:2016} having been scaled to account for pipe diameter while maintaining the Womersley number and maximum Reynolds number found in the human carotid artery. The rapid acceleration and subsequent deceleration produces rich flow physics of physiological relevance, motivating our investigation of the entrance effect on resulting vortical content and wall shear stress patterns within a curved geometry.

Details of the physiological waveform are provided in Table~\ref{t:waveform_stats}. Integration of the numerical velocity profile at the minimum, mean and maximum flow rate gives 1.0, 13.4 and 53.2~$\mathrm{mL~s^{-1}}$, respectively. This produces a peak-to-mean ratio of approximately 4. The corresponding Reynolds numbers are $Re\in\{28, 377, 1\,501\}$, and Dean numbers computed from Eq.~(\ref{e:dean_number_bty}) are $\kappa\in\{10, 143, 567\}$. As noted earlier, the Reynolds number in this formulation of the Dean number is based on bulk flow velocity $\overbar{u}$ and pipe diameter $d$. The reduced velocity $u_{red} = \overbar{u}_{mean} T / d$, where $\overbar{u}_{mean}$ is the mean velocity over the pulse period $T$, is another dimensionless quantity that can be interpreted as the ratio of distance traveled by the mean flow along the pipe to the pipe diameter. For the current waveform $u_{red}=33.2$, signifying that the mean flow travels approximately 33 diameters over the pulse period. Since the length of the $180^\circ$ curved section is $7 \pi d / 2 \approx 11d$, the distance traveled by the mean flow in one period is nearly three times the length of the entire curve. This large value of the reduced velocity indicates that flow structures generated within each pulse period do not interfere with each other~\cite{doorly-sherwin:2009} and we can investigate these structures in isolation without flow disturbance from previous waveform cycles.

\subsection{Numerical scheme}
\label{s:numerical_scheme}

The equations governing a Newtonian fluid through the curved pipe geometry under the pulsatile conditions described in Sec.~\ref{s:cfd_geometry} are the unsteady three-dimensional incompressible Navier-Stokes equations. To numerically solve these equations, we use the artificial compressibility formulation of Chorin~\cite{chorin:1967} in which pressure and velocity are loosely coupled such that an auxiliary system of equations can be written as
\begin{subequations}
    \begin{align}
    \frac{1}{\beta_o}\frac{\partial p}{\partial \tau} + \nabla \cdot \bmm{u} &= 0
    \label{e:ac_continuity}
    \\
    \frac{\partial \bmm{u}}{\partial \tau} + \frac{\partial \bmm{u}}{\partial t} + \bmm{u} \cdot \nabla\bmm{u} &= -\nabla p + \nu \nabla^2 \bmm{u}
    \label{e:ac_momentum}
    \end{align}
\end{subequations}
where artificial/pseudo time derivatives, denoted by $\partial(\cdot)/\partial \tau$, of $p=P(x,y,z,t)/\rho$ and velocity $\bmm{u}=\bmm{u}(x,y,z,t)$ have been added to the continuity and momentum equations, respectively. The constant $\nu$ is the kinematic viscosity, $P(x,y,z,t)$ is the pressure and $\rho$ is the density. The symbol $\beta_o$ is termed the artificial compressibility parameter and is analogous to a relaxation parameter that allows the above system of equations to converge in order to satisfy the divergence-free constraint on the velocity field. Further discussion on this method and details on the eigenstructure of the incompressible Navier-Stokes equations with artificial compressibility can be found in Elsworth and Toro~\cite{elsworth-toro:1992a,elsworth-toro:1992b} and Drikakis and Rider.~\cite{drikakis-rider:2005}

To compute the spatial derivatives of the fluxes, we use the flux reconstruction (FR) scheme of Huynh,~\cite{huynh:2007,huynh:2009} which is a type of discontinuous spectral element method designed to allow for various high-order nodal schemes to be cast within a generalized framework. This method has become increasingly popular over the past decade, spawning further research in discontinuous high-order schemes.~\cite{vincent-castonguay-jameson:2011,jameson-castonguay-vincent:2012,williams:2013,degrazia:2014,mengaldo:2016,romero:2016,cox:2021a,huynh:2020} Details on the current flow solver are provided in Cox et al.~\cite{cox-aiaa:2015,cox-aiaa:2016,cox-liang-plesniak:2016} and Cox.~\cite{cox:2017} A grid spacing $h$ and polynomial $\mathcal{p}$ convergence study was performed~\cite{cox-najjari-plesniak:2019} where we found that a mesh containing a total of 69\,888 hexahedral elements and a $\mathcal{p}=4$ Lagrange polynomial used to approximate the solution in each element provide ample spatial resolution (${\sim}8.7\mathrm{M}$ degrees of freedom) to converge the $L^2$-norm of the error in the wall shear stress to within $0.78\%$ and the velocity magnitude to within $0.25\%$. We use the free 3\nobreakdash-D finite element mesh generator GMSH~\cite{geuzaine-remacle:2009} to create all computational domains.

\begin{figure}[t]
    \captionsetup[subfigure]{labelformat=parens}
    \centering\setcounter{subfigure}{0}
    \includegraphics[width=0.48\textwidth,keepaspectratio]
    {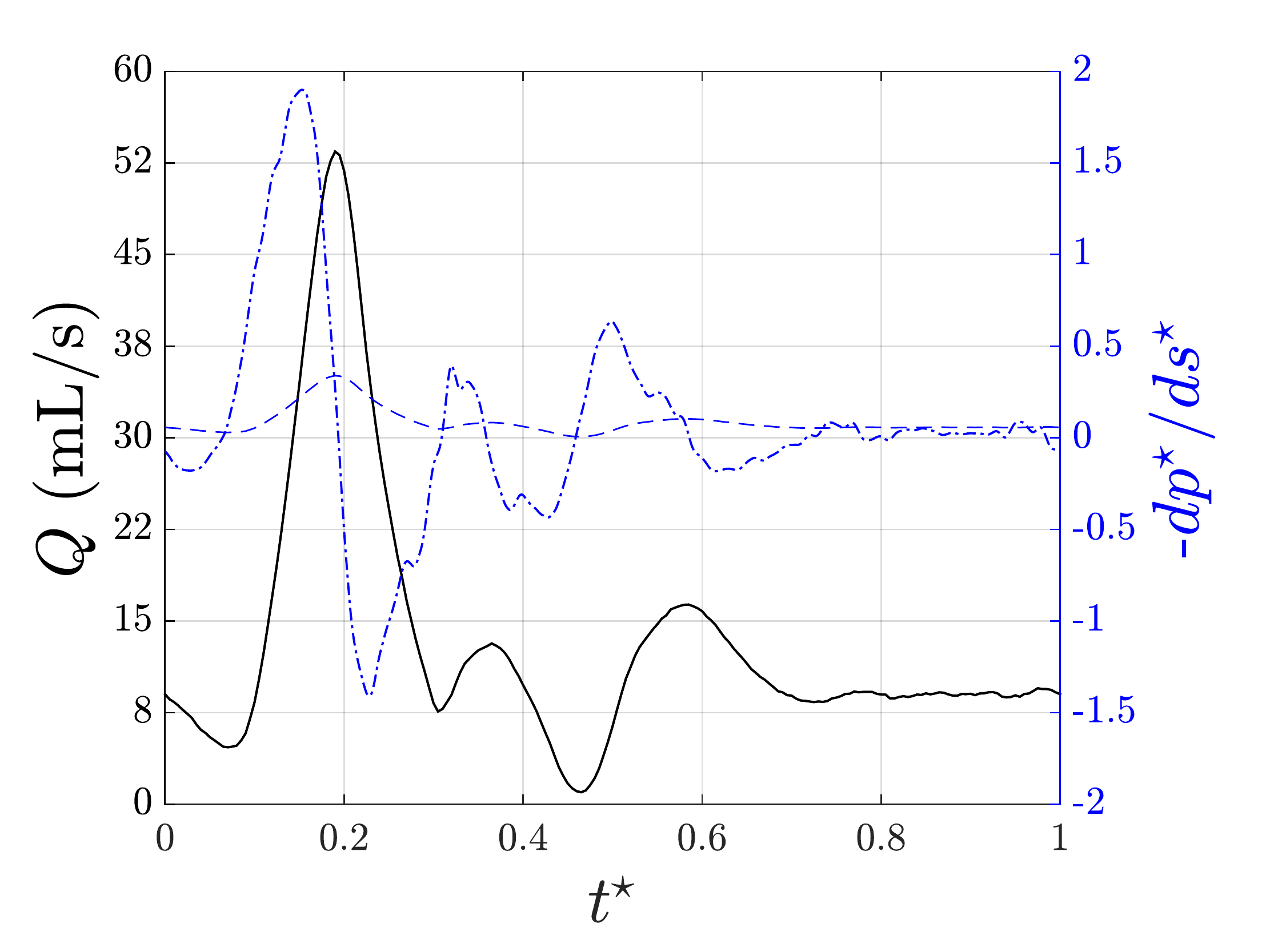}
    
    \caption{Flow rate $Q$ (black solid) ``lagging'' the straight pipe nondimensional pressure gradient $\textnormal{-}\partial p^\star/\partial s^\star$ (blue dash dot) at Womersley number $\alpha=4.22$. The pressure gradient from a pulsatile Poiseuille solution (blue dash) required to produce the same flow rate is shown for comparison.}
    \label{f:pressure_gradient_lead_flowrate}
\end{figure}



\begin{figure*}[t]
    \begin{minipage}[c]{\textwidth}
        \captionsetup[subfigure]{labelformat=parens}
        \centering\setcounter{subfigure}{0}
        \subfloat[]{
            \includegraphics[width=0.44\textwidth,keepaspectratio]
            {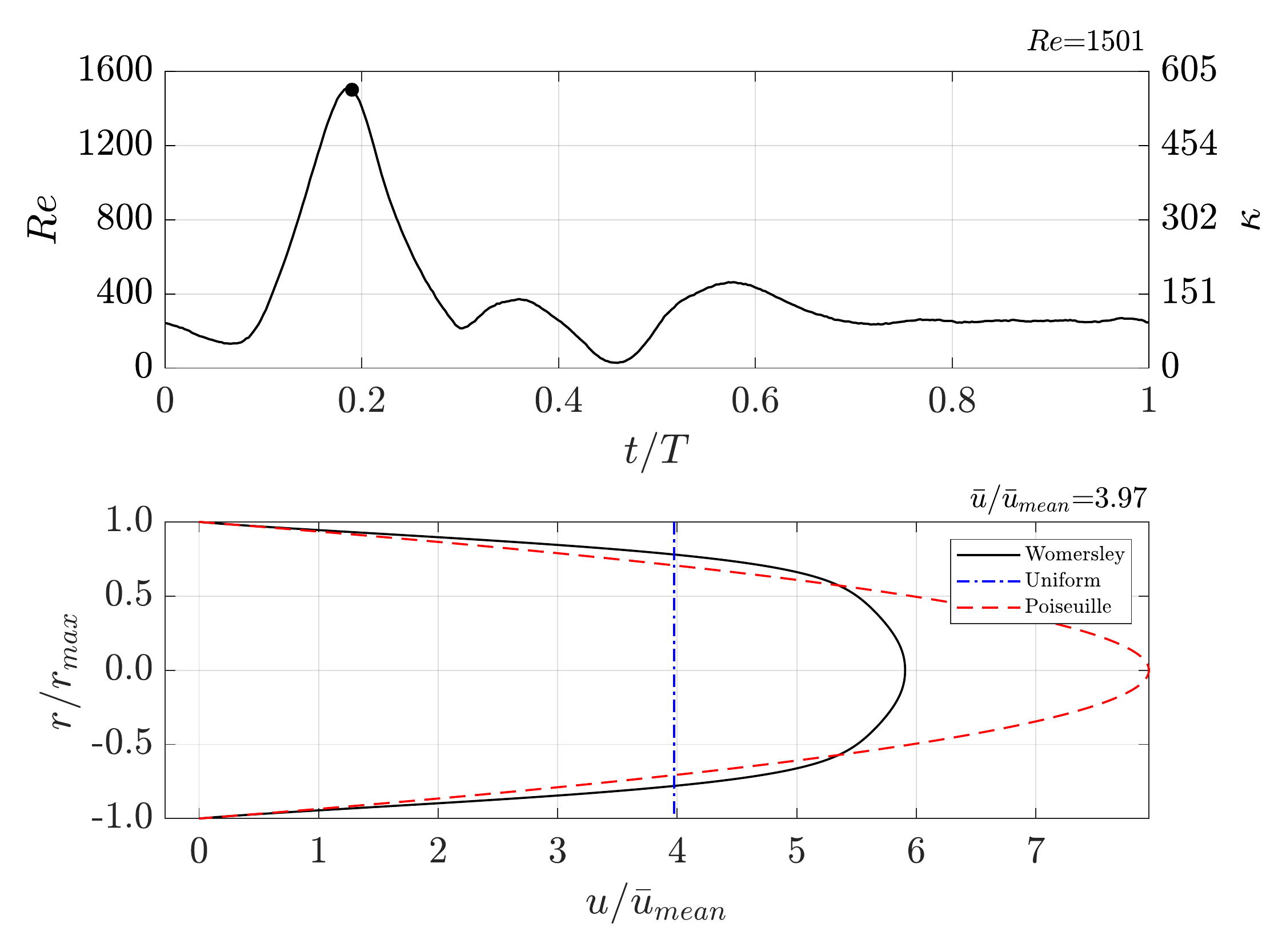}
            \label{f:t14_Re_waveform}
        }
        \subfloat[]{
            \includegraphics[width=0.44\textwidth,keepaspectratio]
            {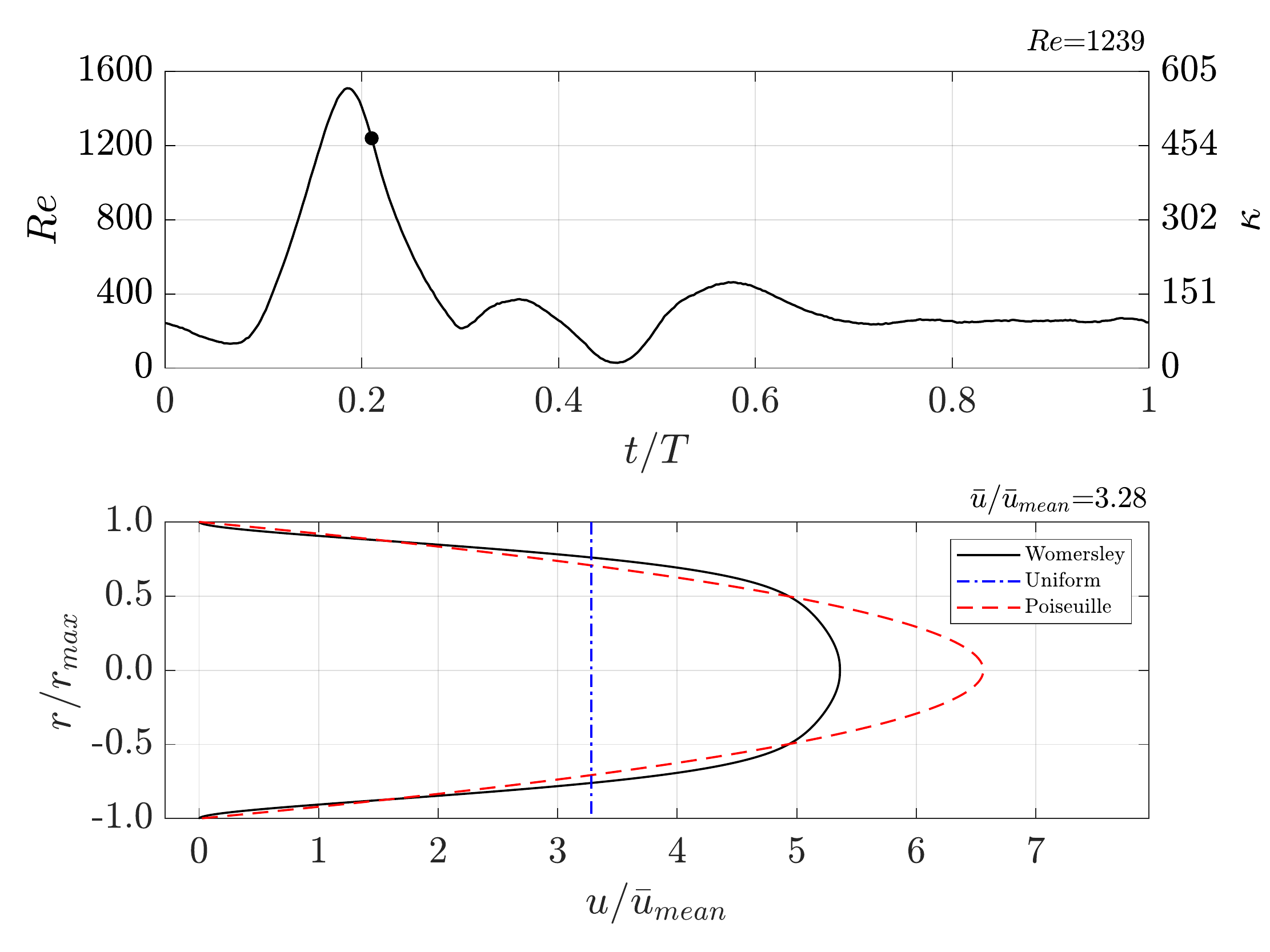}
            \label{f:t15_Re_waveform}
        }
        \\[-0.18in]
        \subfloat[]{
            \includegraphics[width=0.44\textwidth,keepaspectratio]
            {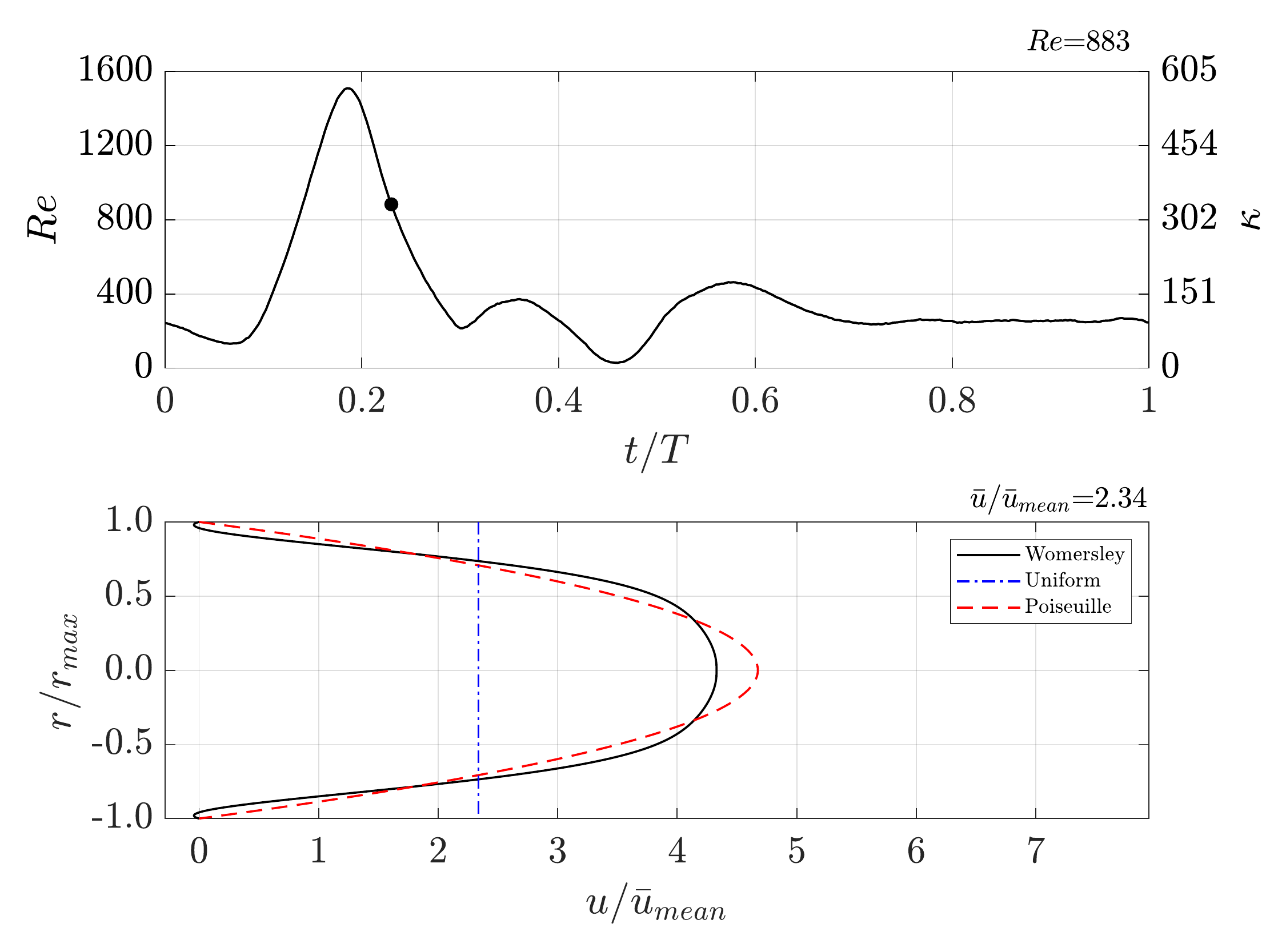}
            \label{f:t16_Re_waveform}
        }
        \subfloat[]{
            \includegraphics[width=0.44\textwidth,keepaspectratio]
            {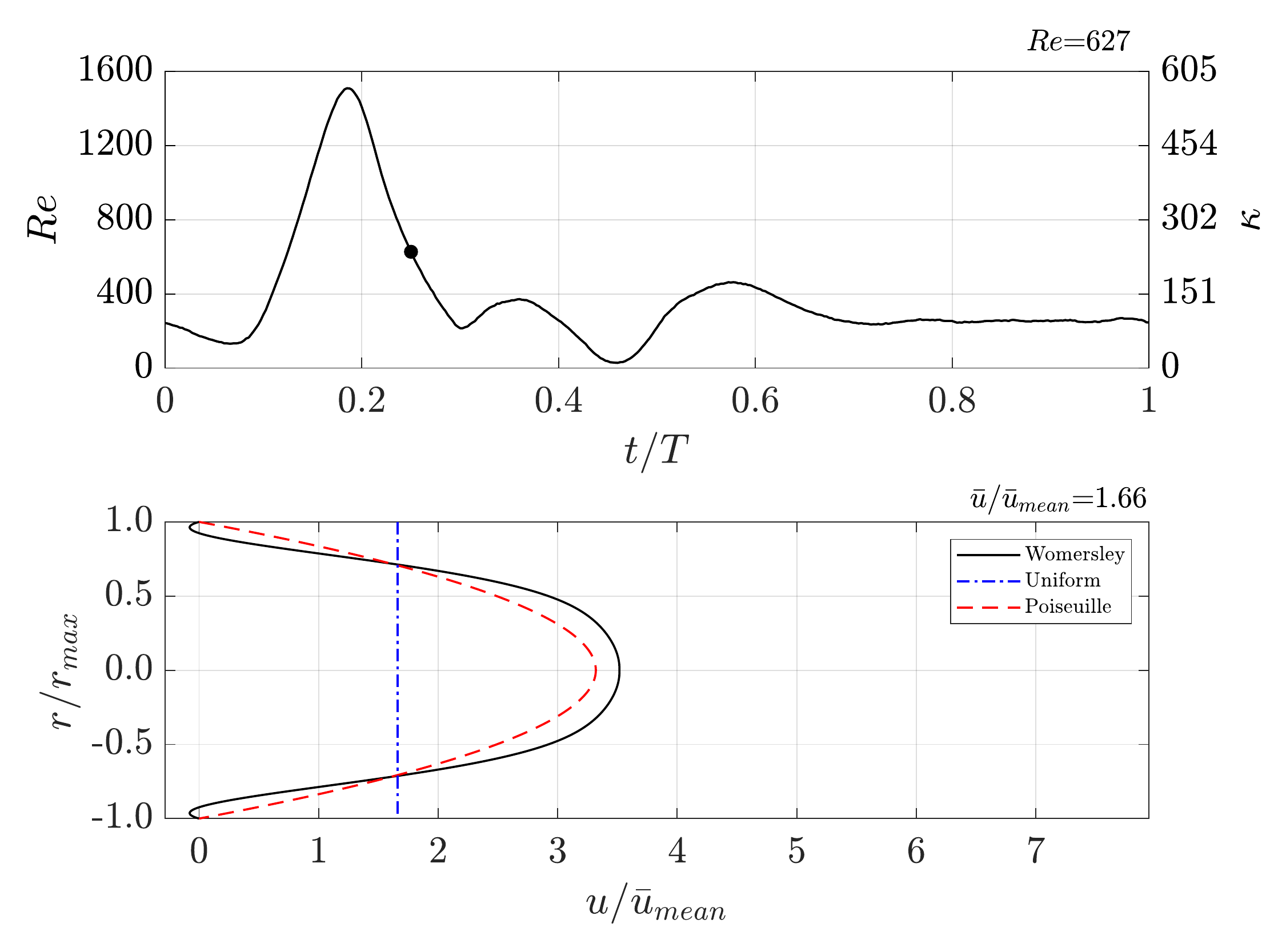}
            \label{f:t17_Re_waveform}
        }
        \\[-0.18in]
        \subfloat[]{
            \includegraphics[width=0.44\textwidth,keepaspectratio]
            {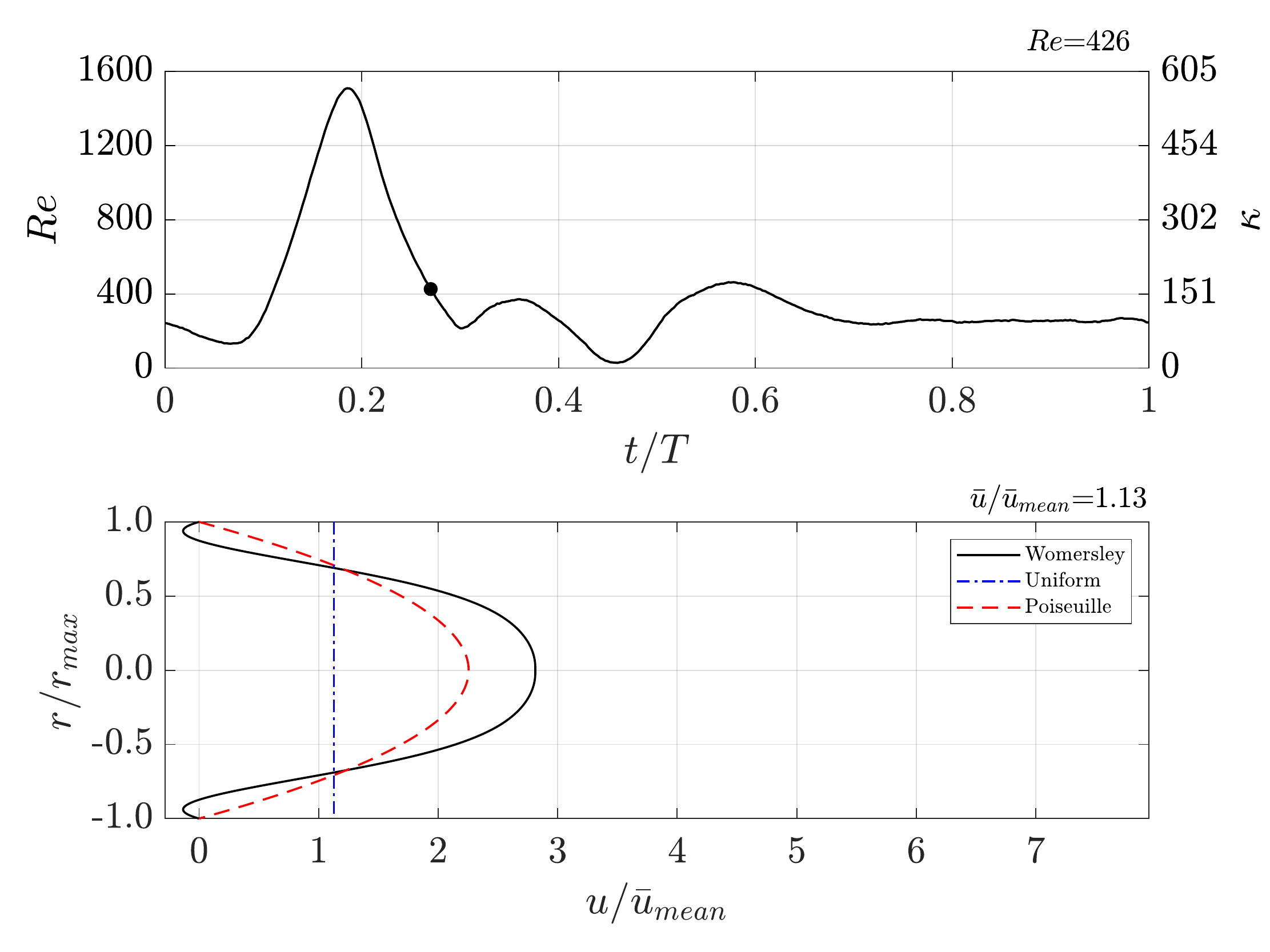}
            \label{f:t18_Re_waveform}
        }
        \subfloat[]{
            \includegraphics[width=0.44\textwidth,keepaspectratio]
            {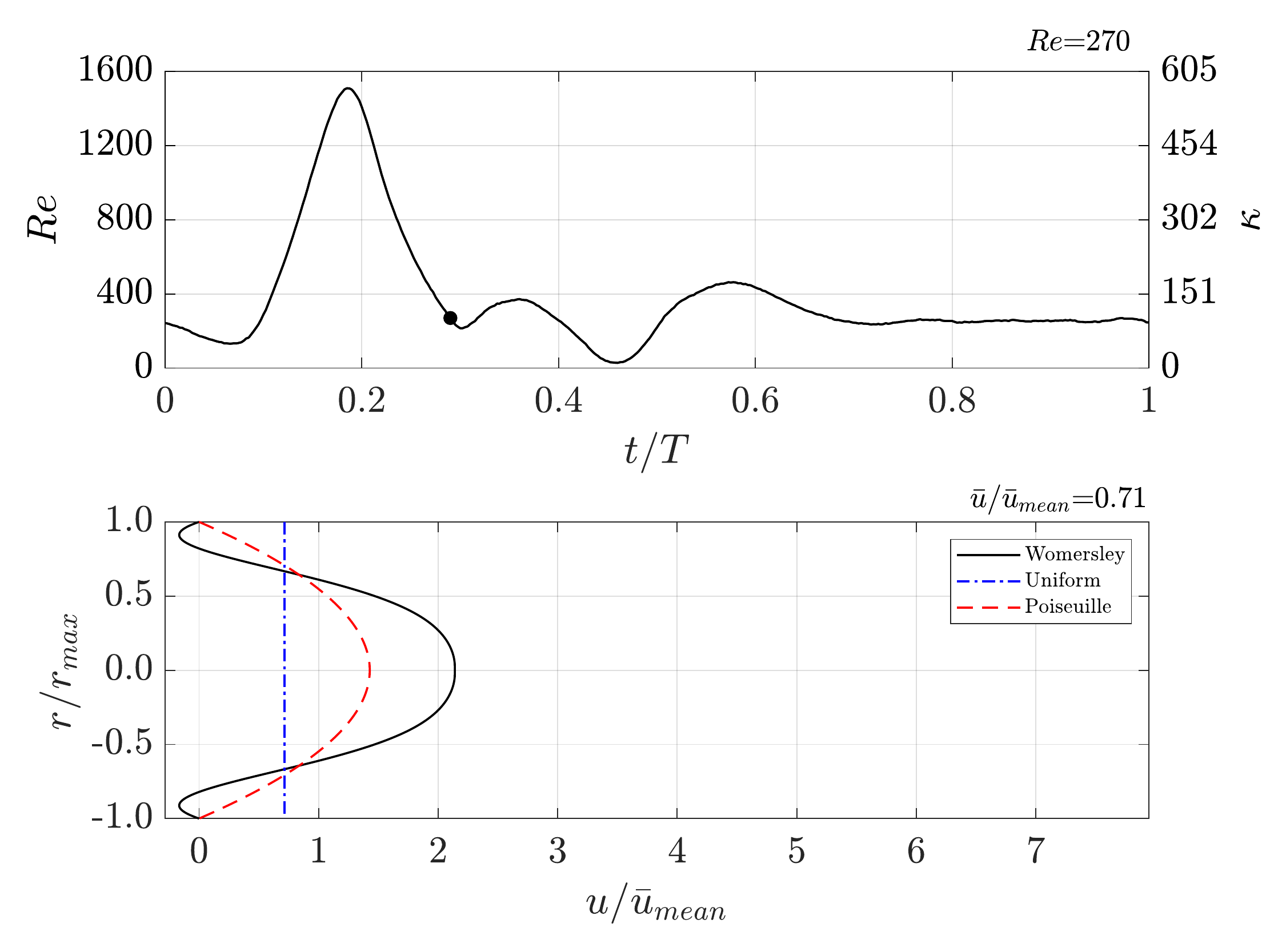}
            \label{f:t19_Re_waveform}
        }
        
        \caption{Nondimensional inlet velocity profile $u/\overbar{u}_{mean}$ for a Womersley entrance condition (black solid) and corresponding Reynolds number $Re$ and Dean number $\kappa$ at nondimensional pulse phase (a) $t^\star=0.19$, (b) $t^\star=0.21$, (c) $t^\star=0.23$, (d) $t^\star=0.25$, (e) $t^\star=0.27$, (f) $t^\star=0.29$. The nondimensional bulk velocity $\overbar{u}/\overbar{u}_{mean}$ corresponds to the uniform entrance condition (blue dash dot). For comparison, a parabolic velocity profile from a Poiseuille solution (red dash) required to produce the same bulk flow is plotted to emphasize the effect of pulsation on velocity profile shape under the current Womersley number $\alpha=4.22$.}
        \label{f:Re_waveform}
    \end{minipage}
\end{figure*}

\subsection{Boundary and initial conditions}
\label{s:curved_pipe_wec_uec}

To address the main research goals presented in Sec.~\ref{s:curved_pipe_introduction}, we perform numerical simulations of pulsatile flow through a curved pipe using two entrance conditions. First, we prescribe a fully developed pulsatile velocity profile to the inlet of the straight section of the pipe slightly upstream ($5d$) to the curvature entrance. This velocity profile is obtained from experimental particle image velocimetry (PIV) performed by Najjari and Plesniak~\cite{najjari-plesniak:2018} and is subsequently referred to as the {\it Womersley entrance condition} (WEC). Fully developed pulsatile flow occurs when the velocity profile does not change along the streamwise direction; the velocity can, however, be a function of both pipe radius and time. Within the curve, however, the flow is developing. The integrated velocity profile produces a flow rate that matches the experiment.
%

The second pulsatile entrance condition studied, motivated by the fact that flow upstream to a curved artery may not be fully developed,~\cite{ku:1997} is one where the flow is undeveloped (i.e. uniform). We refer to this inlet velocity condition as the {\it Uniform entrance condition} (UEC). Under UEC, the pulsatile flow rate is equivalent to that used under WEC. With all else being equal, this allows us to solely study the effect of flow development on the evolution of secondary flow patterns, vortical structures and wall shear stresses.

\subsubsection{Womersley entrance condition}
\label{s:wec}

The numerical solver requires that a velocity field $\bmm{u}$ be specified at the inlet boundary to the computational domain. Under WEC, this velocity field is extracted from PIV data captured upstream of the curvature~\cite{cox-najjari-plesniak:2019} in the straight section of the pipe where the flow is fully developed pulsatile flow, where the only non-zero component of the velocity field is the streamwise velocity. These PIV data were obtained over a discretized set of 100 phases over the entire waveform period. In order to assign the velocity at an arbitrary set of spatial coordinates, we fit a 10th order polynomial to the experimental velocity profile at each phase of the waveform and interpolate between phases to prescribe a velocity field that is a function of time. The initial velocity field at $t^\star=0$ is such that all radial and azimuthal components of velocity are zero and all streamwise components are prescribed the entrance condition throughout the entire geometry.
%

\subsubsection{Uniform entrance condition}
\label{s:uec}

The inlet boundary condition under UEC is applied in the same manner as in Sec.~\ref{s:wec}; however, the streamwise component of the velocity field is set equal to the bulk velocity $\overbar{u}=4Q / (\pi d^2)$ at each phase of the waveform, where $Q=Q(t)$ is the pulsatile flow rate.
%
%
We prescribe a plug profile to the initial velocity field at $t^\star=0$ such that all radial and azimuthal components of velocity are zero and all streamwise components are set equal to the bulk velocity of the pulsatile waveform.
%
%

For both entrance conditions, the initial pressure field is given a uniform value of zero. At the inlet boundary the pressure is extrapolated from the interior solution while the velocity is specified. At the outlet boundary far downstream of the pipe curve, the velocity is extrapolated from the interior solution. Since the incompressible solver permits negative pressure, the pressure at the outlet is given a uniform value of zero for all time; therefore, only relative pressure differences throughout the computational domain are important. Finally, we apply the no-slip boundary condition at the pipe wall. Under the given flow conditions with a high reduced velocity (see Sec.~\ref{s:cfd_geometry}), non-physical effects from the initial conditions disappear after three waveform cycles, at which point we achieve cycle-to-cycle convergence of the flow field. Successful convergence studies under the current flow conditions have been performed~\cite{cox-najjari-plesniak:2019} to conclude that the spatial discretization of the computational domain is sufficient and the numerical solution exhibits both grid and polynomial convergence. We note here that numerical results under the Womersley entrance condition have been thoroughly validated in a previous study by comparing velocity and vorticity fields and vortex trajectories to experimental results.~\cite{cox-najjari-plesniak:2019} The period of the waveform $T=4$~s and the kinematic viscosity of the fluid $\nu=3.5 \times 10^{-6}$~$\mathrm{m^2~s^{-1}}$ were set to match the experimental setup found therein. Although the density $\rho$ is not explicitly needed to run the current flow solver, relevant physiological values of the wall shear stress can be obtained by setting $\rho=1\,060$~$\mathrm{kg~m^{-3}}$---an approximate value for the density of blood plasma. For the given pipe diameter $d=0.0127$~m, the Womersley number is $\alpha=4.22$. 
%
%

\subsection{Pulsatile flow in a pipe}
\label{s:pulsatile_waveform}

\subsubsection{Pressure gradient}
\label{s:pressure_gradient}

It is a well known fact that pulsatile flow through a pipe exhibits a phase lag between flow rate and pressure gradient due to the inertia of the fluid, and that this lag increases with Womersley number, with flow rate ``lagging'' the pressure gradient. That is to say that for a given pipe diameter and fluid viscosity, the higher the waveform frequency $f$ and flow rate $Q$ the more out-of-phase that flow rate and pressure gradient become. Following the description given by Zamir,~\cite{zamir:2000} under pulsatile flow in a straight pipe the flow rate increases gradually as the pressure gradient rises to its peak, with the peak velocity being smaller than that obtained under Poiseuille flow using a constant pressure gradient equal to the peak pressure gradient reached under the pulsatile conditions. This loss in peak flow rate increases with higher Womersley number, such that at very high frequency the fluid hardly moves. On the other hand, the phase lag between flow rate and pressure gradient decreases as the pulsatile frequency decreases, i.e. flow rate becomes more in-phase with pressure gradient if the change in pressure is slow. As the frequency approaches zero, the solution at each phase in a straight pipe approaches a steady Poiseuille flow solution. This is referred to as the quasi-steady Poiseuille solution or ``pulsatile Poiseuille flow.''~\cite{zamir:2015} In a curved pipe, as $f \rightarrow 0$ the solution at each phase approaches the steady state solution for a given Dean number under a Poiseuille entrance condition.

Figure~\ref{f:pressure_gradient_lead_flowrate} demonstrates the aforementioned phase lag between flow rate and nondimensional pressure gradient along the streamwise direction $\textnormal{-}\partial p^\star/\partial s^\star$ under the current Womersley entrance flow condition to the curved pipe. The pressure gradient shown here is computed by our solver using the flux reconstruction methodology discussed in Sec.~\ref{s:numerical_scheme} and extracted from the straight section upstream of the curve. For comparison, we superimpose the pressure gradient from a Poiseuille flow required to produce the same flow rate. This plot shows that $\textnormal{-}\partial p^\star/\partial s^\star$ is positive during flow rate acceleration and negative during deceleration for the current Womersley number $\alpha=4.22$. During mid-acceleration at $t^\star=0.14$ where the flow rate is $Q=31.3~\mathrm{mL~s^{-1}}$, the ratio of pulsatile pressure gradient to a pressure gradient from Poiseuille flow is $\big(\partial p^\star/\partial s^\star\rvert_{W}\big) / \big(\partial p^\star/\partial s^\star\rvert_{P}\big) \approx 9$. Therefore, under the current geometric and pulsatile flow conditions, the pressure gradient at this phase is nine times larger than the pressure gradient required under Poiseuille flow to produce the same flow rate. During mid-deceleration at $t^\star=0.23$, where the flow rate matches that at $t^\star=0.14$, the ratio is $\big(\partial p^\star/\partial s^\star\rvert_{W}\big) / \big(\partial p^\star/\partial s^\star\rvert_{P}\big) \approx -6$, which is 66\% of the magnitude and opposite in sign to the acceleration side.
%
%
Under the current flow conditions, the maximum pressure gradient leading the maximum flow rate of $Q_{max}=53.2~\mathrm{mL~s^{-1}}$ is 1.90. Under a pulsatile Poiseuille flow, the pressure gradient required to produce the same maximum flow rate is 0.34, which is 18\% of that which is needed under the current Womersley number. This description of the pressure gradient and flow rate provides perspective, in relation to Poiseuille flow, on the amount of ``lag'' in the flow rate.

\subsubsection{Inlet velocity profile}
\label{s:inlet_vel_profile}

Figure~\ref{f:Re_waveform} plots velocity profiles under WEC and UEC throughout all of deceleration $0.19 \leq t^\star \leq 0.29$, as well as the bulk velocity $\overbar{u}$ and velocity profile from a Poiseuille solution computed from the same bulk flow value. For a given phase, the parabolic Poiseuille profile corresponds to the Poiseuille pressure gradient shown in Fig.~\ref{f:pressure_gradient_lead_flowrate}. At peak flow rate near $t^\star=0.19$ ($Re=1501$, $\kappa=567$), the ratio of bulk velocity to waveform mean velocity is $\overbar{u} / \overbar{u}_{mean} \approx 4$, and the maximum velocity at the centerline does not reach the maximum of the Poiseuille flow profile due to fluid inertia resulting from the flow rate lagging the pressure gradient. There is a phase during deceleration at which the maximum velocity at the $r/r_{max}=0$ centerline is approximately equal to the Poiseuille flow solution---this occurs at approximately $t^\star=0.24$. At this phase ($Re=745$, $\kappa=282$), the bulk velocity to waveform mean velocity ratio is $\overbar{u} / \overbar{u}_{mean} \approx 2$ and the maximum velocity to mean velocity ratio is $u_{max} / \overbar{u}_{mean} \approx 4$. This means that $u_{max} / \overbar{u} \approx 2$, which matches the result from a Poiseuille solution. Throughout the remainder of the deceleration phase, the maximum velocity exceeds that of the Poiseuille solution. These plots demonstrate the effect of inertia inherent in the pulsatile waveform at $\alpha=4.22$ compared to a pulsatile Poiseuille solution, for which the pressure gradient and flow rate are in-phase. Note that a small region of reverse flow in the velocity profile near the pipe wall begins at $t^\star=0.23$, and that this region grows as the flow decelerates further. Reverse flow, or lack thereof, plays a vital role in the distribution of the wall shear stress during deceleration, especially along the inner wall of the curved artery model. This fact is highlighted in Sec.~\ref{s:wss} in the side-by-side comparison of results between WEC and UEC.

\section{RESULTS AND DISCUSSION}
\label{s:results}

\begin{figure*}[t]
    \captionsetup[subfigure]{labelformat=parens}
    \renewcommand{\fsize}{54mm}
    \centering\setcounter{subfigure}{0}
    \subfloat[]{
        \includegraphics[width=\fsize,height=\fsize,keepaspectratio]
        {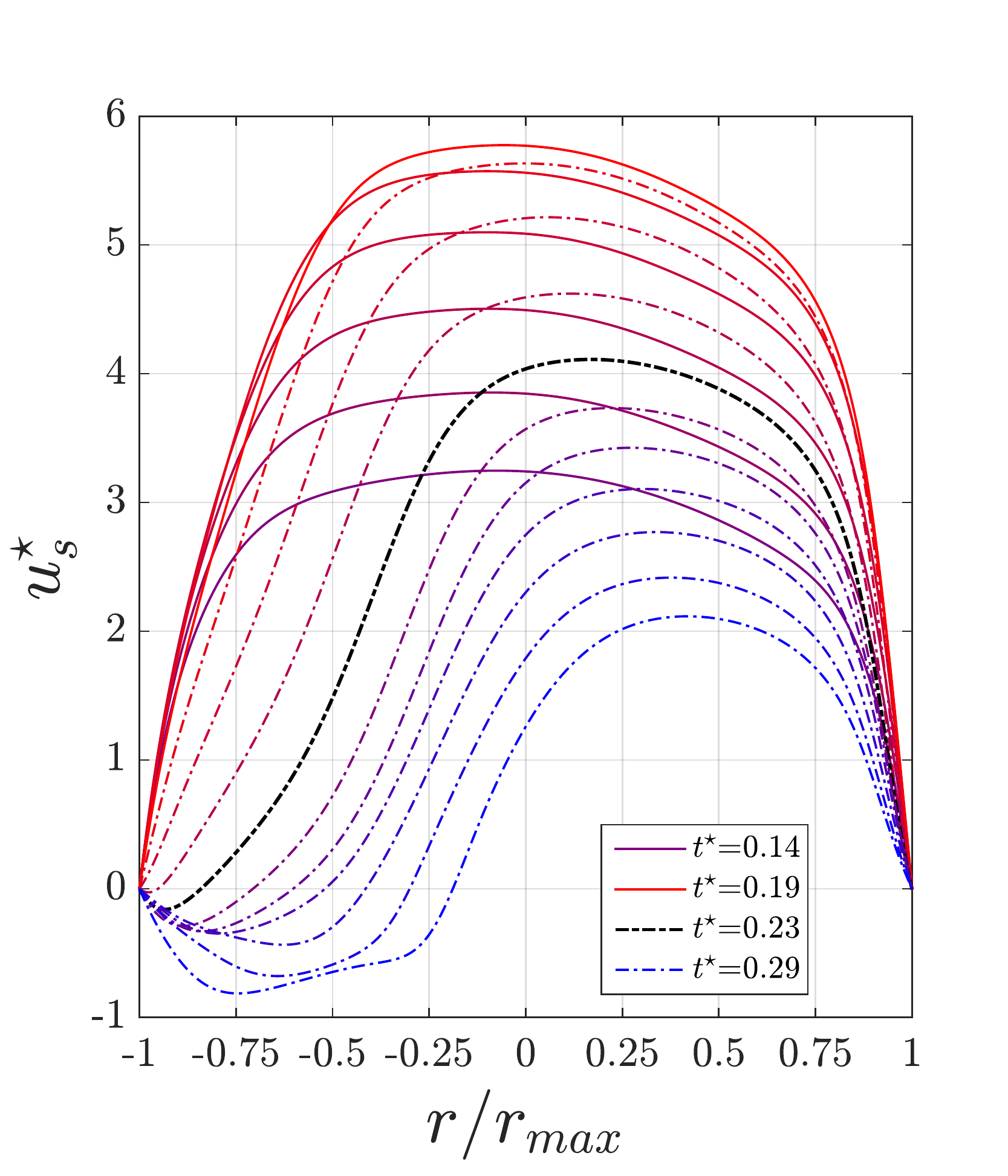}
        \label{f:us_profile_22_womersley_z0}
    }\hspace{-0.25in}
    \subfloat[]{
        \includegraphics[width=\fsize,height=\fsize,keepaspectratio]
        {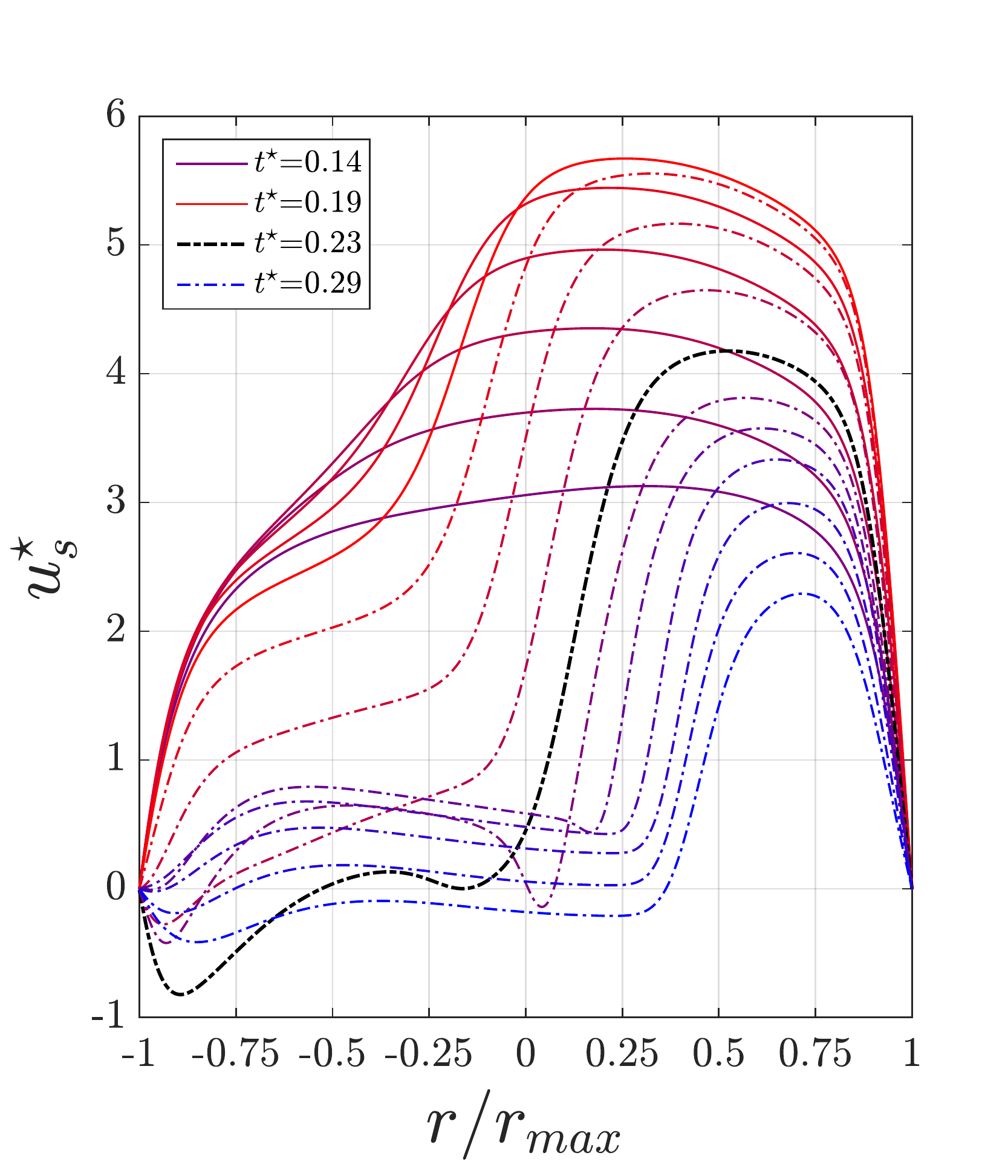}
        \label{f:us_profile_45_womersley_z0}
    }\hspace{-0.25in}
    \subfloat[]{
        \includegraphics[width=\fsize,height=\fsize,keepaspectratio]
        {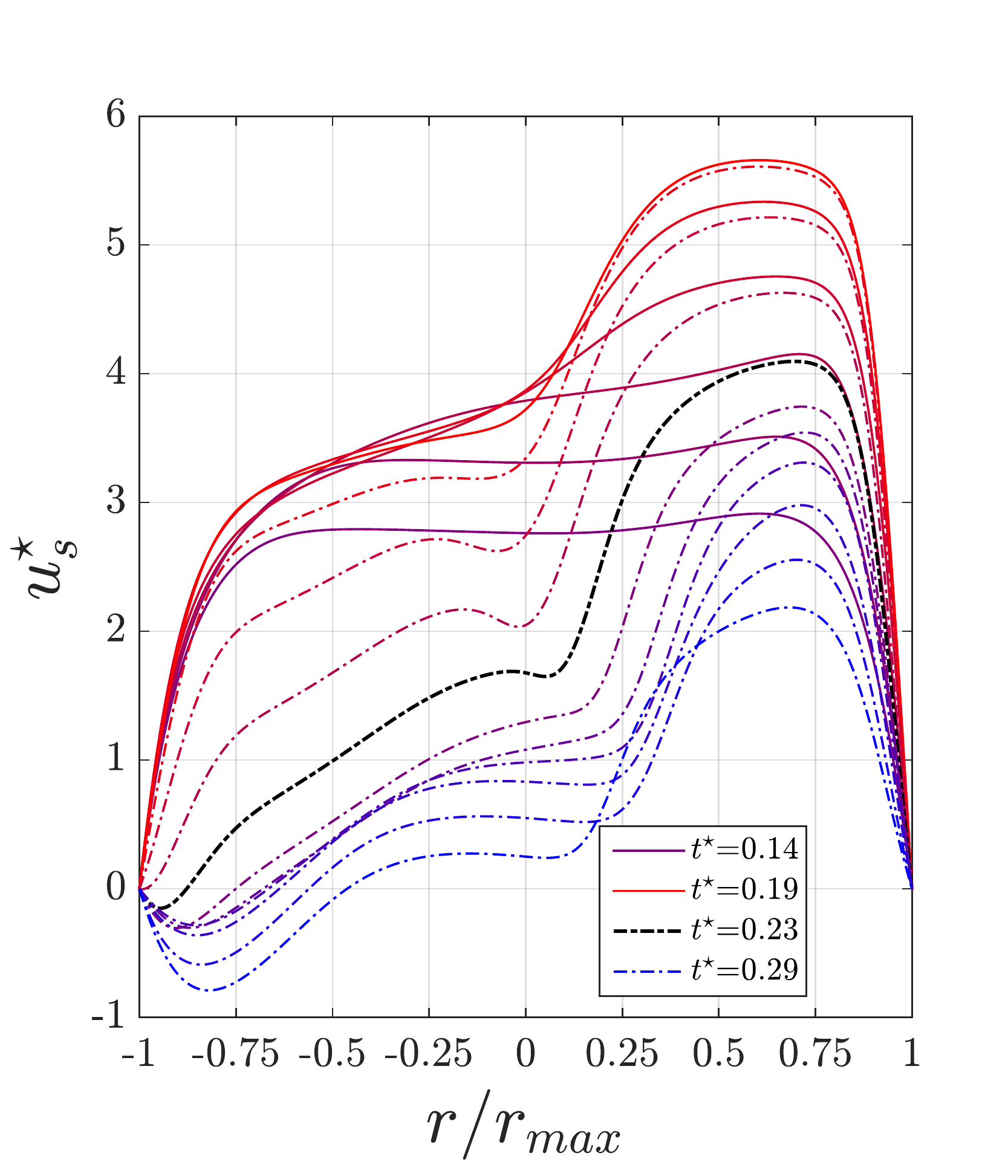}
        \label{f:us_profile_90_womersley_z0}
    }\hspace{-0.25in}
    \subfloat[]{
        \includegraphics[width=\fsize,height=\fsize,keepaspectratio]
        {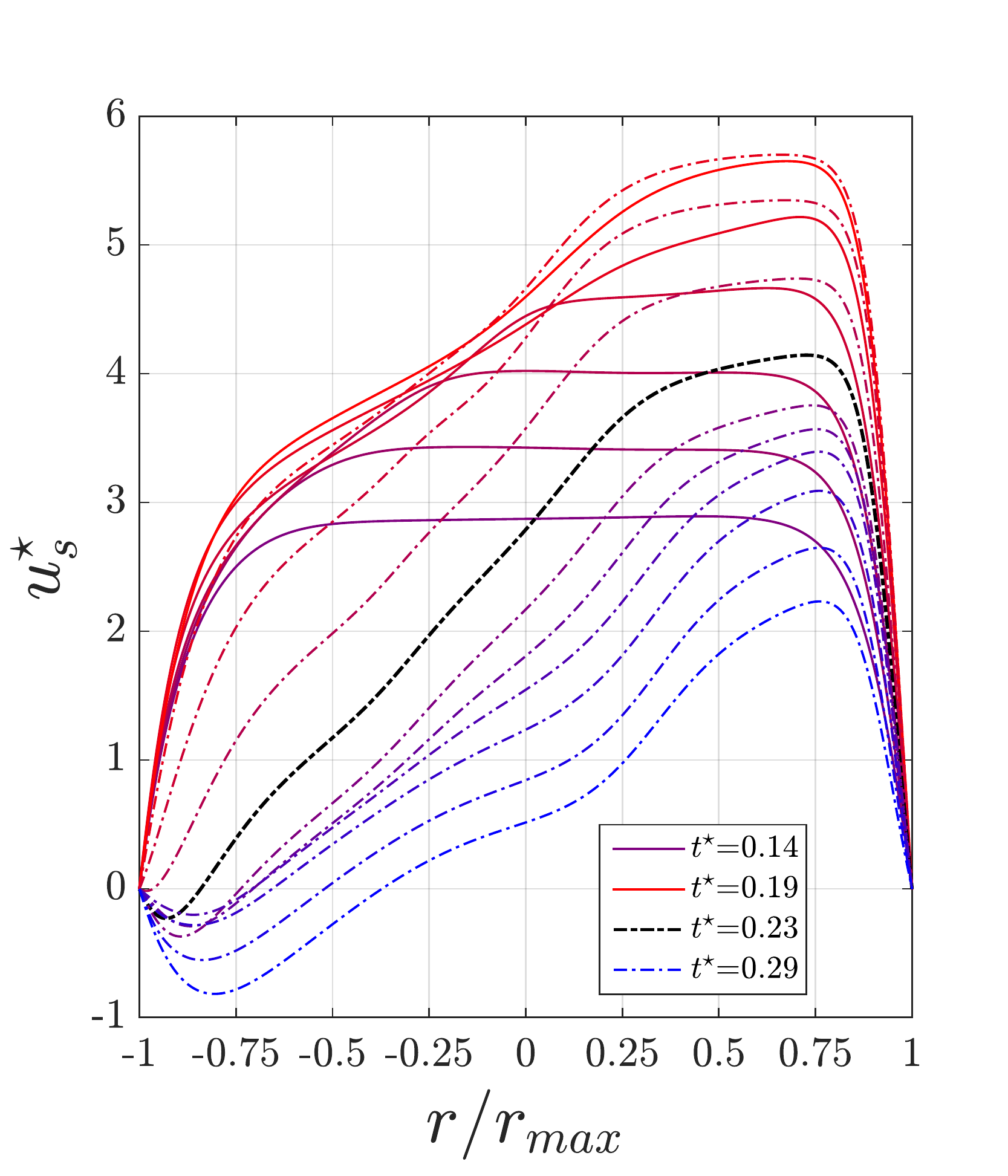}
        \label{f:us_profile_135_womersley_z0}
    }
    \\[-0.15in]
    \subfloat[]{
        \includegraphics[width=\fsize,height=\fsize,keepaspectratio]
        {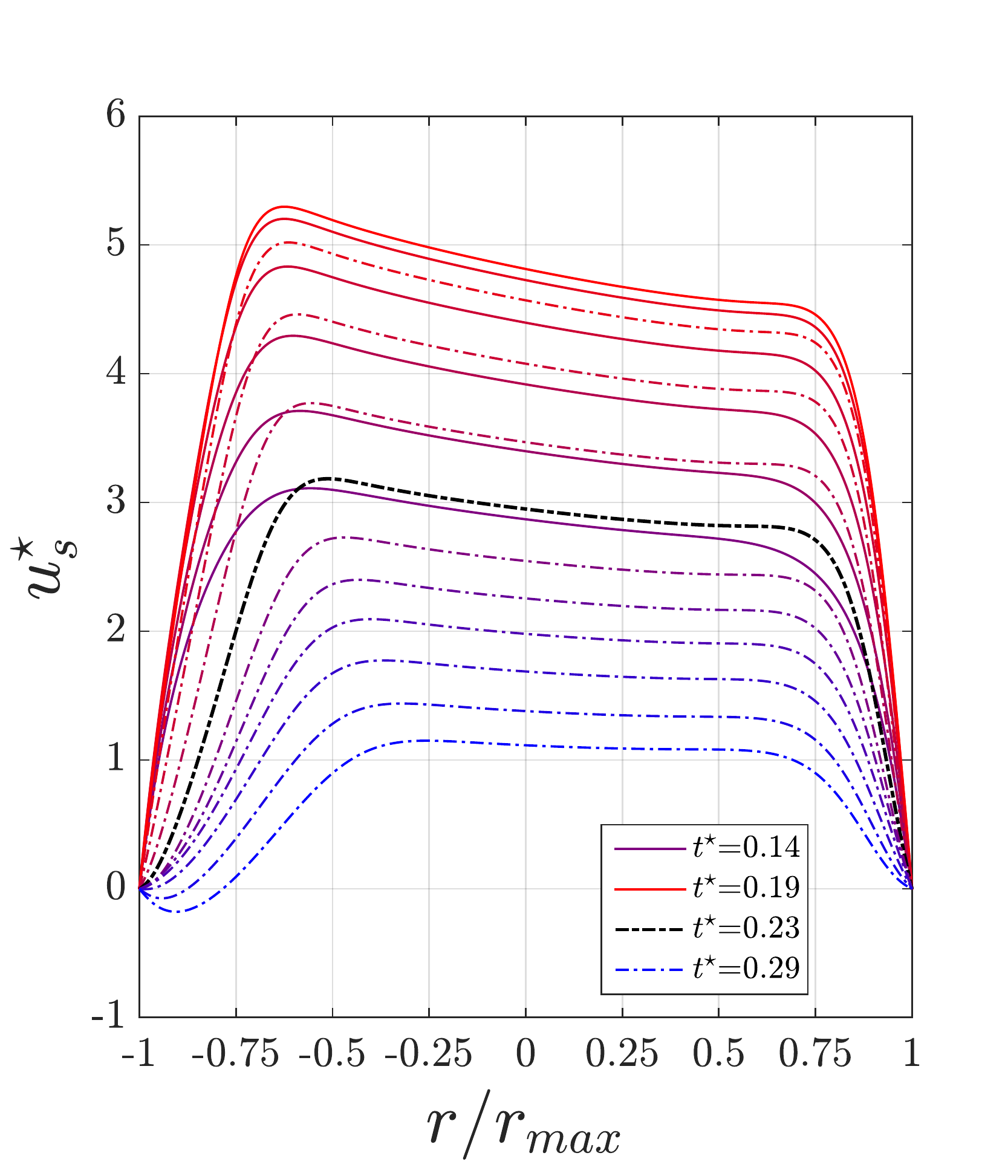}
        \label{f:us_profile_22_plug_z0}
    }\hspace{-0.25in}
    \subfloat[]{
        \includegraphics[width=\fsize,height=\fsize,keepaspectratio]
        {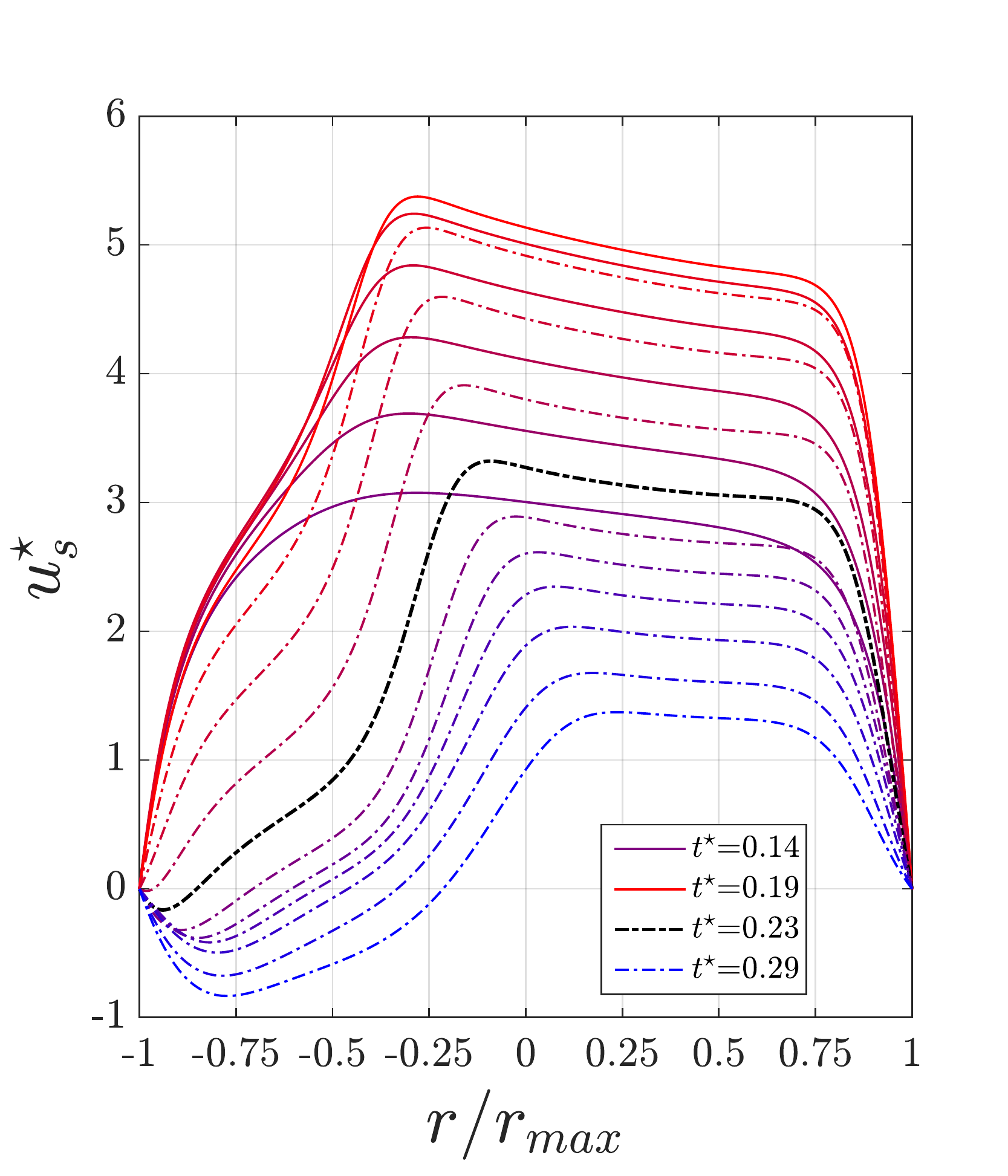}
        \label{f:us_profile_45_plug_z0}
    }\hspace{-0.25in}
    \subfloat[]{
        \includegraphics[width=\fsize,height=\fsize,keepaspectratio]
        {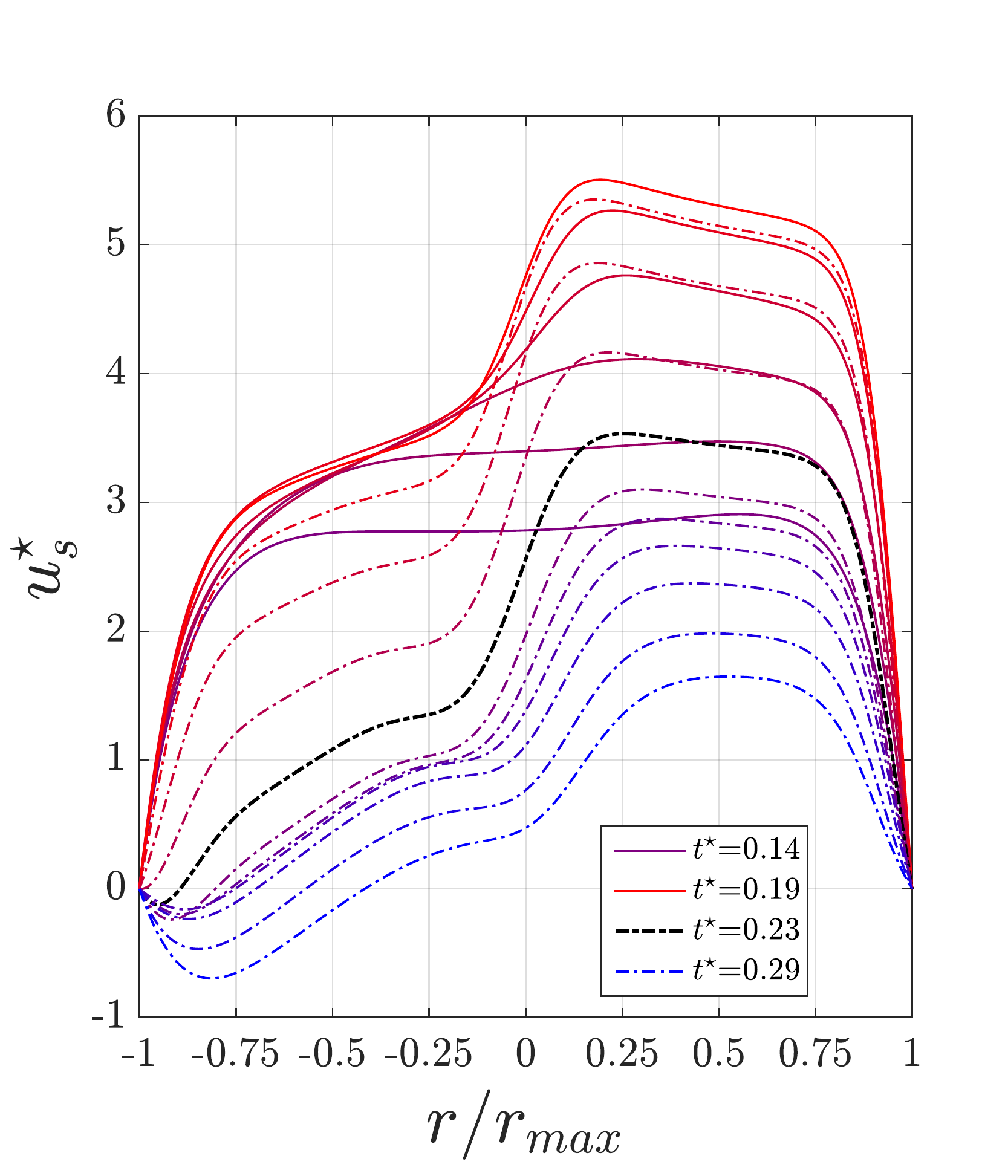}
        \label{f:us_profile_90_plug_z0}
    }\hspace{-0.25in}
    \subfloat[]{
        \includegraphics[width=\fsize,height=\fsize,keepaspectratio]
        {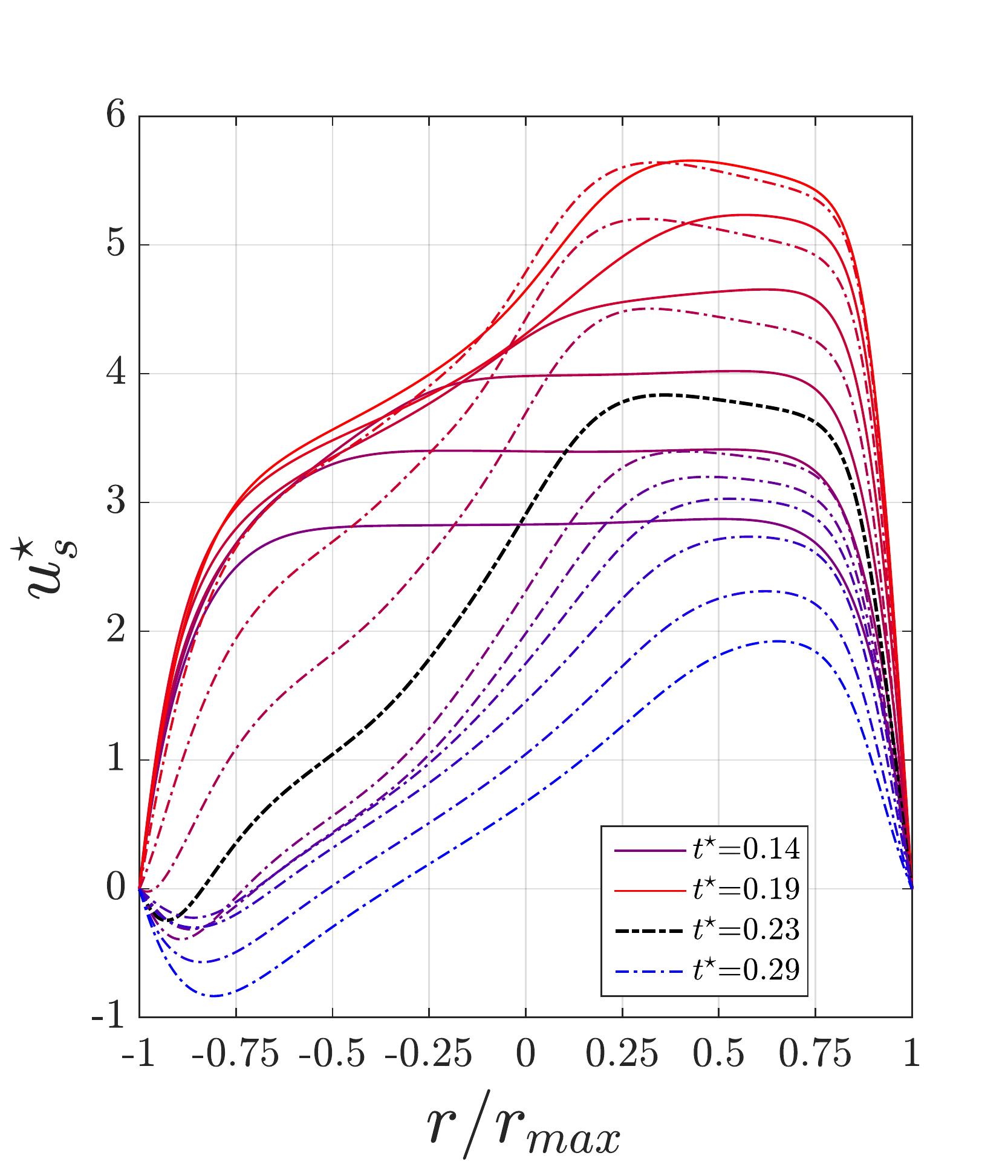}
        \label{f:us_profile_135_plug_z0}
    }
    
    \caption{Womersley (\textit{top}) and Uniform (\textit{bottom}) entrance conditions: profiles of nondimensional streamwise velocity, $u^\star_s$, at $z=0$ plane of symmetry for (a),(e) $\theta=22^\circ$ (b),(f) $\theta=45^\circ$ (c),(g) $\theta=90^\circ$ (d),(h) $\theta=135^\circ$ and $0.14 \leq t^\star \leq 0.29$ showing flow acceleration (solid) and deceleration (dash-dot). Under both entrance conditions, velocity profiles are skewed towards the inner wall at $\phi=22^\circ$ and towards the outer wall as the bulk flow moves downstream and the flow decelerates. Under the uniform entrance condition, the profiles are generally much less skewed towards the outer wall since the flow is less developed near the curvature entrance, which results in more flattened profiles. At $t^\star=0.23$ (black dash-dot), the peak velocity under a uniform condition is smaller than that under a fully developed condition. Ultimately, this lack of outer wall skewness and maximum velocity reduce the intensity of Dean and Lyne-type vortical structures.}
    \label{f:us_profile_plug_z0}
\end{figure*}

\subsection{Flow description}
\label{s:flow_description}

A description of the flow field is provided in this section with particular emphasis on the phase $t^\star=0.23$, where multiple
counter-rotating vortices coexist at the $\phi = 90^\circ$ location. This phase occurs within the deceleration portion of the physiological waveform, during which time transitional secondary flow morphologies of interest~\cite{boiron-deplano-pelissier:2007} emerge. Results of nondimensional streamwise velocity $u^\star_s = u_s / \overbar{u}_{mean}$, secondary velocity $\bmm{u}^\star_{\theta r} = \bmm{u}_{\theta r} / \overbar{u}_{mean}$, pressure $p^\star = P / (\rho \overbar{u}_{mean}^2)$, vorticity $\bmm{\omega}^\star = \bmm{\omega} d / \overbar{u}_{mean}$, vortex identification criterion $\lambda^\star_2 = \lambda_2 d^2 / \overbar{u}_{mean}^2$, and magnitude of the wall shear stress vector $\bmm{\tau}^{\star}_w = \bmm{\tau}_w d / (\mu \overbar{u}_{mean})$ are presented and appropriate comparisons drawn between the fully developed and uniform entrance conditions.

\subsubsection{Primary flow}
\label{s:pulsatile_primary_flow}

Various profiles of streamwise velocity $u^\star_s$ under WEC and UEC during flow acceleration and deceleration for which $0.14 \leq t^\star \leq 0.29$ are provided in Fig.~\ref{f:us_profile_plug_z0}. The velocity profiles are plotted at the $z=0$ plane of symmetry for $\phi\in\{22^\circ$, $45^\circ$, $90^\circ$, $135^\circ\}$. In these figures, flow rate acceleration is denoted by solid lines and deceleration is denoted by dashed lines, with the dashed black line highlighting the specific phase $t^\star=0.23$. This phase is interesting because it represents the point in time where multiple vortical structures coexist at $\phi=90^\circ$. Both sets of profiles exhibit skewness towards the inner wall where $r/r_{max}=-1$ ($r_{max}=d/2$) at $\phi=22^\circ$ during acceleration; this inner wall skewness is more pronounced under UEC. Compared to WEC, we can observe different profile shapes under UEC in the first half of the curve, such as a top-hat shape and inner wall skewness at $\phi=45^\circ$. Under WEC, profile flattening in the core and a double peak can be seen at $\phi=45^\circ$ and $t^\star=0.24$, where a local minimum in the streamwise velocity occurs at the pipe center near $r=0$. This flattening in the core corresponds to strong vortical activity where a majority of fluid motion is in the cross-sectional plane. The reduced peak under UEC at $t^\star=0.24$ produces a smaller centrifugal force and, therefore, a smaller pressure gradient along the radius of curvature. This leads to reduced secondary flow (see Sec.~\ref{s:pulsatile_secondary_flow}) and weaker vortical structures in the core (see Sec.~\ref{s:pulsatile_vortex_id}).

The profile under UEC shows reverse flow occurring along the inner wall at the same phases as WEC for $\phi\in\{45^\circ$, $90^\circ$, $135^\circ\}$, even though no reverse flow is prescribed under a pulsatile uniform entrance condition. At $\phi=22^\circ$, flow reversal does not occur, however, until the end of deceleration where $t^\star > 0.27$. As the fluid moves through the curve and the flow develops, the velocity profiles between UEC and WEC become similar, with both sets of profiles exhibiting skewness towards the outer wall. The fact that skewness in the velocity profile shifts from the inner wall near the entrance to the outer wall as the fluid moves through the curve was predicted by Singh~\cite{singh:1974} in his analysis of steady flow development in a curved tube near the entrance. He showed that initial flow development is influenced by both geometry and entry condition, and that the cross-over in maximum wall shear stress from the inner to outer wall occurs at different toroidal locations downstream of the entrance depending on the entry condition (see Sec.~\ref{s:wss}). Our results confirm Singh's prediction of this cross-over effect.

\renewcommand{\ftype}{womersley}
\renewcommand{\fres}{N5M30_slices_1-29_150dpi}
\renewcommand{\slicetype}{slice}
\renewcommand{\fsize}{27.25mm}

\forloop{il}{2}{\value{il} < 3}{
    
    \ifthenelse{ \equal{\value{il}}{2} }{
        \renewcommand{\fvardir}{uplanar}
        \renewcommand{\fvar}{uplanar}
        \renewcommand{\filetype}{png}
        \renewcommand{\fvartext}{nondimensional secondary velocity magnitude}
        \renewcommand{\fvarlatex}{\lvert \bmm{u}^{\star}_{\theta r} \rvert}
    }{}
    
    \ifthenelse{ \equal{\value{il}}{3} }{
        \renewcommand{\fvardir}{uplanar_vector_old} 
        \renewcommand{\fvar}{uplanar_vector}
        \renewcommand{\filetype}{png}
        \renewcommand{\fvartext}{nondimensional secondary velocity vectors}
        \renewcommand{\fvarlatex}{\bmm{u}^{\star}_{\theta r}}
    }{}
    
    \renewcommand{\dt}{0.020}
    \renewcommand{\ftime}{0.170}
    
    \begin{figure*}[t]
        \captionsetup[subfigure]{labelformat=empty}
        \centering\setcounter{subfigure}{0}
        
        \forloop{jl}{1}{\value{jl} < 6}{
            
            \FPeval{\result}{trunc(\ftime+\dt:3)}
            \FPeval{\myresult}{trunc(\ftime+\dt:2)}
            \renewcommand{\ftime}{\myresult}
            
            \renewcommand{\slice}{22}
            \renewcommand{\fdir}{\ftype/\fres/\fvardir/\slice}
            \subfloat[$(\myresult$, $\slice^\circ)$]{
                \includegraphics[height=\fsize,keepaspectratio]
                    {./figures/wec/\fvar_\slice_\slicetype_\result.\filetype}
            }
            \renewcommand{\slice}{45}
            \renewcommand{\fdir}{\ftype/\fres/\fvardir/\slice}
            \subfloat[$(\myresult$, $\slice^\circ)$]{
                \includegraphics[height=\fsize,keepaspectratio]
                    {./figures/wec/\fvar_\slice_\slicetype_\result.\filetype}
            }
            \renewcommand{\slice}{90}
            \renewcommand{\fdir}{\ftype/\fres/\fvardir/\slice}
            \subfloat[$(\myresult$, $\slice^\circ)$]{
                \includegraphics[height=\fsize,keepaspectratio]
                    {./figures/wec/\fvar_\slice_\slicetype_\result.\filetype}
            }
            \renewcommand{\slice}{135}
            \renewcommand{\fdir}{\ftype/\fres/\fvardir/\slice}
            \subfloat[$(\myresult$, $\slice^\circ)$]{
                \includegraphics[height=\fsize,keepaspectratio]
                    {./figures/wec/\fvar_\slice_\slicetype_\result.\filetype}
            }
        }
        
        \caption{Womersley entrance condition: \fvartext~$\fvarlatex$ at $(t^\star$,$\phi)$. Different secondary flow patterns emerge along the curved cross-sections due to varying degrees of flow development, centrifugal forcing and flow reversal. At $(t^\star,\phi)=(0.25,45^\circ)$, increased secondary flow in the core is due to large velocities from the upper and lower half of the pipe competing at the plane of symmetry, causing an outward jet-like motion. At $(t^\star,\phi)=(0.23,90^\circ)$, two intense, circular secondary flow patterns occur---one above and one below the plane of symmetry---which correspond to the deformed Dean vortex pair.}
        \label{f:womersley_uplanar}
    \end{figure*}
}
\renewcommand{\ftype}{plug}
\renewcommand{\fres}{N5M30_slices_1-29_150dpi}
\renewcommand{\slicetype}{slice}
\renewcommand{\fsize}{27.25mm}

\forloop{il}{2}{\value{il} < 3}{
    
    \ifthenelse{ \equal{\value{il}}{2} }{
        \renewcommand{\fvardir}{uplanar}
        \renewcommand{\fvar}{uplanar}
        \renewcommand{\filetype}{png}
        \renewcommand{\fvartext}{nondimensional secondary velocity magnitude}
        \renewcommand{\fvarlatex}{\lvert \bmm{u}^{\star}_{\theta r} \rvert}
    }{}
    
    \ifthenelse{ \equal{\value{il}}{3} }{
        \renewcommand{\fvardir}{uplanar_vector_old} 
        \renewcommand{\fvar}{uplanar_vector}
        \renewcommand{\filetype}{png}
        \renewcommand{\fvartext}{nondimensional secondary velocity vectors}
        \renewcommand{\fvarlatex}{\bmm{u}^{\star}_{\theta r}}
    }{}
    
    \renewcommand{\dt}{0.020}
    \renewcommand{\ftime}{0.170}
    
    \begin{figure*}[t]
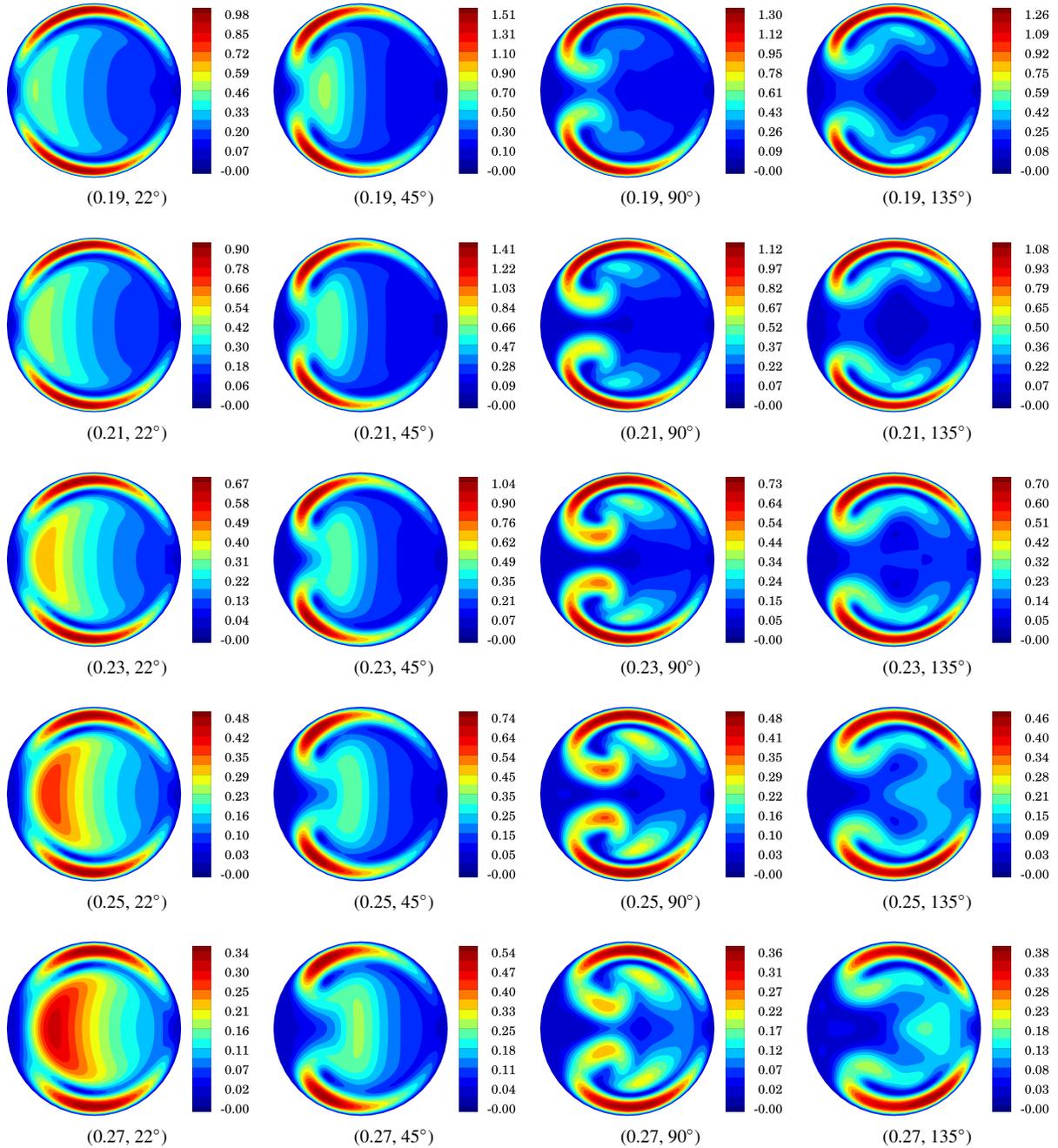

        \captionsetup[subfigure]{labelformat=empty}
        \centering\setcounter{subfigure}{0}
        
        \forloop{jl}{1}{\value{jl} < 6}{
            
            \FPeval{\result}{trunc(\ftime+\dt:3)}
            \FPeval{\myresult}{trunc(\ftime+\dt:2)}
            \renewcommand{\ftime}{\myresult}
            
            \renewcommand{\slice}{22}
            \renewcommand{\fdir}{\ftype/\fres/\fvardir/\slice}
            \subfloat[$(\myresult$, $\slice^\circ)$]{
                \includegraphics[height=\fsize,keepaspectratio]
                    {./figures/uec/\fvar_\slice_\slicetype_\result.\filetype}
            }
            \renewcommand{\slice}{45}
            \renewcommand{\fdir}{\ftype/\fres/\fvardir/\slice}
            \subfloat[$(\myresult$, $\slice^\circ)$]{
                \includegraphics[height=\fsize,keepaspectratio]
                    {./figures/uec/\fvar_\slice_\slicetype_\result.\filetype}
            }
            \renewcommand{\slice}{90}
            \renewcommand{\fdir}{\ftype/\fres/\fvardir/\slice}
            \subfloat[$(\myresult$, $\slice^\circ)$]{
                \includegraphics[height=\fsize,keepaspectratio]
                    {./figures/uec/\fvar_\slice_\slicetype_\result.\filetype}
            }
            \renewcommand{\slice}{135}
            \renewcommand{\fdir}{\ftype/\fres/\fvardir/\slice}
            \subfloat[$(\myresult$, $\slice^\circ)$]{
                \includegraphics[height=\fsize,keepaspectratio]
                    {./figures/uec/\fvar_\slice_\slicetype_\result.\filetype}
            }
        }
        
        \caption{Uniform entrance condition: \fvartext~$\fvarlatex$ at $(t^\star$,$\phi)$. Different secondary flow patterns emerge along the curved cross-sections due to varying degrees of flow development, centrifugal forcing and flow reversal, although the lack of flow development at the entrance to the curvature inhibits secondary motion in the core and growth of interior flow vortices (see Sec.~\ref{s:pulsatile_vortex_id}).}
        \label{f:uniform_uplanar}
    \end{figure*}
}

\subsubsection{Secondary flow}
\label{s:pulsatile_secondary_flow}

\begin{figure*}[t]
    \captionsetup[subfigure]{labelformat=parens}
    \renewcommand{\fsize}{65mm}
    \centering\setcounter{subfigure}{0}
    \subfloat[]{
        \includegraphics[height=\fsize,keepaspectratio]
        {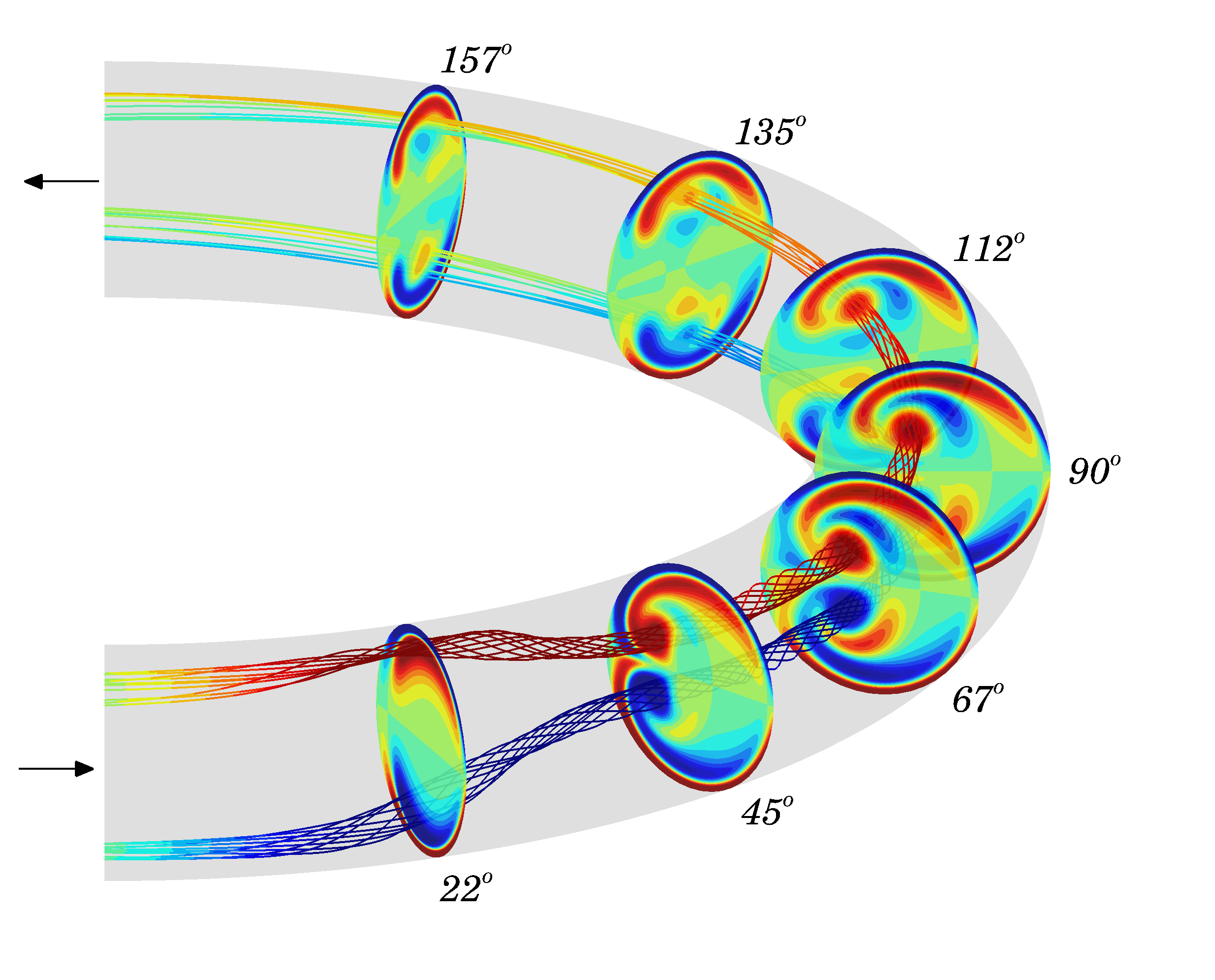}
    }
    \subfloat[]{
        \includegraphics[height=\fsize,keepaspectratio]
        {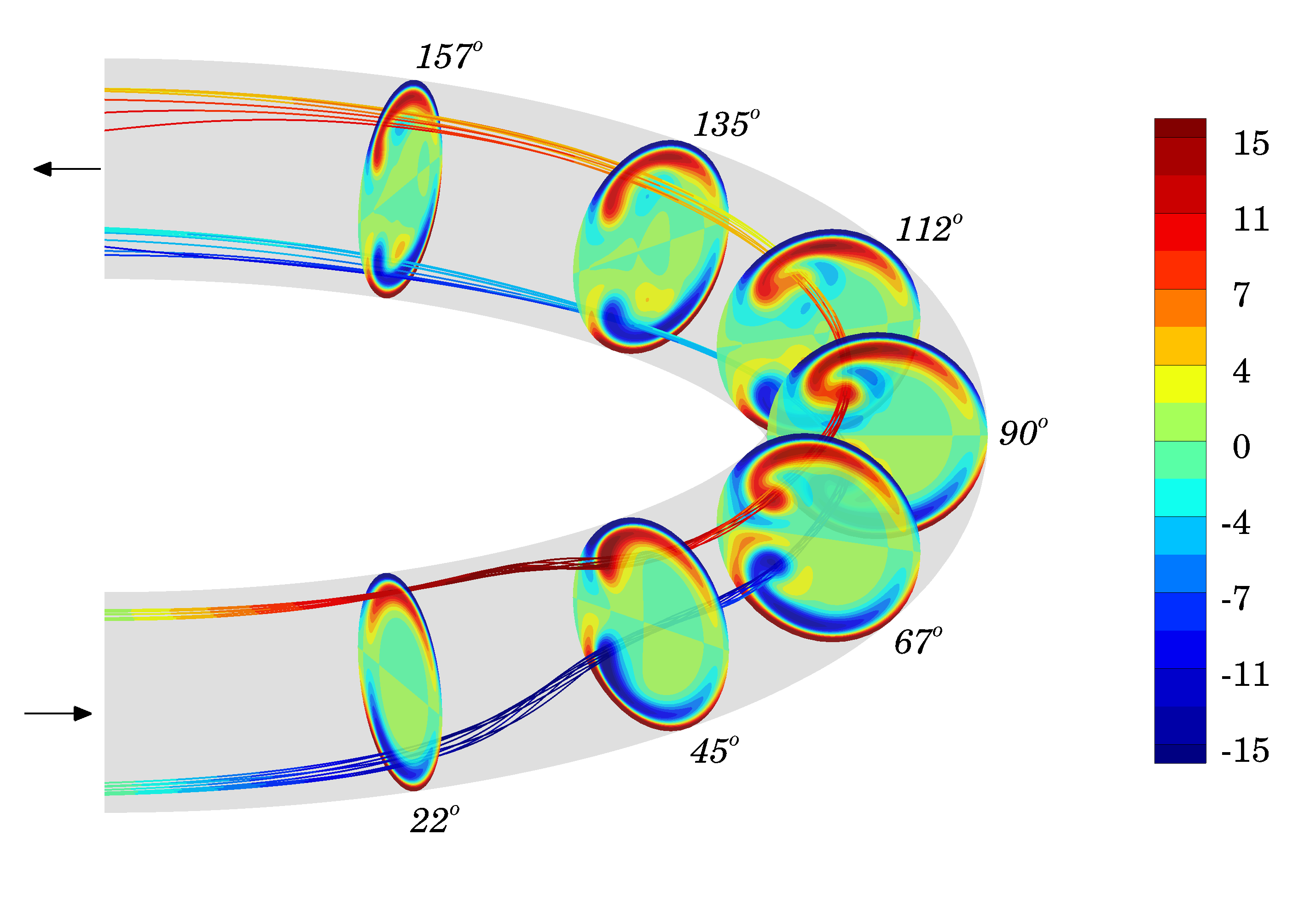}
    }
    
    \caption{Womersley and Uniform entrance conditions: streamlines associated with the ``head'' of deformed Dean colored by nondimensional streamwise vorticity $\omega_s^\star$ at mid-deceleration phase $t^\star=0.23$, highlighting the helical nature of streamlines under (a) WEC compared to the more straight streamlines under (b) UEC. Helical streamlines result from secondary flow, which is stronger under a fully developed pulsatile entrance condition. In turn, this secondary flow produces a larger streamwise vorticity and more intense vortical structures. Stronger secondary flow also improves convection of $z$ vorticity (not shown) away from the inner wall and into the core region of the flow, ultimately intensifying the Lyne-type vortex shown in (a) at $\phi \in \{67^\circ,90^\circ,112^\circ\}$.}
    \label{f:streamlines_vorticity_t23}
\end{figure*}

To maintain the momentum balance between the centrifugal force and the pressure gradient, slower-moving fluid particles must move along paths whose radii of curvature are smaller than those of faster-moving particles. This leads to the onset of a secondary flow. Cross-sections of nondimensional secondary velocity magnitude $|\bmm{u}^\star_{\theta r}|$ are shown in Figs.~\ref{f:womersley_uplanar} and~\ref{f:uniform_uplanar}. Under both entrance conditions, during acceleration and deceleration we see increased velocities along the upper and lower walls where the velocity vectors (not shown) point inward (toward the inner wall). In the interior, the secondary flow is outward (toward the outer wall). Generally, this inward-outward motion is seen during acceleration and deceleration throughout the various cross-sections, although different secondary flow patterns emerge along the streamwise direction due to varying degrees of flow development, centrifugal forcing and flow reversal. Comparing results between WEC and UEC at $\phi=45^\circ$ and $\phi=90^\circ$, we see more secondary motion under WEC in the interior/core of the flow, away from the wall. At $(t^\star,\phi)=(0.25,45^\circ)$, we observe that secondary motion of the fluid is larger in the interior near the plane of symmetry than along the walls. This increase in secondary flow is due to large velocities from the upper and lower half of the pipe competing at the plane of symmetry, causing an outward jet-like motion to appear. At $(t^\star,\phi)=(0.23,90^\circ)$, we see a pair of circular regions of strong secondary flow, one above and one below the plane of symmetry, which correspond to the deformed Dean vortex pair. Such increased secondary flows are not as prevalent under UEC, and we can conclude that a more uniform entrance condition inhibits secondary motion, ultimately affecting growth of any interior flow vortices (see Sec.~\ref{s:pulsatile_vortex_id}).

\subsubsection{Vorticity}
\label{s:pulsatile_vorticity}

\begin{figure*}[t]
    \captionsetup[subfigure]{labelformat=parens}
    \renewcommand{\fsize}{35mm}
    \centering\setcounter{subfigure}{0}
    \subfloat[D]{
        \includegraphics[width=\fsize,keepaspectratio]
        {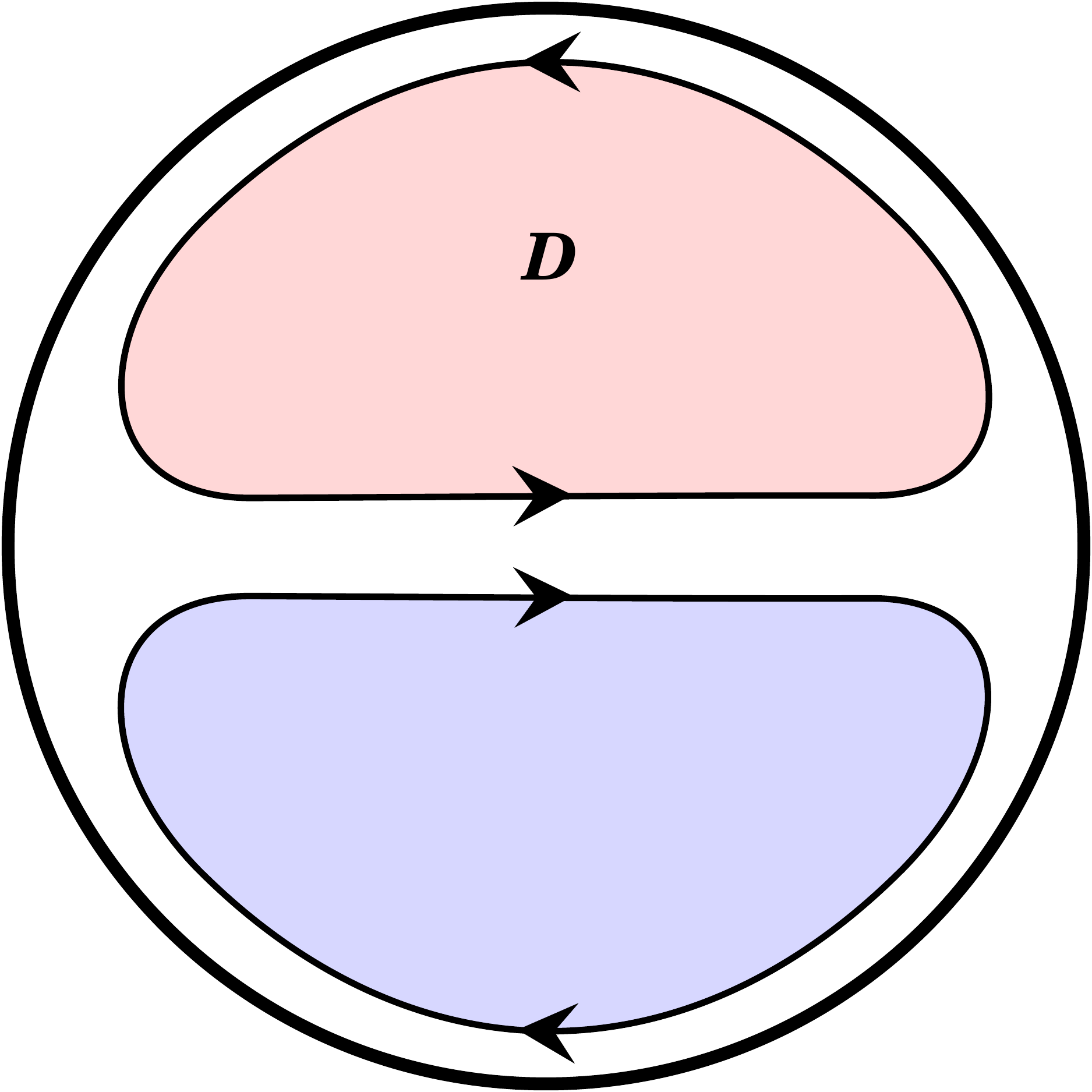}
        \label{f:D_schematic}
    }\hspace{0.2in}
    \subfloat[DD]{
        \includegraphics[width=\fsize,keepaspectratio]
        {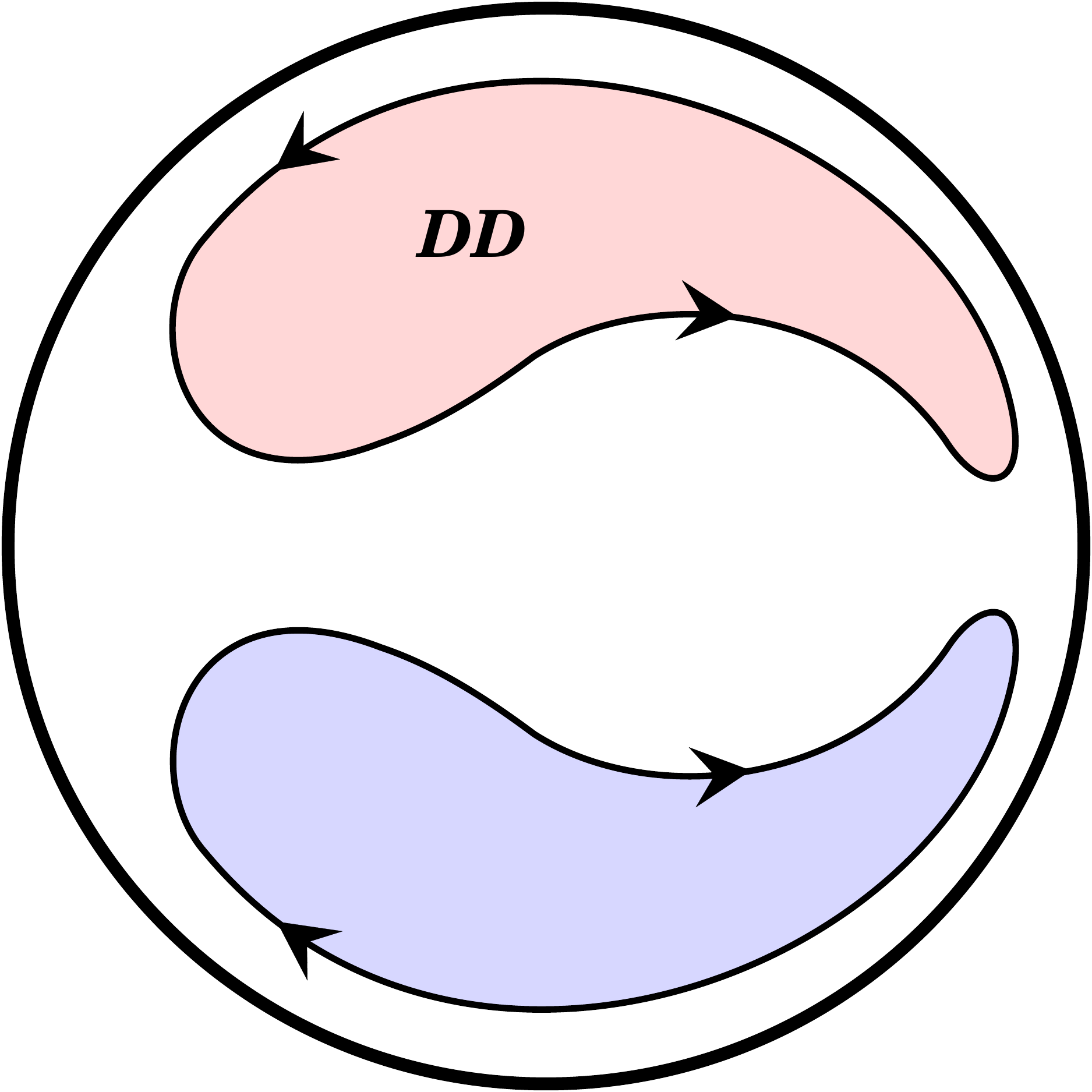}
        \label{f:DD_schematic}
    }\hspace{0.2in}
    \subfloat[DD + LT]{
        \includegraphics[width=\fsize,keepaspectratio]
        {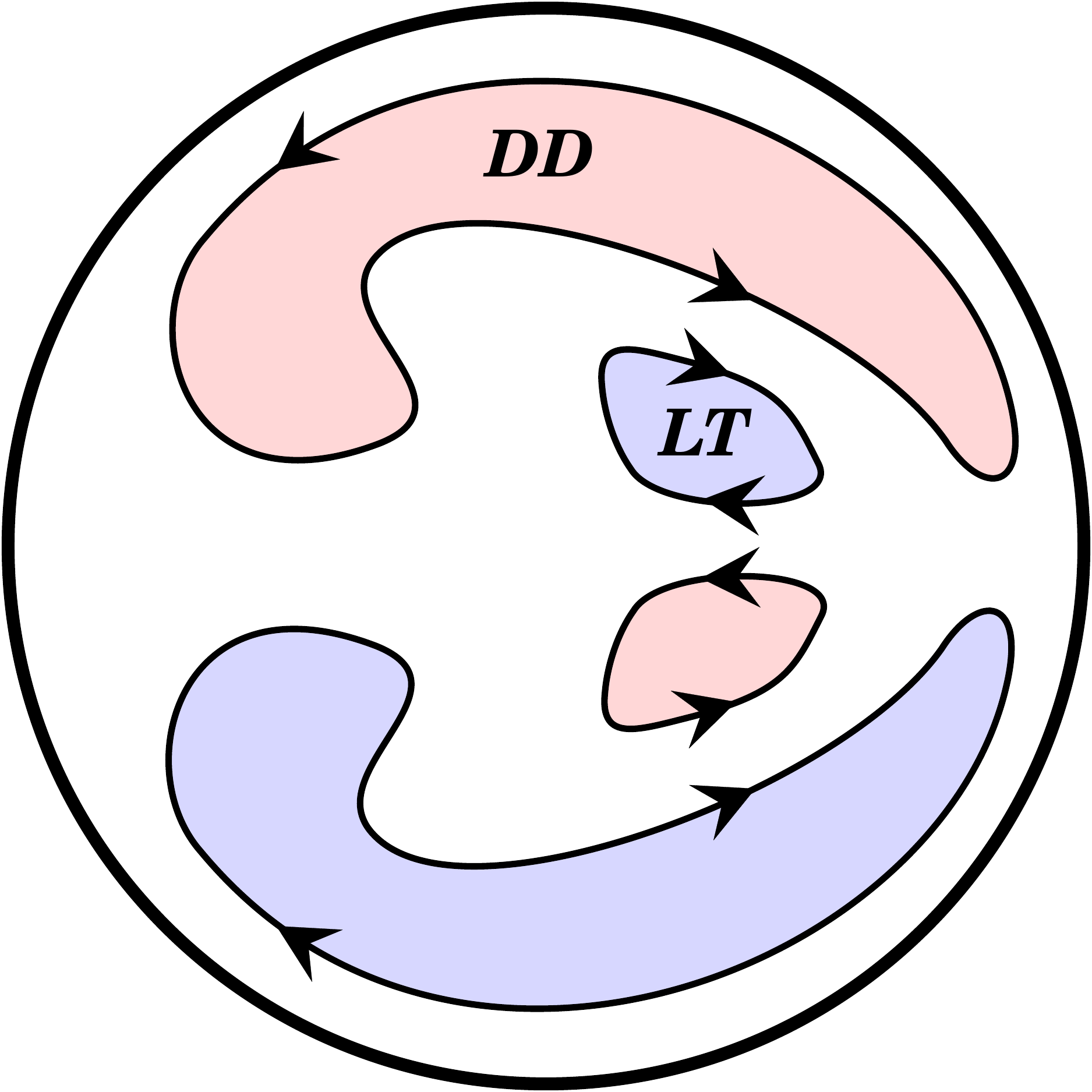}
        \label{f:DD_LT_schematic}
    }\hspace{0.2in}
    \subfloat[DD + SD + LT]{
        \includegraphics[width=\fsize,keepaspectratio]
        {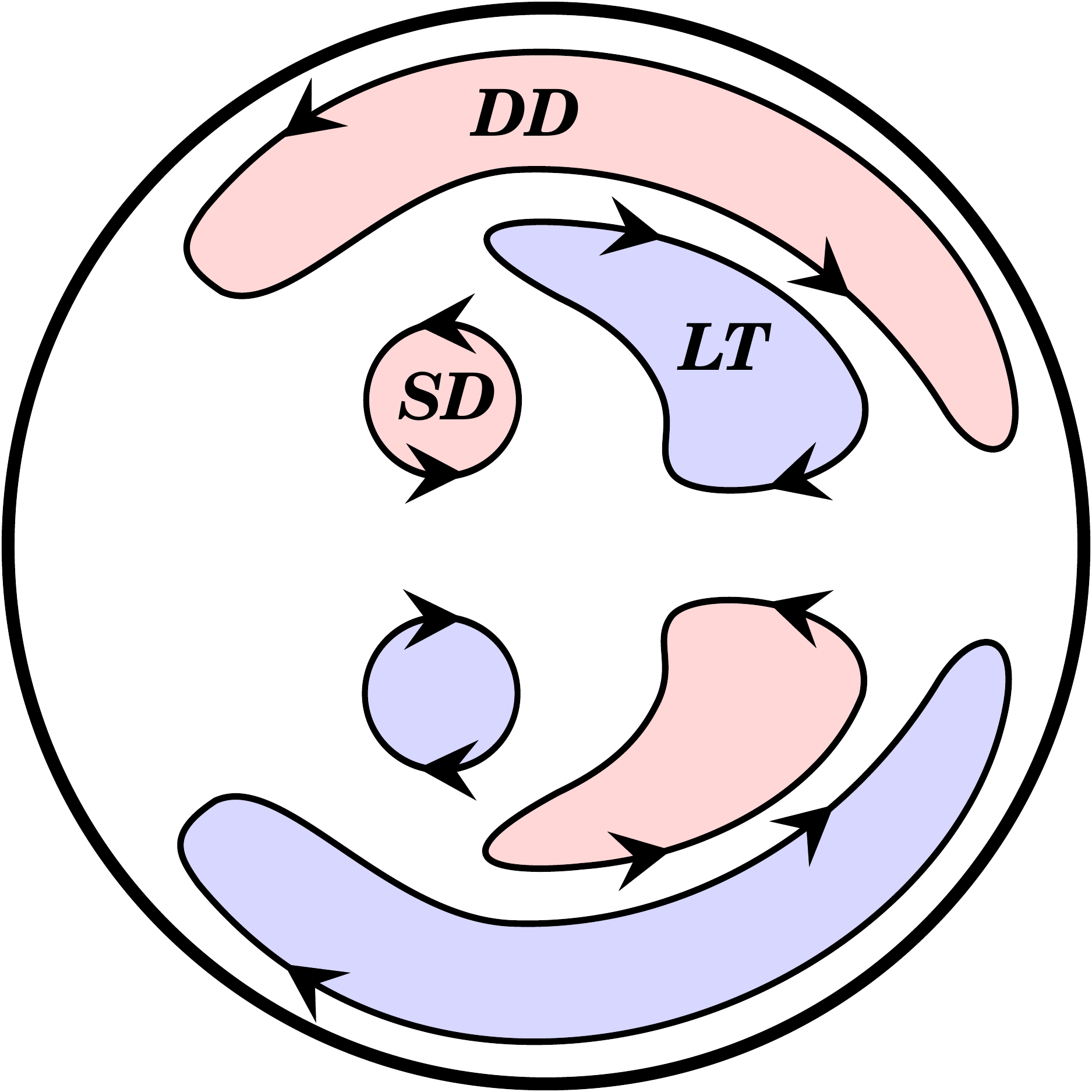}
        \label{f:DD_SD_LT_schematic}
    }
    
    \caption{Schematic of Dean (D), deformed Dean (DD), split Dean (SD) and Lyne-type (LT) vortex pairs viewed from upstream; clockwise and counter-clockwise rotation of the secondary flow are indicated by the arrows.}
    \label{f:vortex_pairs_schematic}
\end{figure*}

Vorticity, mathematically defined as the curl of the velocity field $\bmm{\omega} = \nabla \times \bmm{u}$, is a measure of solid-body rotation of fluid elements and can be used to further understand the viscous flow physics under the entrance conditions studied in this work. We provide results of the nondimensional streamwise component of vorticity $\omega^\star_s=\omega_s d/\overbar{u}_{mean}$ for various cross-sectional planes $\phi$ in Fig.~\ref{f:streamlines_vorticity_t23} for both entrance conditions. In these cross-sectional views, we use the right-hand-screw rule with thumb pointing upstream to signify positive vorticity in all planes of $\phi$. Various degrees of secondary motion can produce either one, two or three pairs of circulation, where the secondary streamlines exhibit a circular pattern. The appearance of these counter-rotating pairs at a given phase or location along the curve depends on the inlet flow condition, the Womersley number and the Dean number. For the current physiological inflow conditions, we can observe four distinct patterns of the secondary flow~\cite{sudo-sumida-yamane:1992} that are depicted in Fig.~\ref{f:vortex_pairs_schematic}. The first is the well known {\it Dean} (D) circulation in Fig.~\ref{f:vortex_pairs_schematic}\subref{f:D_schematic}, which shows a pair of counter-rotating ``cells''. The second pattern in Fig.~\ref{f:vortex_pairs_schematic}\subref{f:DD_schematic}, shows a more ``comma-like'' shape and is denoted {\it deformed Dean} (DD) circulation. The third pattern in Fig.~\ref{f:vortex_pairs_schematic}\subref{f:DD_LT_schematic} shows both deformed Dean and {\it Lyne-type} (LT) circulation. We adopt the term {\it Lyne-type} here due to the fact that this region of secondary motion rotates in the opposite direction to a Dean/deformed Dean cell, similar to the original Lyne rotation. However, we use the term Lyne-type loosely under pulsatile developing flow conditions since Lyne circulation was originally described under fully developed high-frequency oscillatory viscous flow where the viscous effects are confined to a thin layer along the wall while the interior core of the flow is inviscid.~\cite{lyne:1970}

Development of secondary boundary layers along the upper and lower walls are a result of the fluid entering these layers near the outer wall and exiting at the inner wall. At a high Dean number, or high Reynolds number for a given curvature ratio, the centrifugal force due to the curvature leads to an increased circumferential velocity where more fluid is ``sucked'' into the boundary layer at the outer wall.~\cite{berger-talbot-yao:1983} This results in the boundary layer being thinner near the outer wall and thicker near the inner wall. As the Dean number increases, the boundary layer at the inner wall thickens further. Our results confirm this description under both entrance conditions while also demonstrating eruption of the secondary boundary layer at the inner wall.~\cite{doorly-sherwin:2009} This inner wall layer eruption feeds vorticity to the Lyne-type system, whose streamwise vorticity is opposite that of Dean-type vortices. However, the amount of secondary flow indicated by helical streamlines and the subsequent vortex formation are strongly dependent upon the type of entrance condition. Figure~\ref{f:streamlines_vorticity_t23} shows that a fully developed entrance condition supports boundary layer growth and strong secondary flow whereas a uniform entrance condition inhibits boundary layer growth and secondary flow. The streamlines associated with the ``head'' of deformed Dean vorticity---which eventually separates and is referred to herein as split Dean (SD) vorticity---at mid-deceleration highlight the strong helical pattern of the flow under the fully developed pulsatile entrance condition compared to the weak helical pattern under a uniform condition. In turn, the secondary flow produces larger streamwise vorticity and more intense vortical structures. Furthermore, stronger secondary flow improves convection of $z$ vorticity away from the inner wall and into the core region of the flow, ultimately intensifying the Lyne-type vortex under WEC at $\phi \in \{67^\circ,90^\circ,112^\circ\}$.

In Figs.~\ref{f:womersley_omegan} and \ref{f:uniform_omegan}, we plot the velocity surfaces colored by streamwise vorticity under WEC and UEC during the first half of deceleration ($0.19 \leq t^\star \leq 0.25$) at $\phi=90^\circ$, where (I), (O), (U) and (L) denote the inner, outer, upper and lower walls, respectively (see Fig.\ref{f:curved_pipe_mesh_geometry_N4} for geometry and orientation). These images display the changing shape of the velocity profile as the flow starts to decelerate. At $t^\star=0.23$, the phase at which multiple vortex patterns coexist, the isometric view shows the separated ``head'' of deformed Dean (a.k.a. SD) in the core of the flow where the velocity profile is flattened. This can be observed under both WEC and UEC, although the former produces larger SD vorticity that is coincident with a locally flattened velocity profile. Under UEC, the ``head'' of deformed Dean vorticity does not intensify as much as it does under WEC. This lack of SD formation can be attributed to the fact that the entrance flow is not developed, producing a skewed but flattened velocity profile (see Fig.~\ref{f:us_profile_plug_z0}), with peak velocity occurring closer to the pipe center and smaller than the result from a fully developed entrance condition. This decreased velocity from UEC produces a smaller centrifugal force and consequently a smaller pressure gradient sets up across the pipe cross-section, thereby driving the secondary motion of the fluid at a slower velocity than that which occurs under a fully developed condition. Therefore, we conclude that the inviscid core formation under a developed entrance condition, coupled with strong secondary flow, plays a large role in the mechanism that promotes full separation of the ``head'' under WEC---a dynamic of the flow that does not occur under UEC.

{\color{gray}
}

\renewcommand{\ftype}{womersley}
\renewcommand{\fres}{N5M30_slices_1-29_150dpi}
\renewcommand{\fvarlatex}{\omega^{\star}_s}

\renewcommand{\fsize}{41mm}
\renewcommand{\fsizeiso}{52mm}
\renewcommand{\dt}{0.020}
\renewcommand{\ftime}{0.170}
    
\begin{figure*}[t]
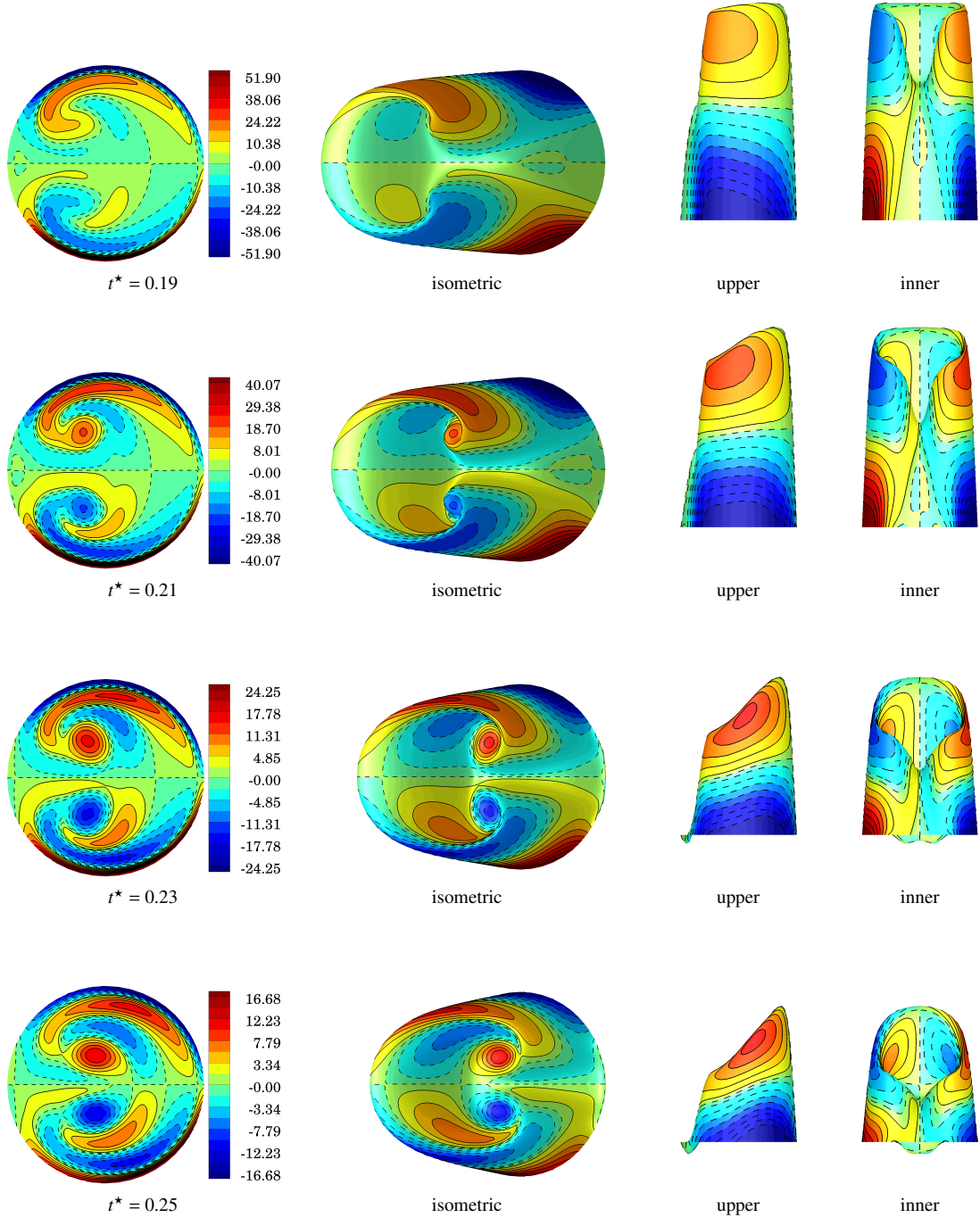

    \begin{minipage}[c]{\textwidth}
        \captionsetup[subfigure]{labelformat=empty}
        \centering\setcounter{subfigure}{0}
    
        \forloop{jl}{1}{\value{jl} < 5}{
            
            \FPeval{\result}{trunc(\ftime+\dt:3)}
            \FPeval{\myresult}{trunc(\ftime+\dt:2)}
            \renewcommand{\ftime}{\myresult}
            \renewcommand{\slice}{90}
            
            \vspace{-0.15in}
            \subfloat[$t^\star=\myresult$]{
                \includegraphics[width=\fsize,height=\fsize,keepaspectratio]
                    {./figures/wec/omegan_\slice_slice_\result.png}
            }
            \subfloat[isometric]{
                \includegraphics[width=\fsizeiso,height=\fsizeiso,keepaspectratio]
                    {./figures/wec/omegan_\slice_surf1_\result.png}
            }
            \subfloat[upper]{
                \includegraphics[width=\fsize,height=\fsize,keepaspectratio]
                    {./figures/wec/omegan_\slice_surf2_\result.png}
            }
            \subfloat[inner]{
                \includegraphics[width=\fsize,height=\fsize,keepaspectratio]
                    {./figures/wec/omegan_\slice_surf3_\result.png}
            }
        }
        
        \caption{Womersley entrance condition: velocity surface of $u^\star_s$ at $\phi=90^\circ$ during deceleration colored by nondimensional streamwise vorticity $\fvarlatex$ visualized from three difference perspectives: \textit{isometric}, \textit{upper} and \textit{inner}. The \textit{upper} perspective shows the inner/outer walls on the left/right, while the \textit{inner} perspective shows the lower/upper walls on the left/right, respectively. At $t^\star=0.23$, where multiple pairs of vortices coexist, the isometric view shows the separated ``head" of deformed Dean where the velocity profile is flattened in the small inviscid core and the vorticity is largely due to rotation of the flow in the cross-sectional plane.}
        \label{f:womersley_omegan}
    \end{minipage}
\end{figure*}
\renewcommand{\ftype}{plug}
\renewcommand{\fres}{N5M30_slices_1-29_150dpi}
\renewcommand{\fvarlatex}{\omega^{\star}_s}

\renewcommand{\fsize}{41mm}
\renewcommand{\fsizeiso}{52mm}
\renewcommand{\dt}{0.020}
\renewcommand{\ftime}{0.170}

\begin{figure*}[t]
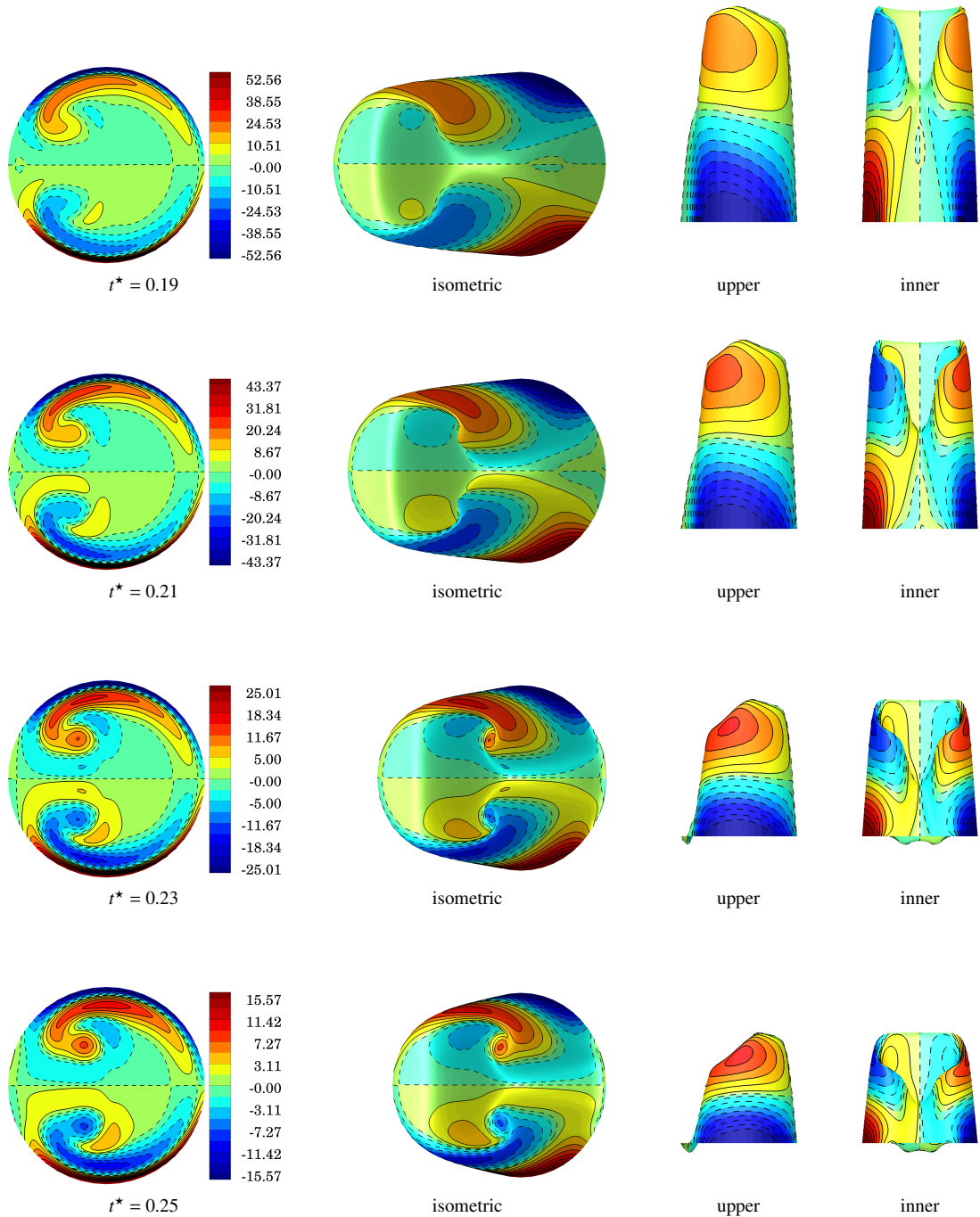

    \begin{minipage}[c]{\textwidth}
        \captionsetup[subfigure]{labelformat=empty}
        \centering\setcounter{subfigure}{0}
        
        \forloop{jl}{1}{\value{jl} < 5}{
            
            \FPeval{\result}{trunc(\ftime+\dt:3)}
            \FPeval{\myresult}{trunc(\ftime+\dt:2)}
            \renewcommand{\ftime}{\myresult}
            \renewcommand{\slice}{90}
            
            \vspace{-0.15in}
            \subfloat[$t^\star=\myresult$]{
                \includegraphics[width=\fsize,height=\fsize,keepaspectratio]
                    {./figures/uec/omegan_\slice_slice_\result.png}
            }
            \subfloat[isometric]{
                \includegraphics[width=\fsizeiso,height=\fsizeiso,keepaspectratio]
                    {./figures/uec/omegan_\slice_surf1_\result.png}
            }
            \subfloat[upper]{
                \includegraphics[width=\fsize,height=\fsize,keepaspectratio]
                    {./figures/uec/omegan_\slice_surf2_\result.png}
            }
            \subfloat[inner]{
                \includegraphics[width=\fsize,height=\fsize,keepaspectratio]
                    {./figures/uec/omegan_\slice_surf3_\result.png}
            }
        }
            
        \caption{Uniform entrance condition: velocity surface of $u^\star_s$ at $\phi=90^\circ$ during deceleration colored by nondimensional streamwise vorticity $\fvarlatex$ visualized from three difference perspectives: \textit{isometric}, \textit{upper} and \textit{inner}. The \textit{upper} perspective shows the inner/outer walls on the left/right, while the \textit{inner} perspective shows the lower/upper walls on the left/right, respectively. The suppressed deformed-Dean ``head" formation is attributed to the lack of flow development at the entrance, producing a more flattened velocity profile and weaker secondary flow in the core.}
        \label{f:uniform_omegan}
    \end{minipage}
\end{figure*}

\subsection{Vortex identification: Dean and Lyne-type vortices}
\label{s:pulsatile_vortex_id}

In Sec.~\ref{s:pulsatile_vorticity} we describe the streamwise vorticity fields associated with DD, SD and LT circulation under the two different entrance conditions. In this section we compute the $\lambda_2$-criterion~\cite{jeong-hussain:1995} to identify actual vortical structures under WEC (Fig.~\ref{f:womersley_lambda2}) and UEC (Fig.~\ref{f:uniform_lambda2}) and correlate them to the vorticity field. Regions of large/small negative values of $\lambda_2^\star$ signify strong/weak vortical activity. The value $\lambda_2$ is the second eigenvalue of $\bmm{S}^2+\bmm{R}^2$, where $\bmm{S}=1/2~(\nabla\bmm{u}+\nabla\bmm{u}^T)$ represents the strain-rate tensor and $\bmm{R}=1/2~(\nabla\bmm{u}-\nabla\bmm{u}^T)$ the rotation tensor. Velocity gradients in the strain-rate and rotation tensors are directly computed from the flux reconstruction methodology described in Sec.~\ref{s:numerical_scheme} in order to retain the solution gradient accuracy afforded by the numerical scheme. We specifically note the difference in results under the two entrance conditions at $45^\circ$ and $90^\circ$. Overall, vortical intensity is highest near peak flow rate at $\phi=45^\circ$ where $\lambda^\star_2=-195$ at $t^\star=0.21$ (WEC) and $\lambda^\star_2=-149$ at $t^\star=0.19$ (UEC). As the flow decelerates, the vortical strength reduces by 86\%.
    
Under WEC, the SD vortical structure separates from DD at $(t^\star,\phi)=(0.23,90^\circ)$. Most of the vortical activity is dumped into SD, where $\lambda_2^\star \approx -61$. What remains of DD is half this amount ($\lambda_2^\star \approx -30$). After separation occurs, the SD vortex drifts towards the outer wall and diminishes while the flow continues to decelerate. This loss in vortex strength is attributed to viscous diffusion, which plays a larger role as the flow rate drops by a factor of seven throughout the deceleration phase.
    
Under UEC at $\phi=45^\circ$ and $\phi=90^\circ$, all vortical content is confined to DD, and the formation and full separation of the SD vortex from DD does not occur as it does under WEC. Since the maximum streamwise velocity is smaller with an undeveloped entrance condition, the secondary flow in the interior of the pipe is smaller. This reduced secondary flow suppresses growth of the SD vortex, reduces convection of positive $z$ vorticity away from the inner wall towards the outer wall which, in turn, leads to weak LT formation as indicated by $\lambda_2$ and streamwise vorticity (see Fig.~\ref{f:streamlines_vorticity_t23}). We note that since LT vorticity can be seen in both steady and pulsatile flows, it is considered a curvature effect. These results supplement previous analyses of LT vortex formation due to convection of $z$ vorticity.~\cite{najjari-cox-plesniak:2019,cox-najjari-plesniak:2019}

Isosurfaces of $\lambda^\star_2=-24$ colored by streamwise vorticity are shown in Fig.~\ref{f:lambda2_vorticity_isosurface} and depict the vortical structure thats forms at $t^\star=0.23$ under both WEC and UEC. It is important to note that a vortical structure displayed by an isosurface does not strictly define the extent of the entire structure---the isosurface is merely a threshold set to visualize the vortical structure. The helical streamlines displayed in Fig.~\ref{f:streamlines_vorticity_t23}, resulting from secondary flow, intensify streamwise vorticity and lead to a larger vortical structure and separation of the deformed Dean ``head'' in Fig.~\ref{f:lambda2_vorticity_isosurface}\subref{f:a} under a fully developed pulsatile entrance condition as opposed to that shown in Fig.~\ref{f:lambda2_vorticity_isosurface}\subref{f:b} under a uniform entrance condition.

\renewcommand{\ftype}{womersley}
\renewcommand{\fres}{N5M30_slices_1-29_150dpi}
\renewcommand{\fsize}{29mm}

\forloop{il}{5}{\value{il} < 6}{
    
    \renewcommand{\slicetype}{slice}
    
    \ifthenelse{ \equal{\value{il}}{5} }{
        \renewcommand{\ilflag}{1}
        \renewcommand{\fvardir}{l2}
        \renewcommand{\fvar}{l2}
        \renewcommand{\filetype}{png}
        \renewcommand{\fvartext}{vortex identification with}
        \renewcommand{\fvarlatex}{\lambda^{\star}_2<0}
    }{}
    
    \renewcommand{\dt}{0.020}
    \renewcommand{\ftime}{0.170}
    
    \begin{figure*}[t]
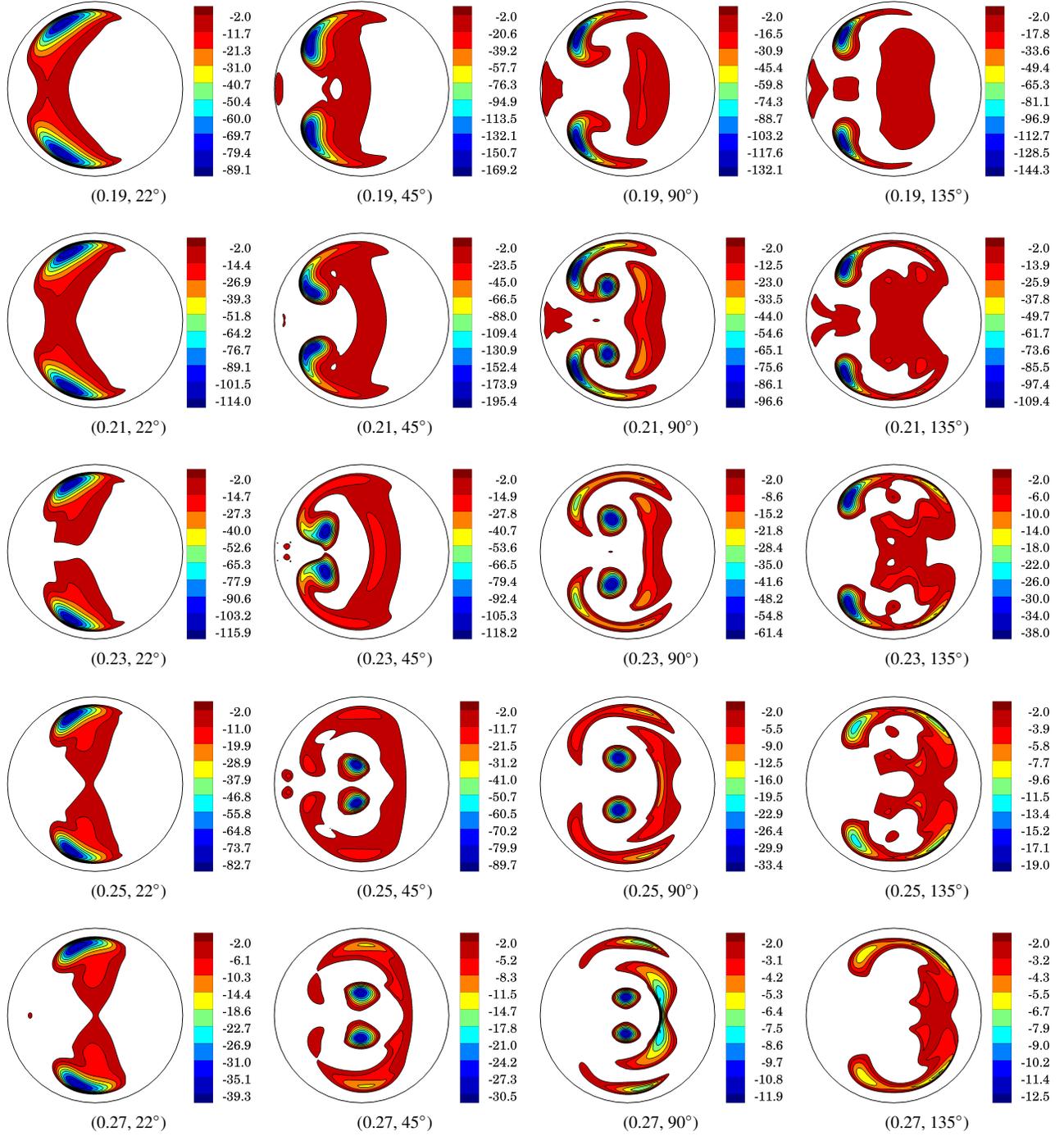

        \begin{minipage}[c]{\textwidth}
            \captionsetup[subfigure]{labelformat=empty}
            \centering\setcounter{subfigure}{0}
            
            \forloop{jl}{1}{\value{jl} < 6}{
                
                \FPeval{\result}{trunc(\ftime+\dt:3)}
                \FPeval{\myresult}{trunc(\ftime+\dt:2)}
                \renewcommand{\ftime}{\myresult}
                
                \renewcommand{\slice}{22}
                \renewcommand{\fdir}{\ftype/\fres/\fvardir/\slice}
                \subfloat[$(\myresult$, $\slice^\circ)$]{
                    \includegraphics[height=\fsize,keepaspectratio]
                        {./figures/wec/\fvar_\slice_\slicetype_\result.\filetype}
                }
                \renewcommand{\slice}{45}
                \renewcommand{\fdir}{\ftype/\fres/\fvardir/\slice}
                \subfloat[$(\myresult$, $\slice^\circ)$]{
                    \includegraphics[height=\fsize,keepaspectratio]
                        {./figures/wec/\fvar_\slice_\slicetype_\result.\filetype}
                }
                \renewcommand{\slice}{90}
                \renewcommand{\fdir}{\ftype/\fres/\fvardir/\slice}
                \subfloat[$(\myresult$, $\slice^\circ)$]{
                    \includegraphics[height=\fsize,keepaspectratio]
                        {./figures/wec/\fvar_\slice_\slicetype_\result.\filetype}
                }
                \renewcommand{\slice}{135}
                \renewcommand{\fdir}{\ftype/\fres/\fvardir/\slice}
                \subfloat[$(\myresult$, $\slice^\circ)$]{
                    \includegraphics[height=\fsize,keepaspectratio]
                        {./figures/wec/\fvar_\slice_\slicetype_\result.\filetype}
                }
            }
            
            \caption{Womersley entrance condition: \fvartext~$\fvarlatex$ at $(t^\star,\phi)$. Regions of large/small negative values of $\lambda_2^\star$ signify strong/weak vortical activity. Vortical intensity is highest at $(t^\star,\phi)=(0.21,45^\circ)$ just after peak flow rate. The ``head" of deformed Dean first separates at $(t^\star,\phi)=(0.23,90^\circ)$, where Dean and Lyne-type vortical structures coexist, and convects into the interior at both $45^\circ$ and $90^\circ$. Loss in vortical intensity is attributed to viscous diffusion, which plays a larger role as the flow rate drops by a factor of seven throughout the deceleration phase.}
            \label{f:womersley_lambda2}
        \end{minipage}
    \end{figure*}
}
\renewcommand{\ftype}{plug}
\renewcommand{\fres}{N5M30_slices_1-29_150dpi}
\renewcommand{\fsize}{29mm}

\forloop{il}{5}{\value{il} < 6}{
    
    \renewcommand{\slicetype}{slice}
    
    \ifthenelse{ \equal{\value{il}}{5} }{
        \renewcommand{\ilflag}{1}
        \renewcommand{\fvardir}{l2}
        \renewcommand{\fvar}{l2}
        \renewcommand{\filetype}{png}
        \renewcommand{\fvartext}{vortex identification with}
        \renewcommand{\fvarlatex}{\lambda^{\star}_2<0}
    }{}

    \renewcommand{\dt}{0.020}
    \renewcommand{\ftime}{0.170}
    
    \begin{figure*}[t]
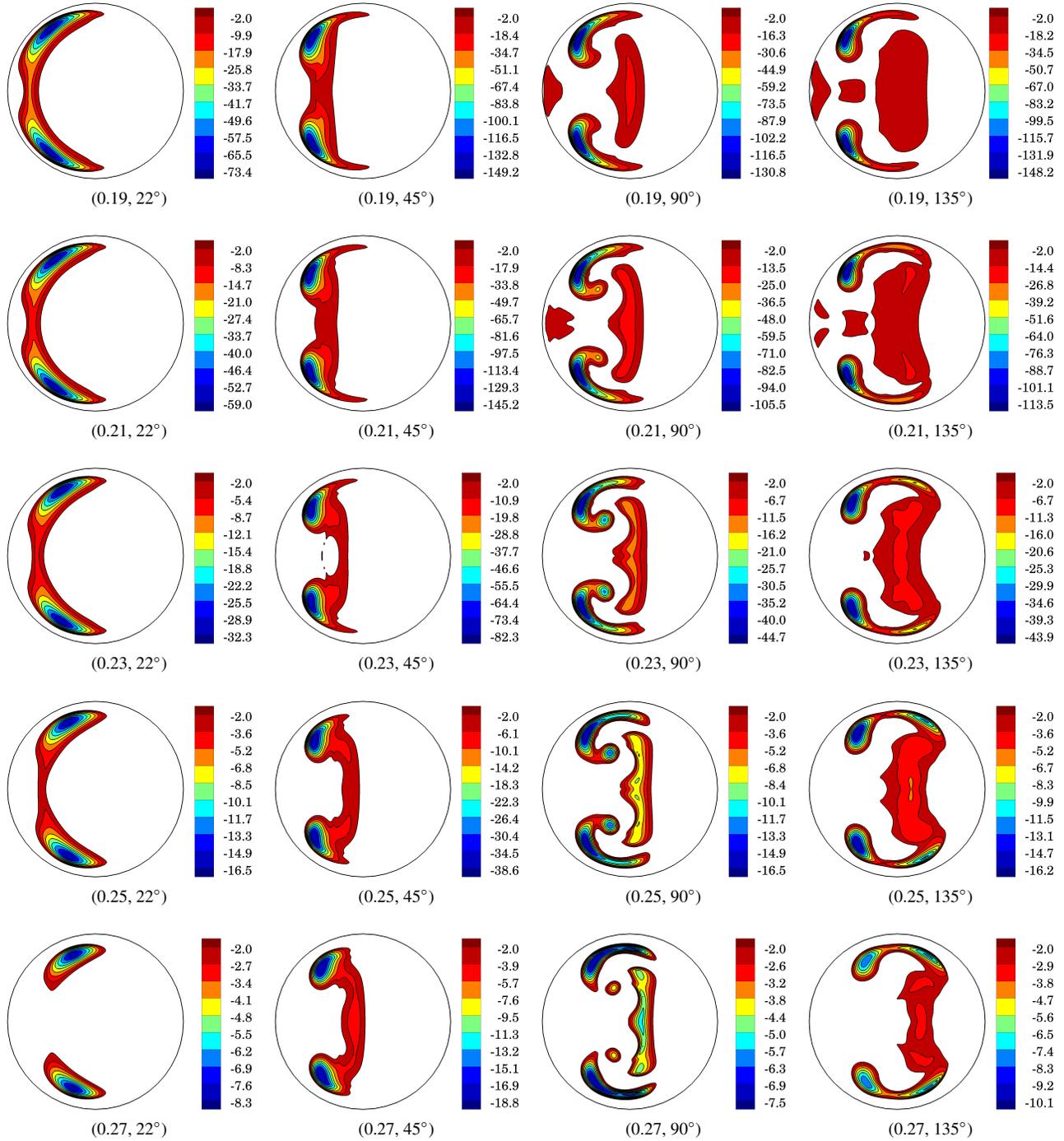

        \begin{minipage}[c]{\textwidth}
            \captionsetup[subfigure]{labelformat=empty}
            \centering\setcounter{subfigure}{0}
            
            \forloop{jl}{1}{\value{jl} < 6}{
                
                \FPeval{\result}{trunc(\ftime+\dt:3)}
                \FPeval{\myresult}{trunc(\ftime+\dt:2)}
                \renewcommand{\ftime}{\myresult}
                
                \renewcommand{\slice}{22}
                \renewcommand{\fdir}{\ftype/\fres/\fvardir/\slice}
                \subfloat[$(\myresult$, $\slice^\circ)$]{
                    \includegraphics[height=\fsize,keepaspectratio]
                        {./figures/uec/\fvar_\slice_\slicetype_\result.\filetype}
                }
                \renewcommand{\slice}{45}
                \renewcommand{\fdir}{\ftype/\fres/\fvardir/\slice}
                \subfloat[$(\myresult$, $\slice^\circ)$]{
                    \includegraphics[height=\fsize,keepaspectratio]
                        {./figures/uec/\fvar_\slice_\slicetype_\result.\filetype}
                }
                \renewcommand{\slice}{90}
                \renewcommand{\fdir}{\ftype/\fres/\fvardir/\slice}
                \subfloat[$(\myresult$, $\slice^\circ)$]{
                    \includegraphics[height=\fsize,keepaspectratio]
                        {./figures/uec/\fvar_\slice_\slicetype_\result.\filetype}
                }
                \renewcommand{\slice}{135}
                \renewcommand{\fdir}{\ftype/\fres/\fvardir/\slice}
                \subfloat[$(\myresult$, $\slice^\circ)$]{
                    \includegraphics[height=\fsize,keepaspectratio]
                        {./figures/uec/\fvar_\slice_\slicetype_\result.\filetype}
                }
            }
            
            \caption{Uniform entrance condition: \fvartext~$\fvarlatex$ at $(t^\star,\phi)$. Intense vortical content is confined to deformed Dean, which is strongest at peak flow rate where $(t^\star,\phi)=(0.19,45^\circ)$. Formation and full separation of the deformed Dean ``head" does not occur as it does under a fully developed entrance condition. The smaller maximum streamwise velocity and its outward shift (see Fig.~\ref{f:us_profile_90_plug_z0}) under a uniform entrance condition lead to a reduced secondary flow that inhibits growth of the ``head" at $(t^\star,\phi)=(0.23,90^\circ)$, where $\lambda^{\star}_2$ is half the value found in Fig.~\ref{f:womersley_lambda2}.}
            \label{f:uniform_lambda2}
        \end{minipage}
    \end{figure*}
}

\begin{figure*}[t]
    \captionsetup[subfigure]{labelformat=parens}
    \renewcommand{\fsize}{65mm}
    \centering\setcounter{subfigure}{0}
    \subfloat[]{
        \includegraphics[height=\fsize,keepaspectratio]
            {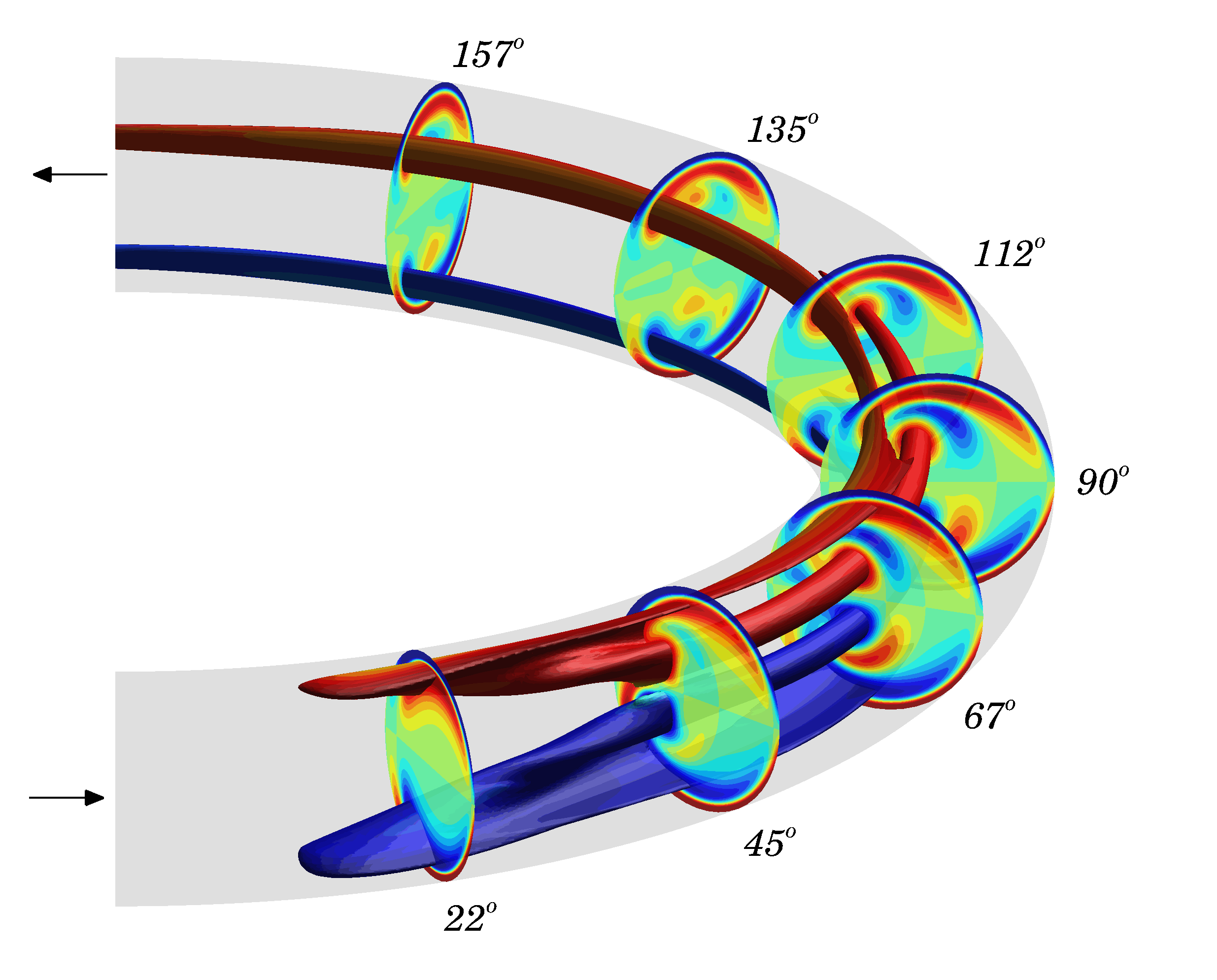}
        \label{f:a}
    }
    \subfloat[]{
        \includegraphics[height=\fsize,keepaspectratio]
            {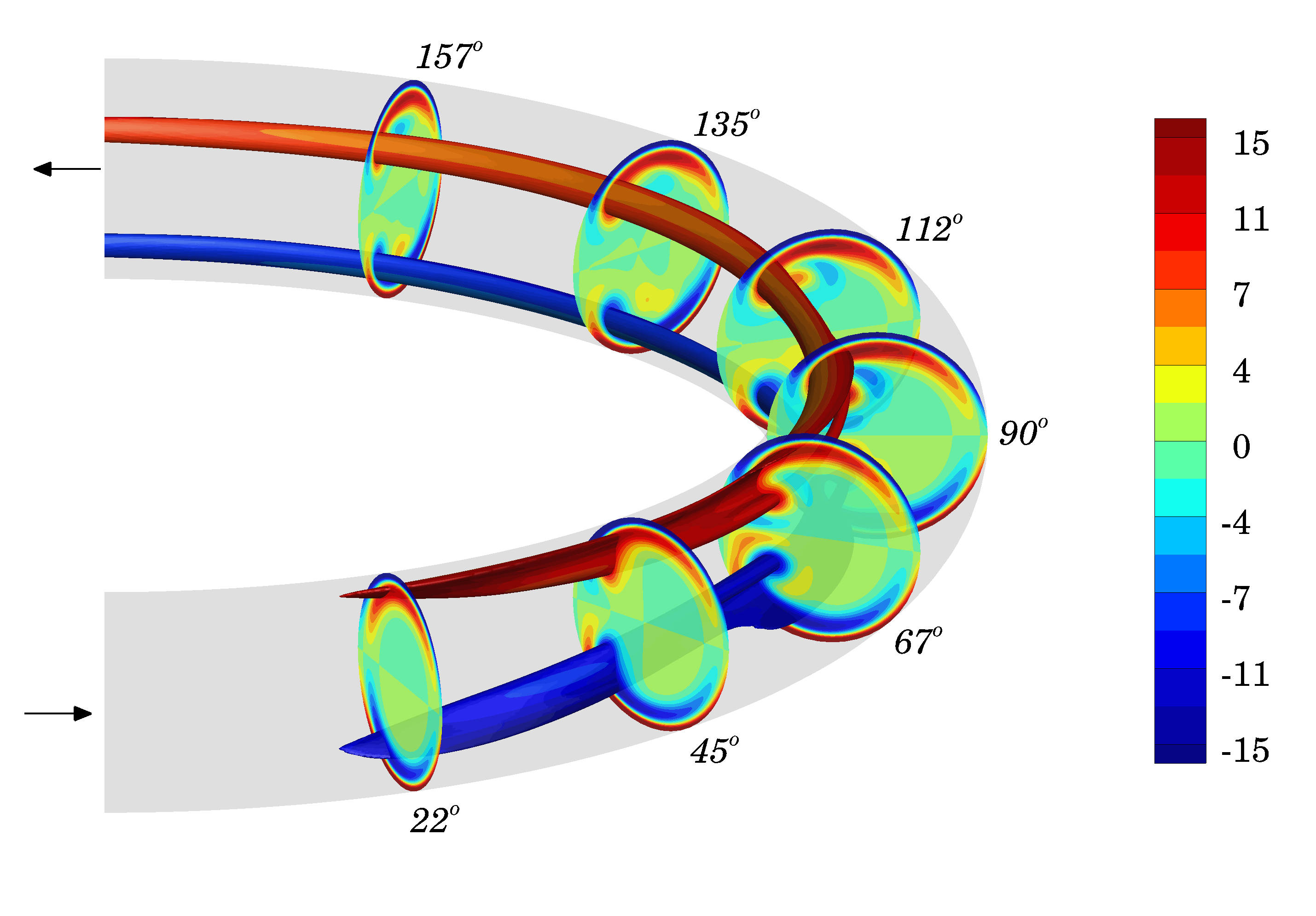}
        \label{f:b}
    }
    
    \caption{Womersley and Uniform entrance conditions: isosurfaces of $\lambda^\star_2=-24$ colored by non-dimensional streamwise vorticity $\omega^\star_s$ at mid-deceleration phase $t^\star=0.23$ under a) WEC and b) UEC. The helical streamlines displayed in Fig.~\ref{f:streamlines_vorticity_t23}, resulting from secondary flow, intensify streamwise vorticity and lead to a larger vortical structure and separation of the deformed-Dean ``head'' in a).}
    \label{f:lambda2_vorticity_isosurface}
\end{figure*}

\section{Wall shear stress}
\label{s:wss}

\begin{figure}[t]
    \centering\setcounter{subfigure}{0}
    \includegraphics[width=0.47\textwidth,keepaspectratio]
        {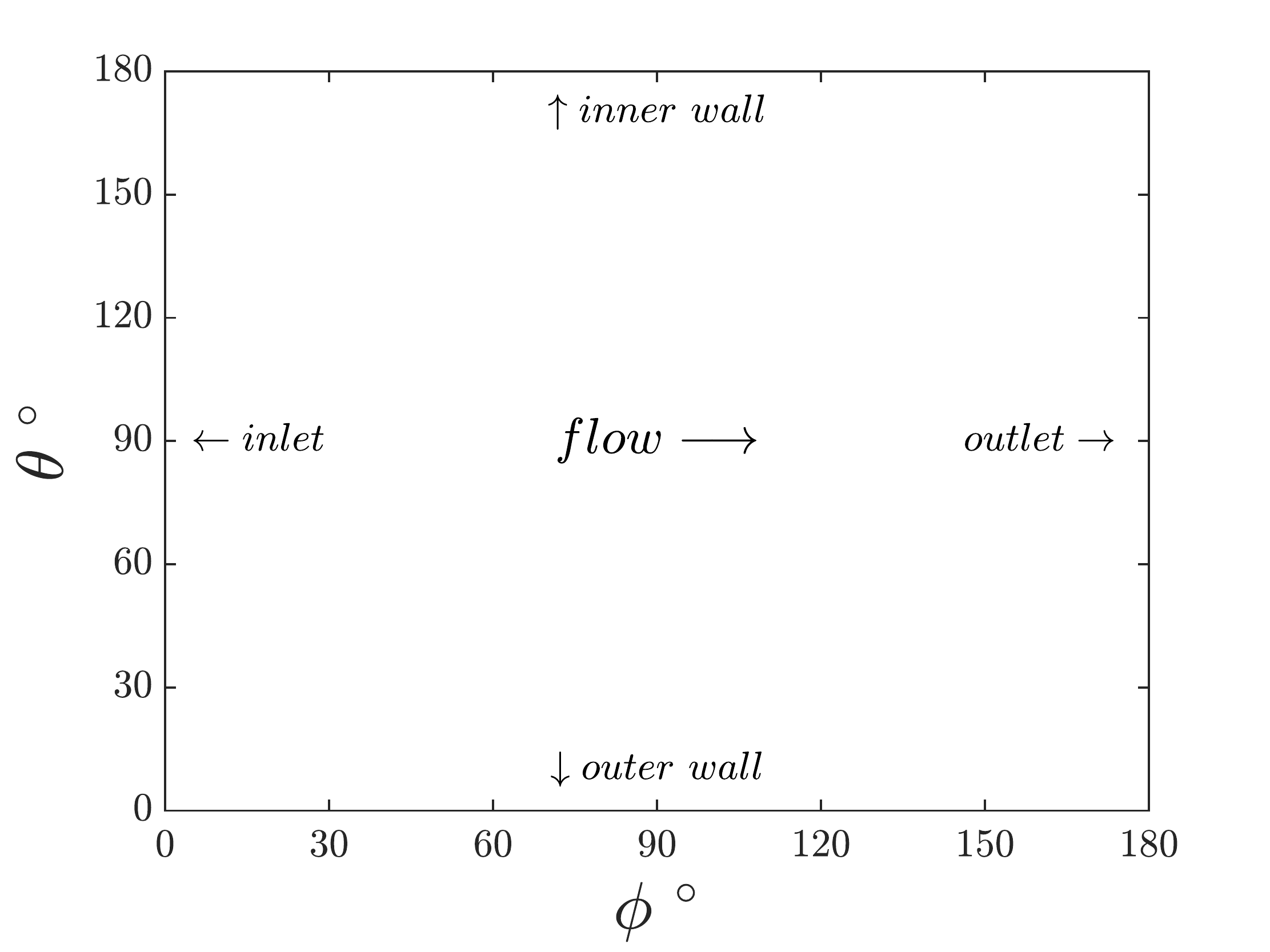}
    \label{f:wss_mapping}
    
    \caption{Two-dimensional mapping of the upper wall of the curved section depicted in Fig.~\ref{f:curved_pipe_mesh_geometry_N4}, indicating positive primary flow direction and location of the inlet, outlet and inner/outer wall. This mapping is used to visualize distribution of the instantaneous wall shear stress vector $\bm{\tau}^\star_w$ over the entire surface.}
    \label{f:wss_axis}
\end{figure}
\begin{figure*}[t]
    \begin{minipage}[c]{\textwidth}
        \centering\setcounter{subfigure}{0}
        \subfloat[WEC: $t^\star=0.19$]{
            \includegraphics[width=0.48\textwidth,keepaspectratio]
            {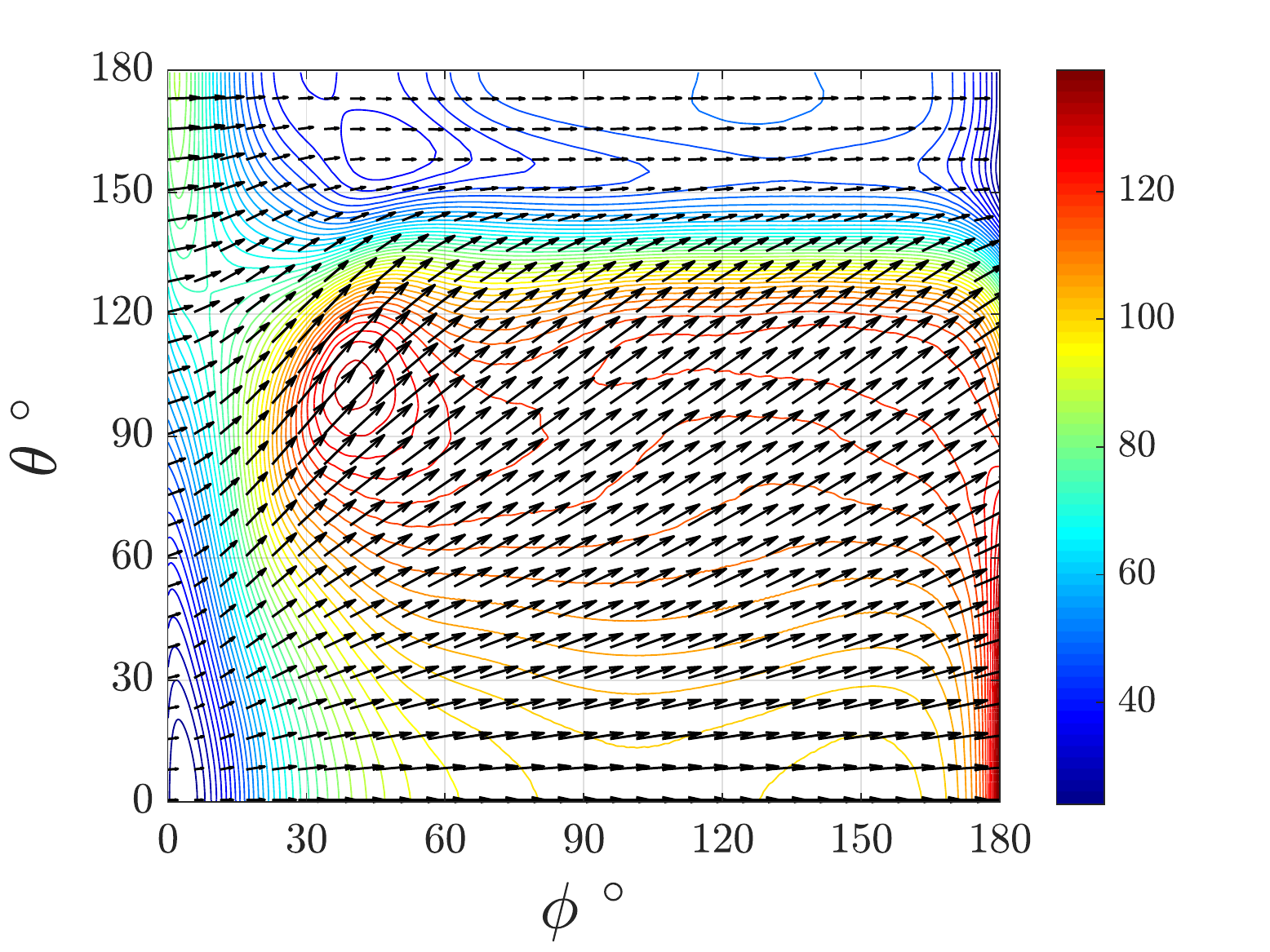}
            \label{f:wss_wec_uec:a}
        }
        \subfloat[UEC: $t^\star=0.19$]{
            \includegraphics[width=0.48\textwidth,keepaspectratio]
            {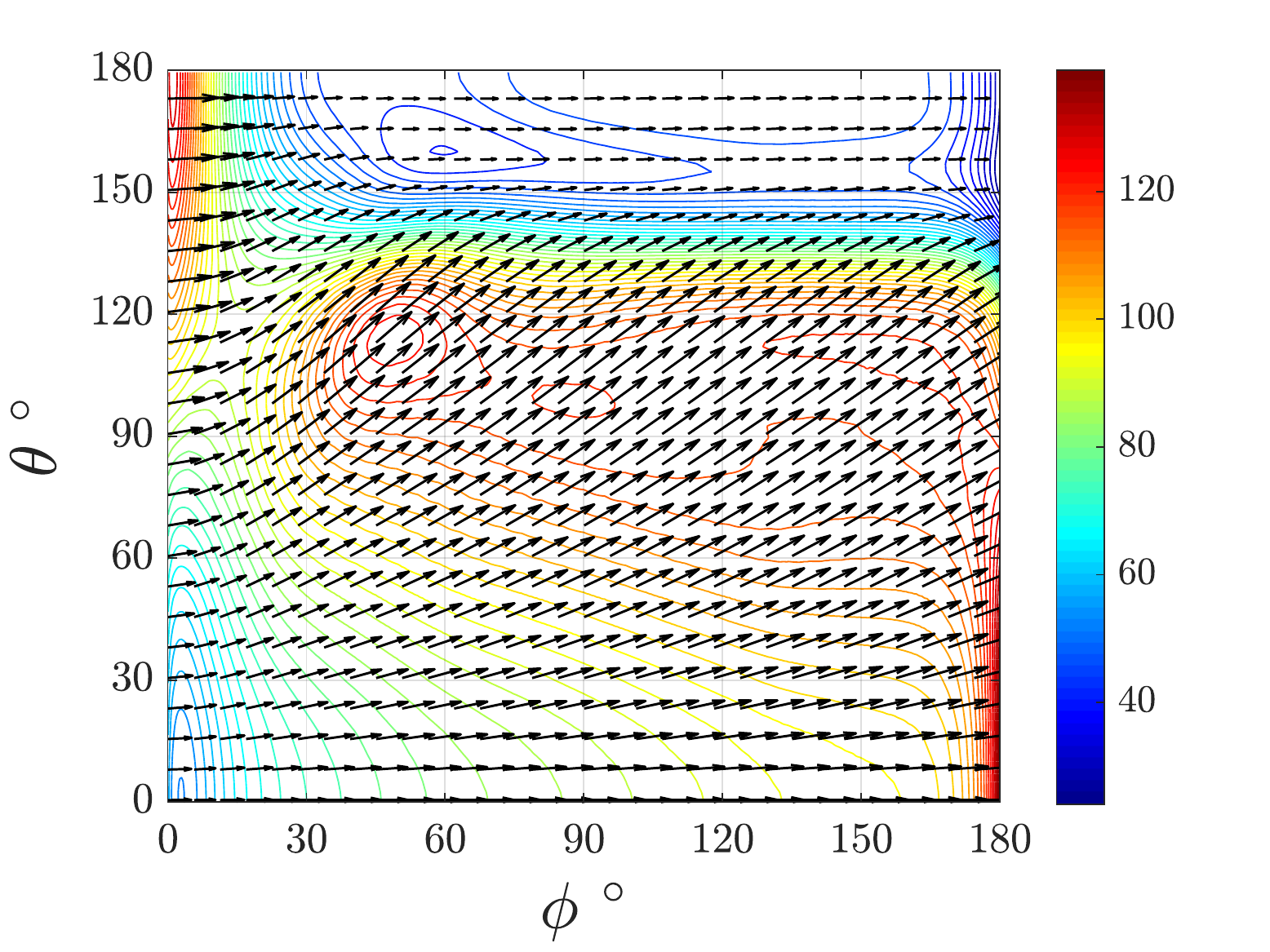}
            \label{f:wss_wec_uec:g}
        }
        \\[-0.15in]
        \subfloat[WEC: $t^\star=0.21$]{
            \includegraphics[width=0.48\textwidth,keepaspectratio]
            {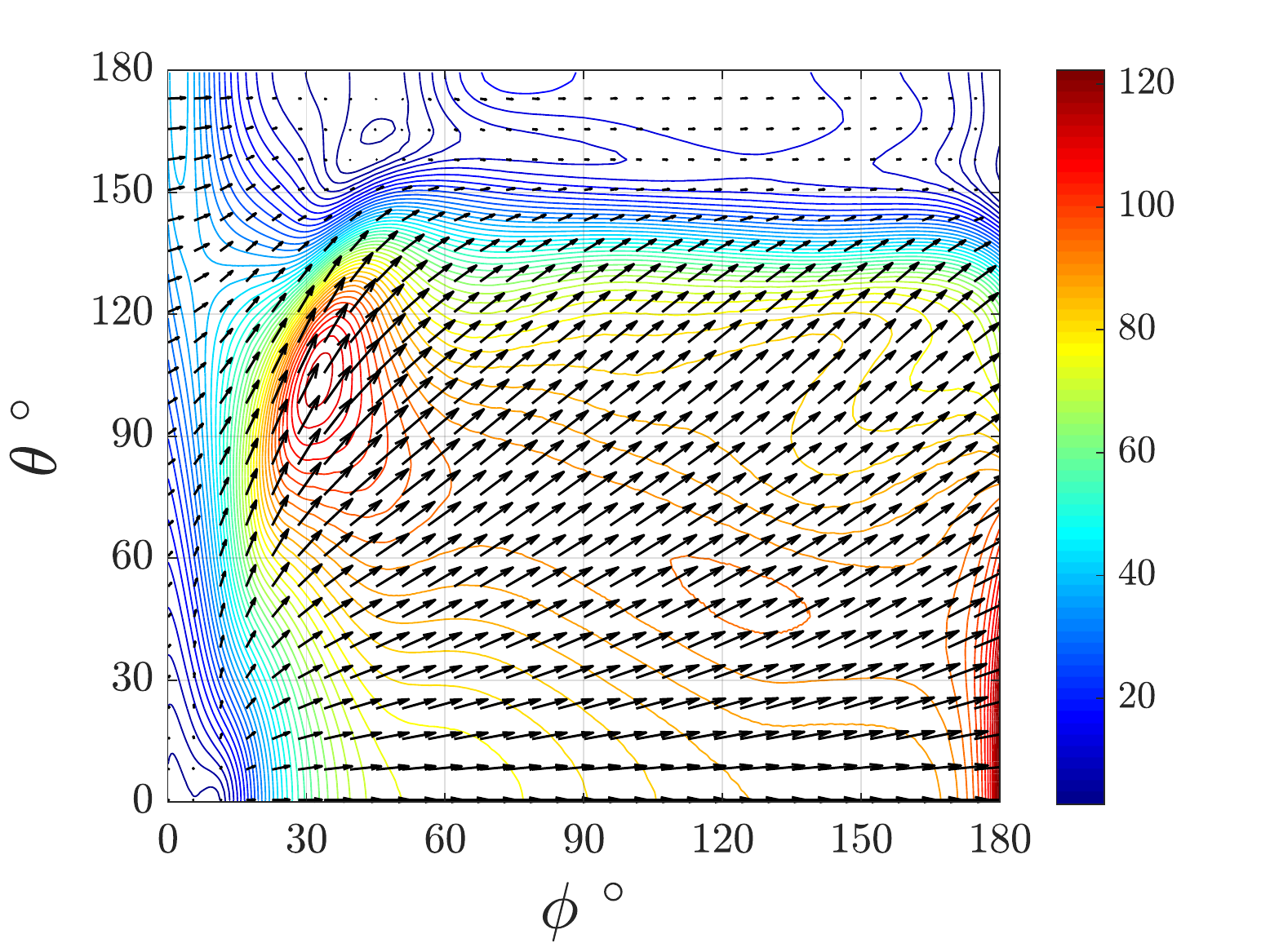}
        }
        \subfloat[UEC: $t^\star=0.21$]{
            \includegraphics[width=0.48\textwidth,keepaspectratio]
            {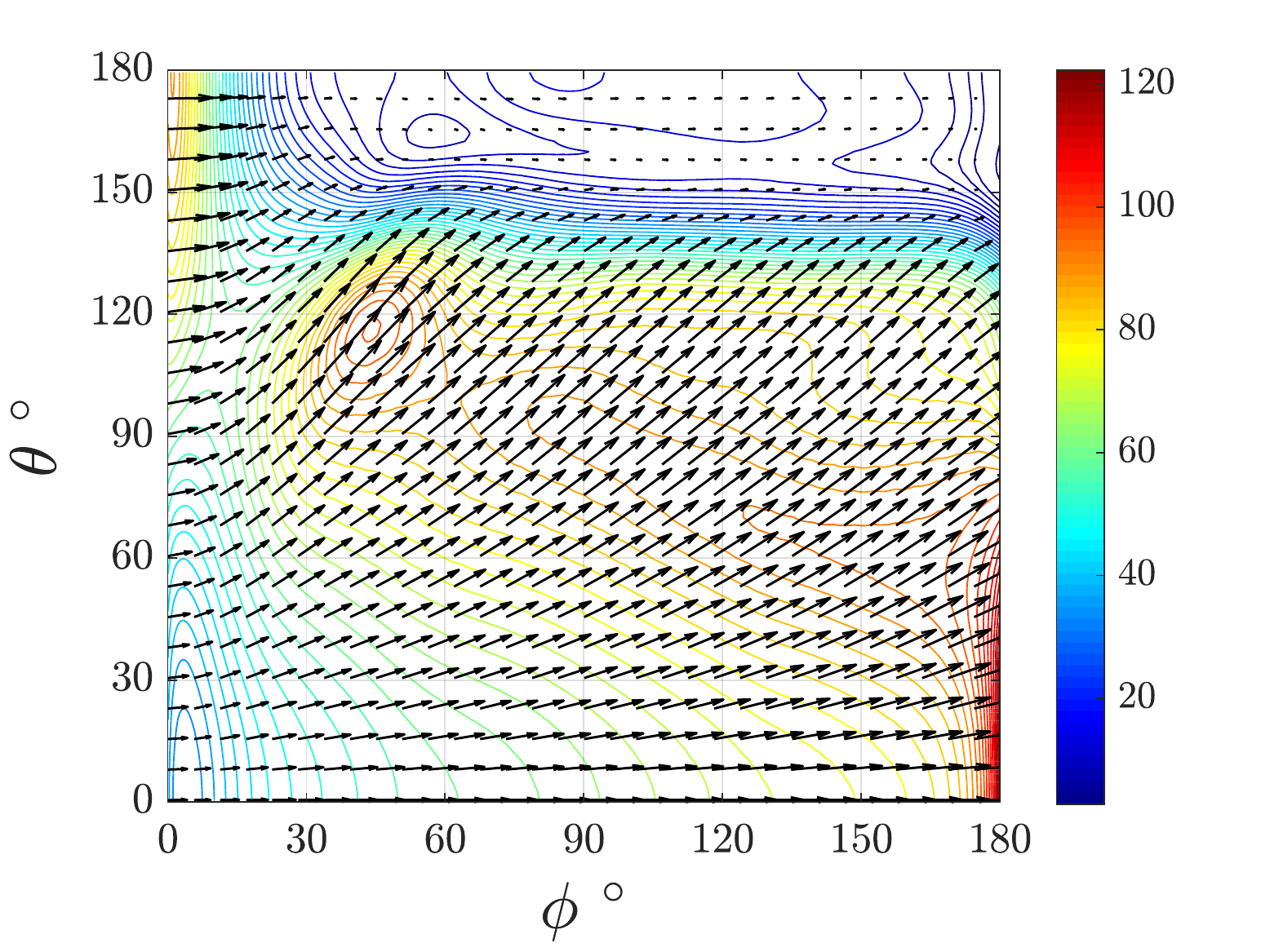}
        }
        
        \caption{Instantaneous wall shear stress $\bmm{\tau}^\star_w$ vector map of upper wall ($0^\circ \leq \theta \leq 180^\circ$, $0^\circ \leq \phi \leq 180^\circ$) during deceleration ($0.19 \leq t^\star \leq 0.29$); WEC (\textit{left}), UEC (\textit{right}). Contour colors signify magnitude. Larger values of inwardly angled WSS vectors occur under WEC between the inner and outer wall where secondary flow is strongest and deformed Dean vortices are more intense than those that form under UEC. Local production of WSS at the inner wall is due to increased flow reversal, which is higher under WEC.}
        \label{f:wss_wec_uec}
    \end{minipage}
\end{figure*}
\begin{figure*}[t]
    \begin{minipage}[c]{\textwidth}
        \ContinuedFloat
        \subfloat[WEC: $t^\star=0.23$]{
            \includegraphics[width=0.48\textwidth,keepaspectratio]
            {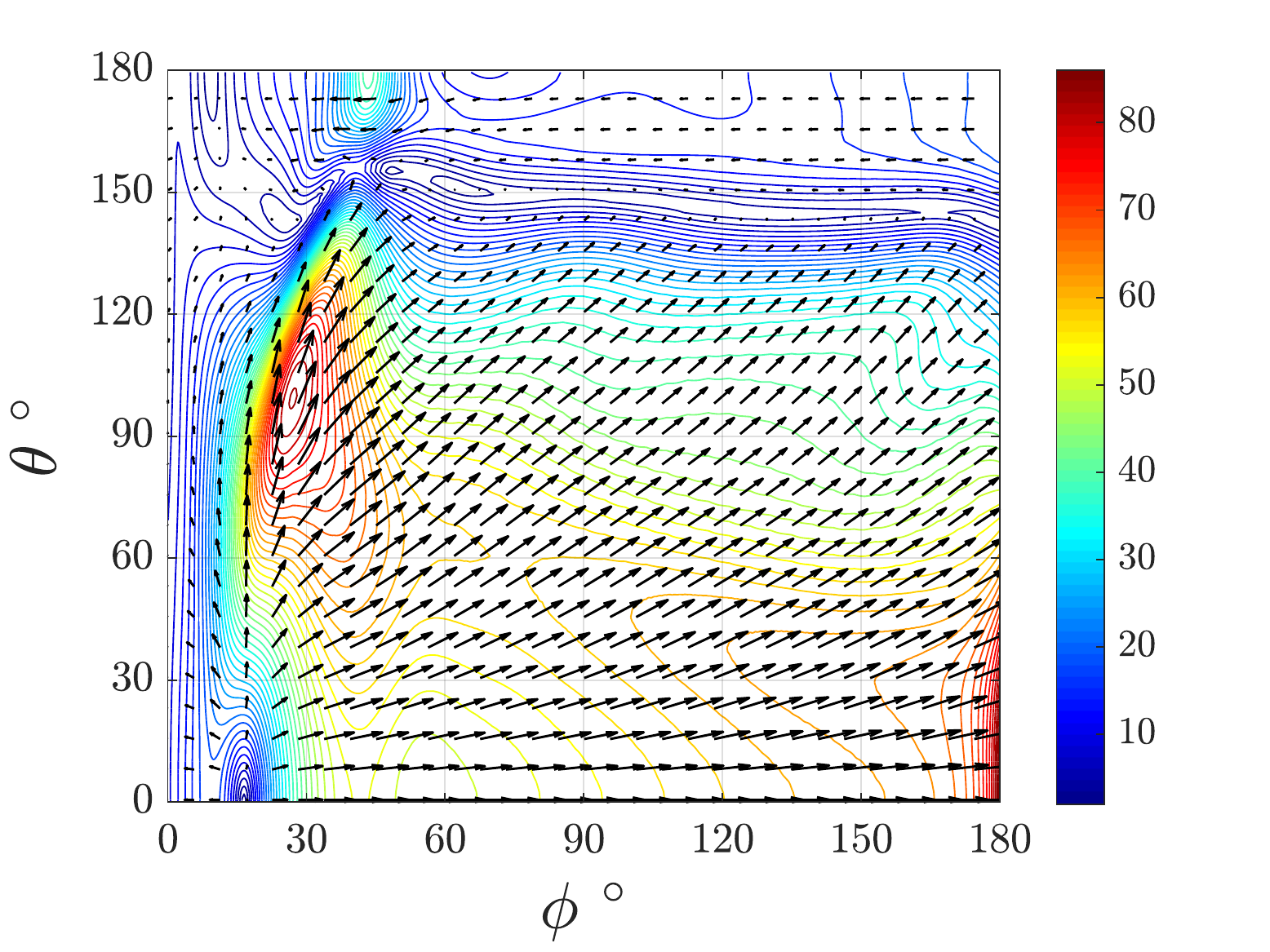}
        }
        \subfloat[UEC: $t^\star=0.23$]{
            \includegraphics[width=0.48\textwidth,keepaspectratio]
            {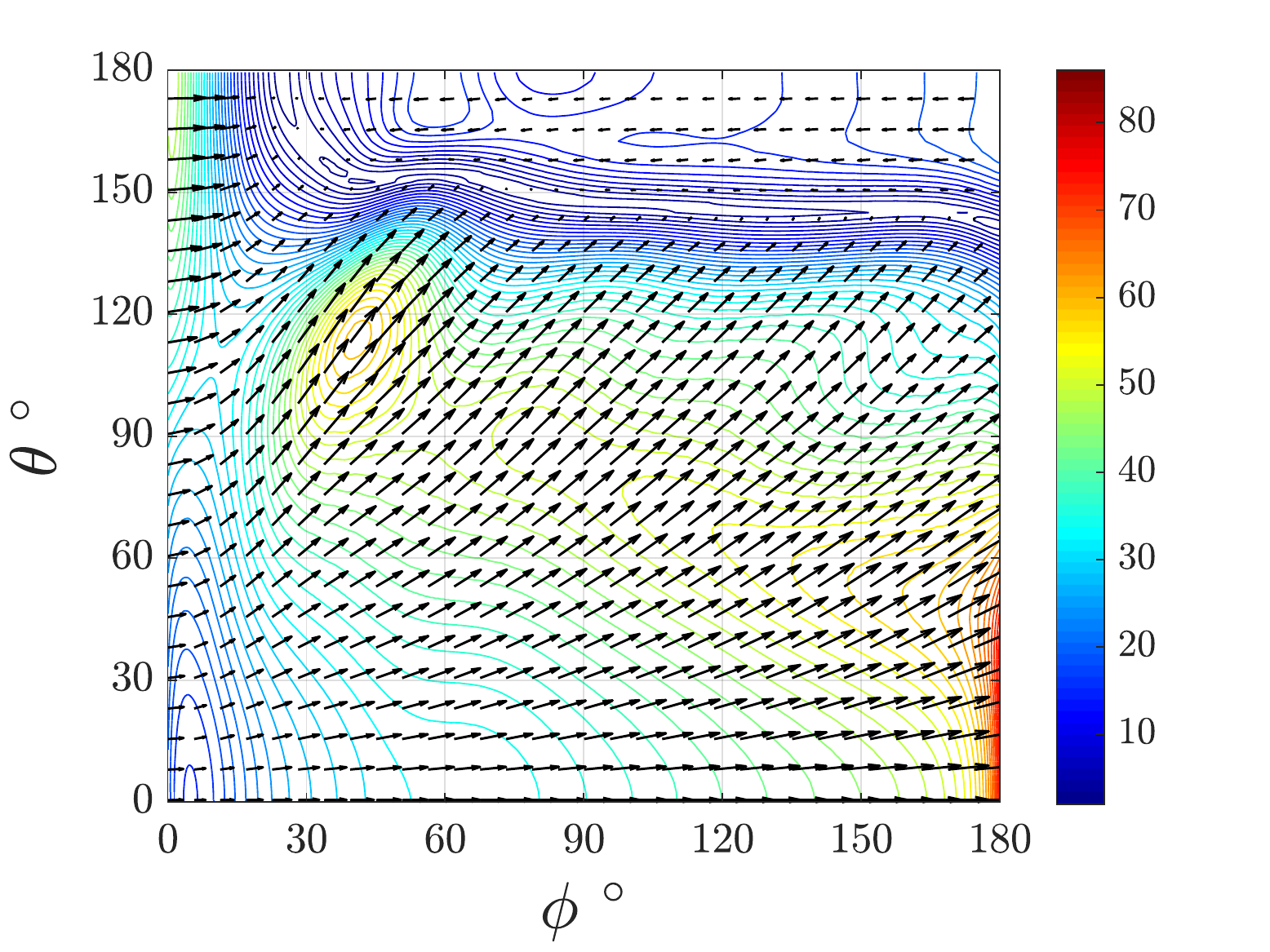}
        }
        \\[-0.15in]
        \subfloat[WEC: $t^\star=0.25$]{
            \includegraphics[width=0.48\textwidth,keepaspectratio]
            {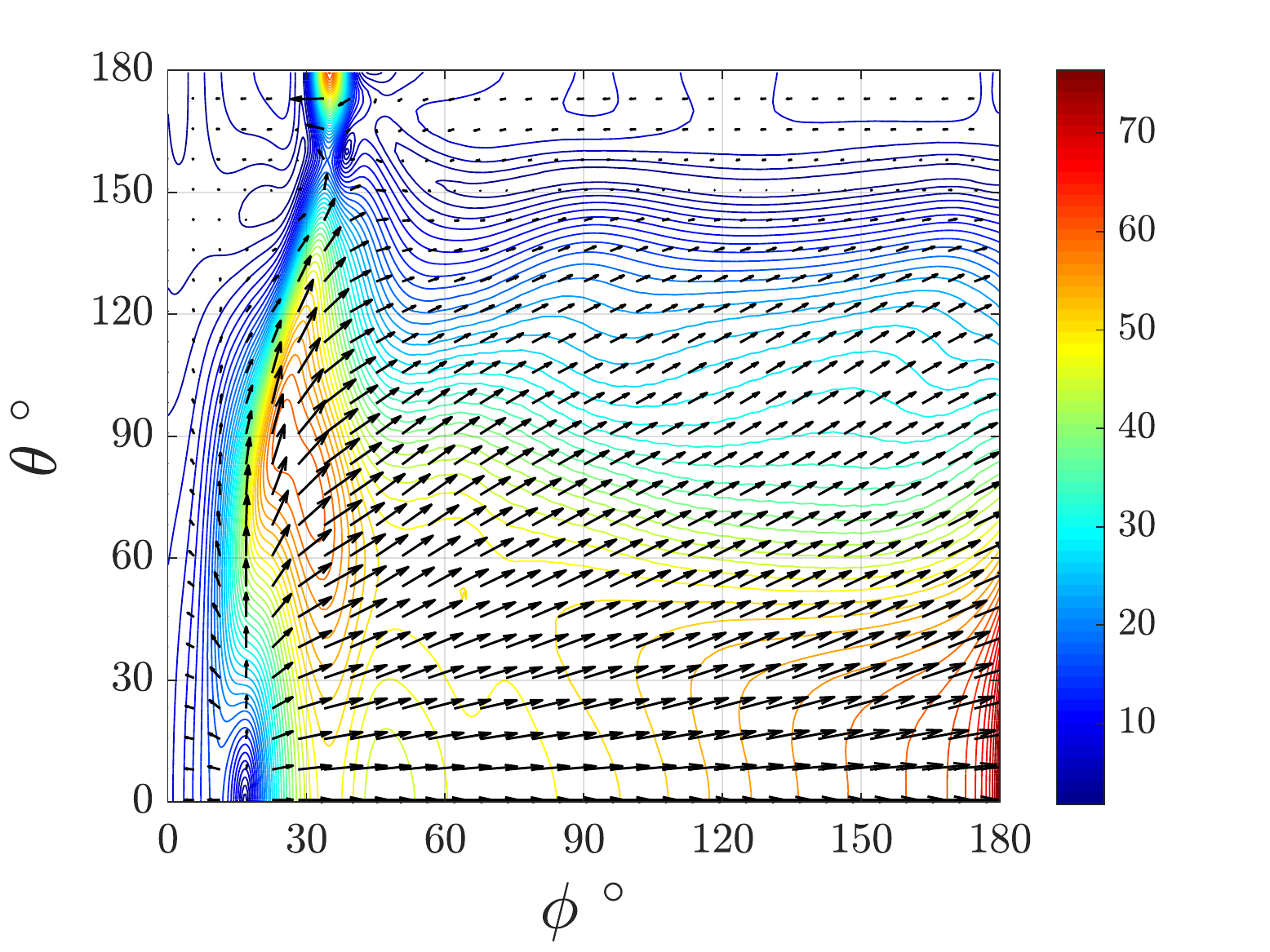}
        }
        \subfloat[UEC: $t^\star=0.25$]{
            \includegraphics[width=0.48\textwidth,keepaspectratio]
            {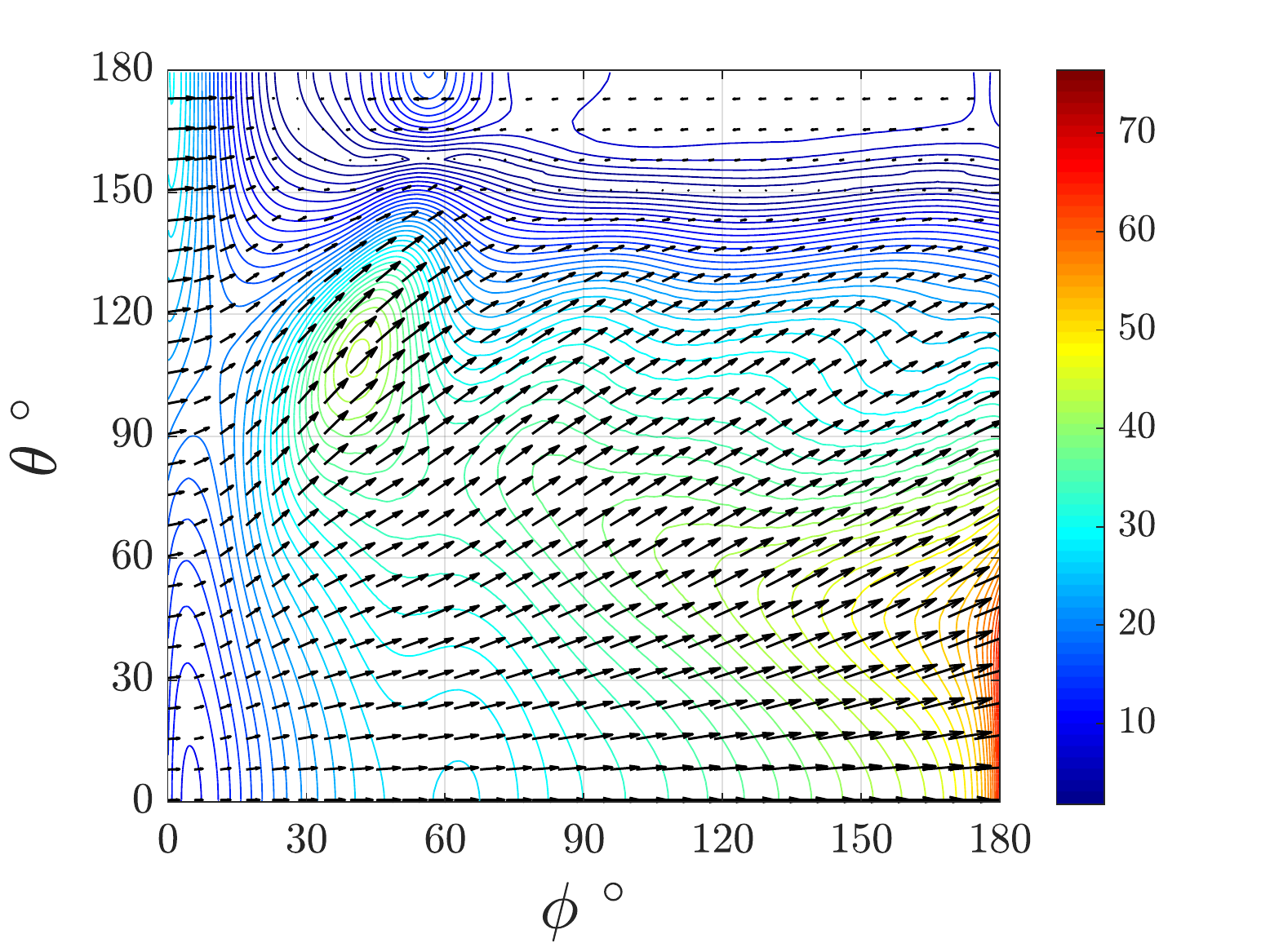}
        }
        \caption[]{({\it Continued})}
    \end{minipage}
\end{figure*}
\begin{figure*}[t]
    \begin{minipage}[c]{\textwidth}
        \ContinuedFloat
        \subfloat[WEC: $t^\star=0.27$]{
            \includegraphics[width=0.48\textwidth,keepaspectratio]
            {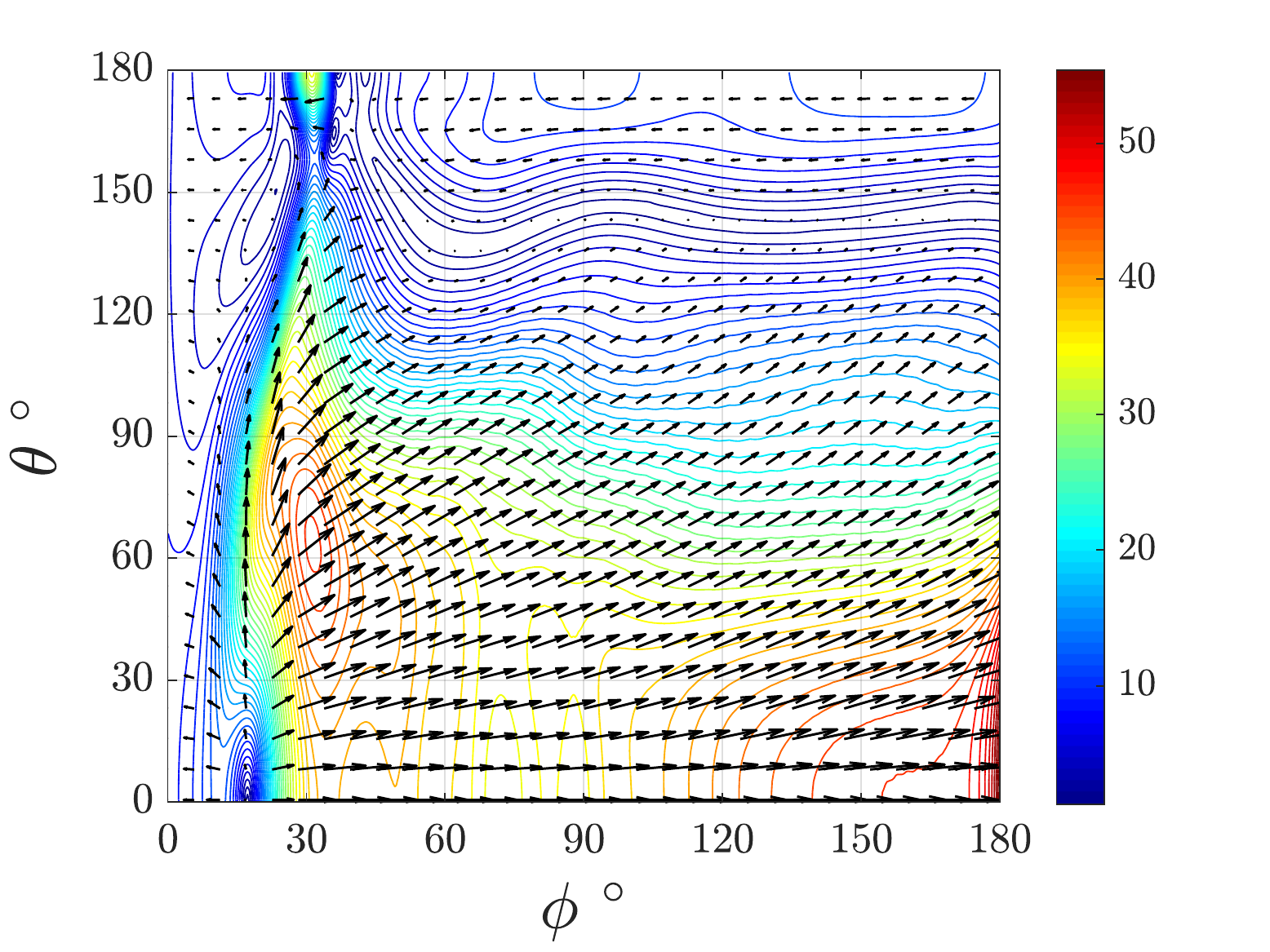}
        }
        \subfloat[UEC: $t^\star=0.27$]{
            \includegraphics[width=0.48\textwidth,keepaspectratio]
            {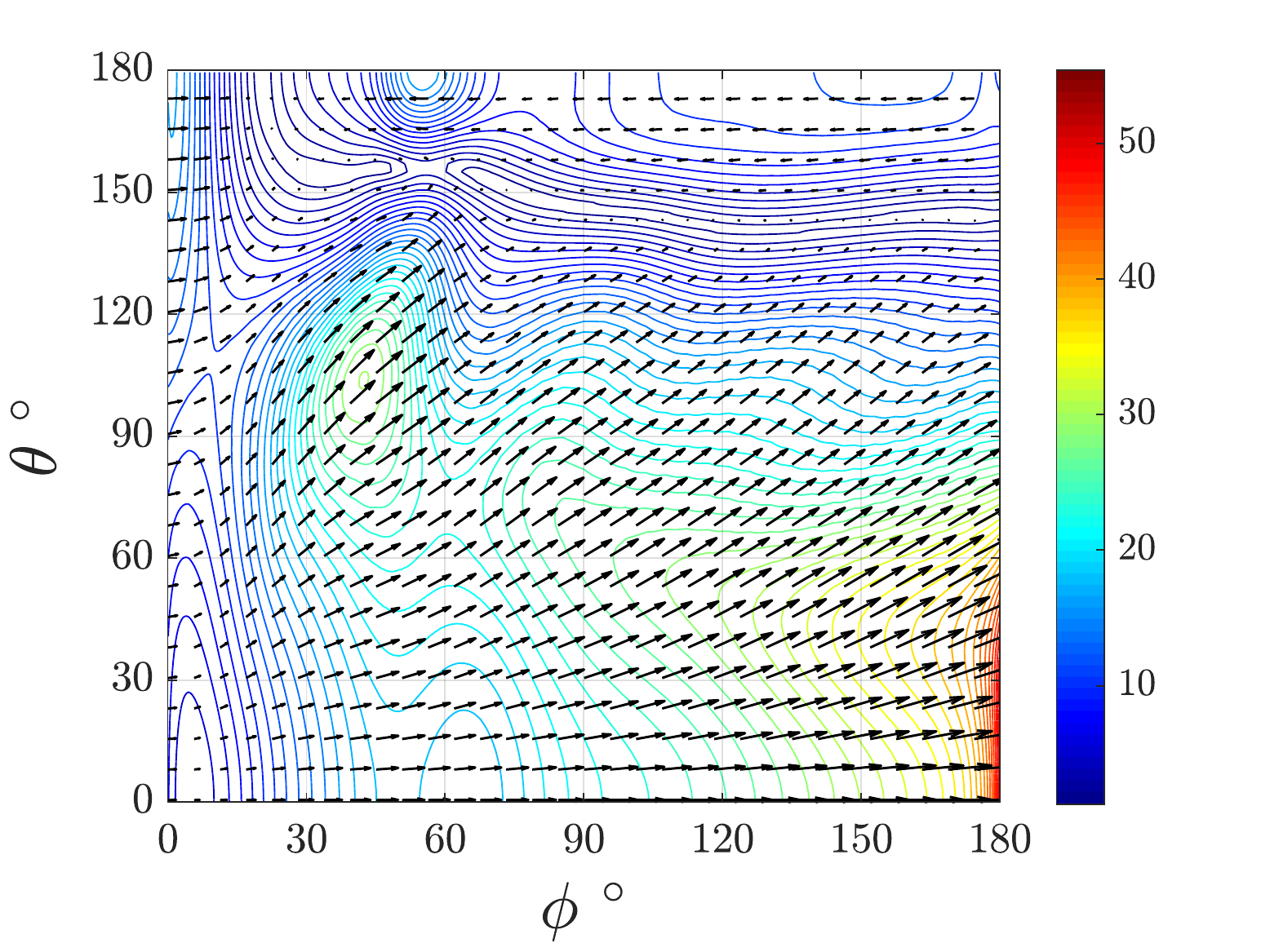}
        }
        \\[-0.15in]
        \subfloat[WEC: $t^\star=0.29$]{
            \includegraphics[width=0.48\textwidth,keepaspectratio]
            {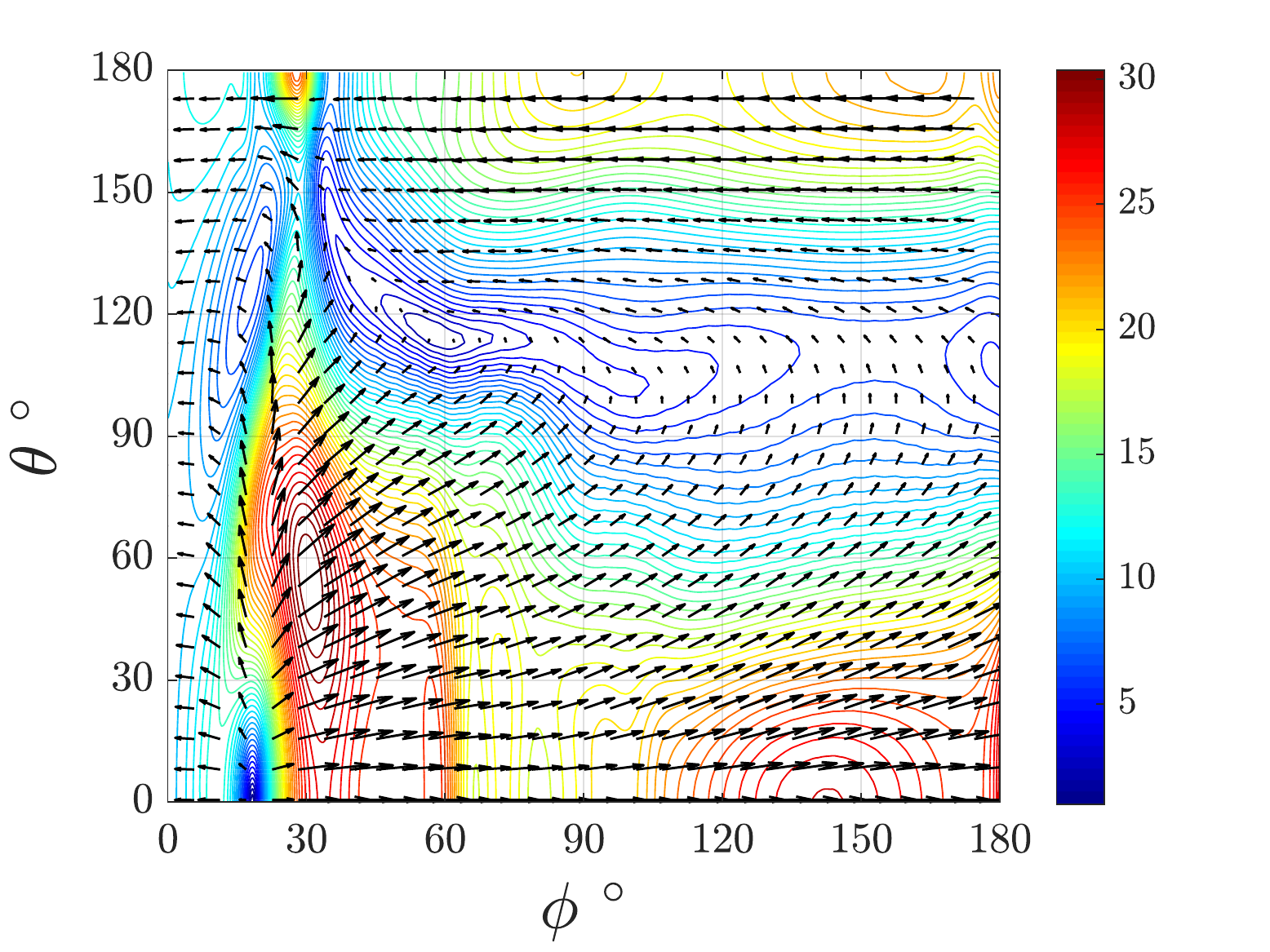}
        }
        \subfloat[UEC: $t^\star=0.29$]{
            \includegraphics[width=0.48\textwidth,keepaspectratio]
            {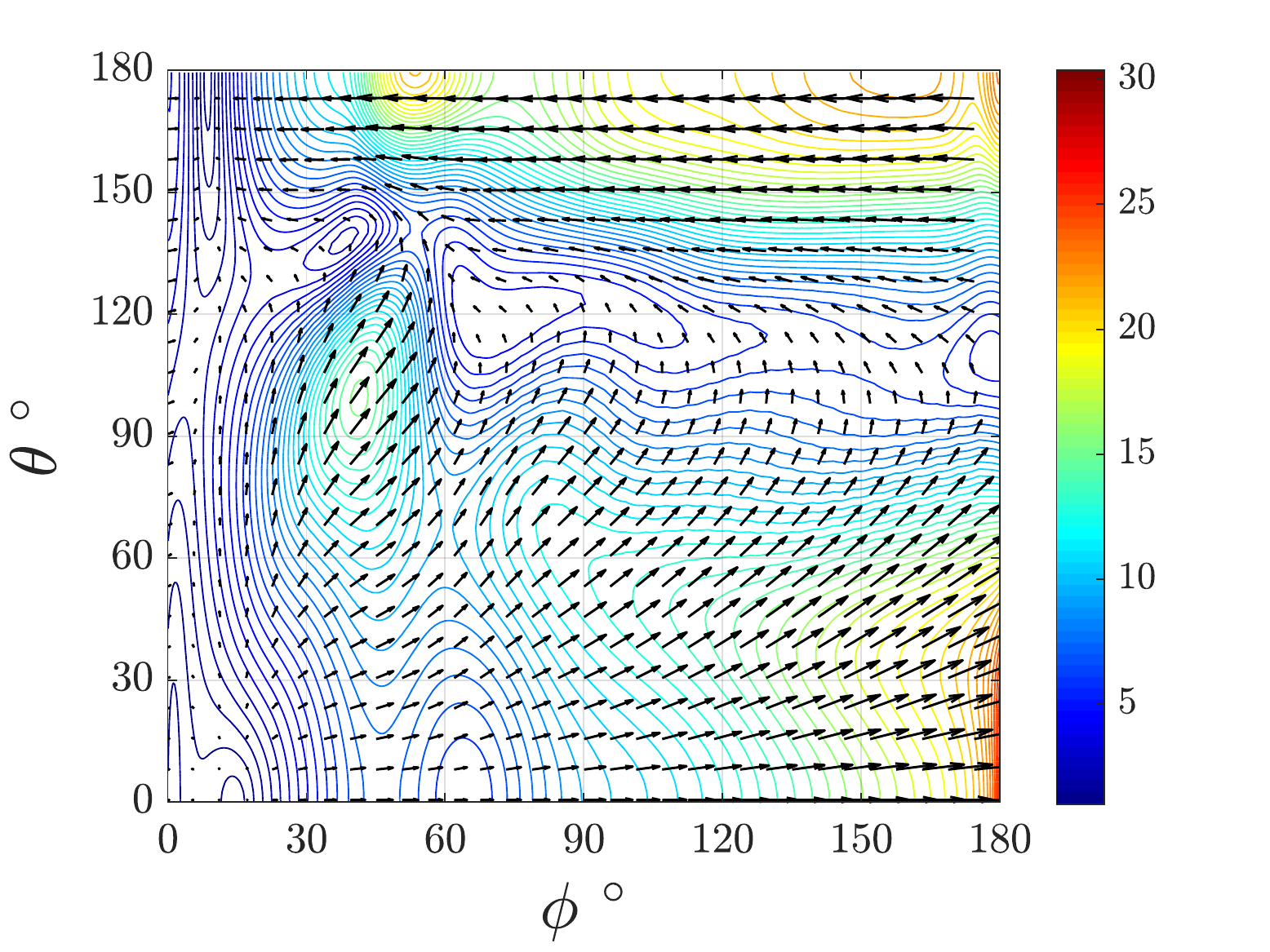}
        }
        
        \caption[]{({\it Continued})}
    \end{minipage}
\end{figure*}
\begin{figure*}[t]
    \centering\setcounter{subfigure}{0}
    \subfloat[]{
        \includegraphics[width=0.48\textwidth,keepaspectratio]
        {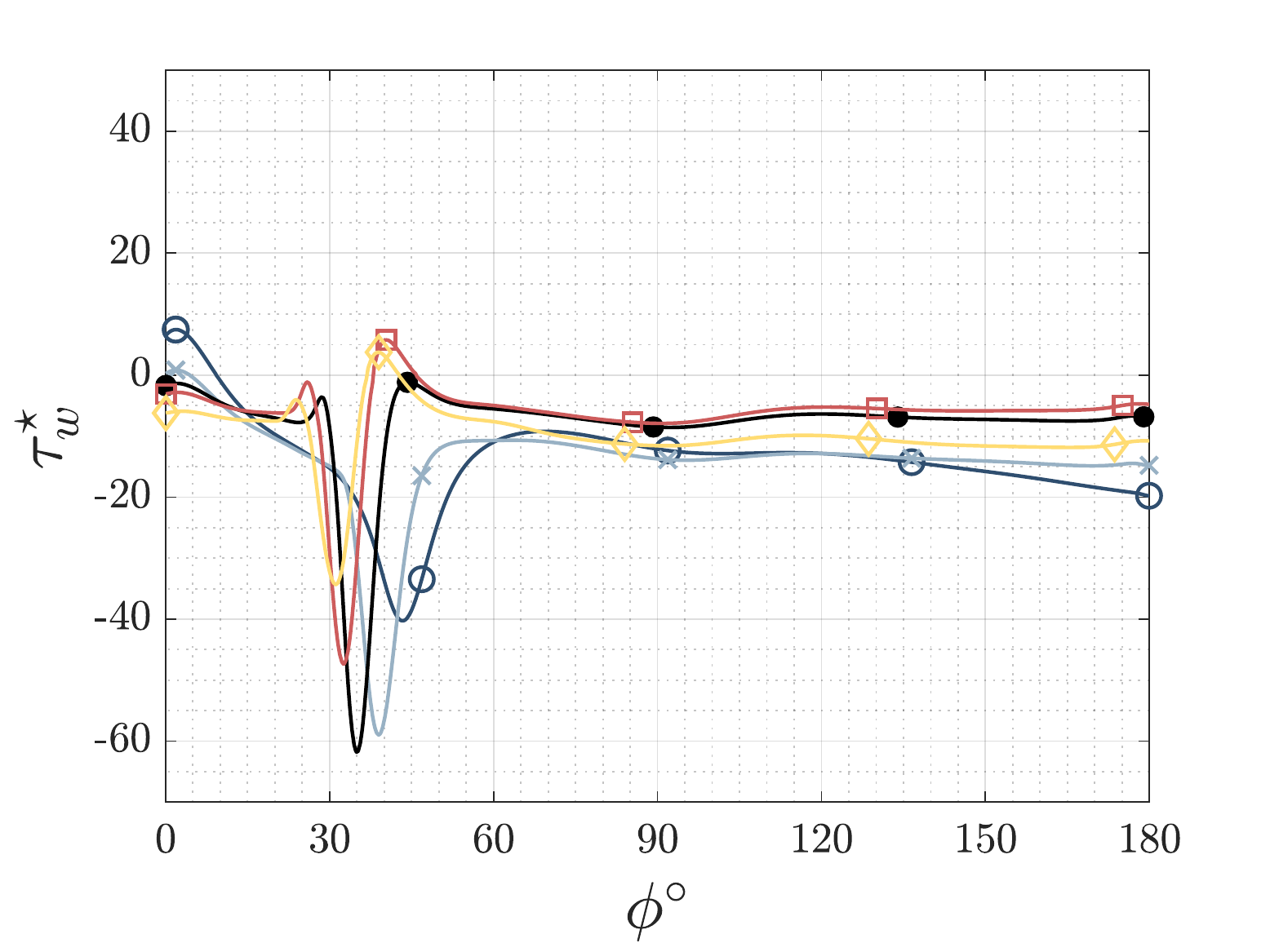}
    }
    \subfloat[]{
        \includegraphics[width=0.48\textwidth,keepaspectratio]
        {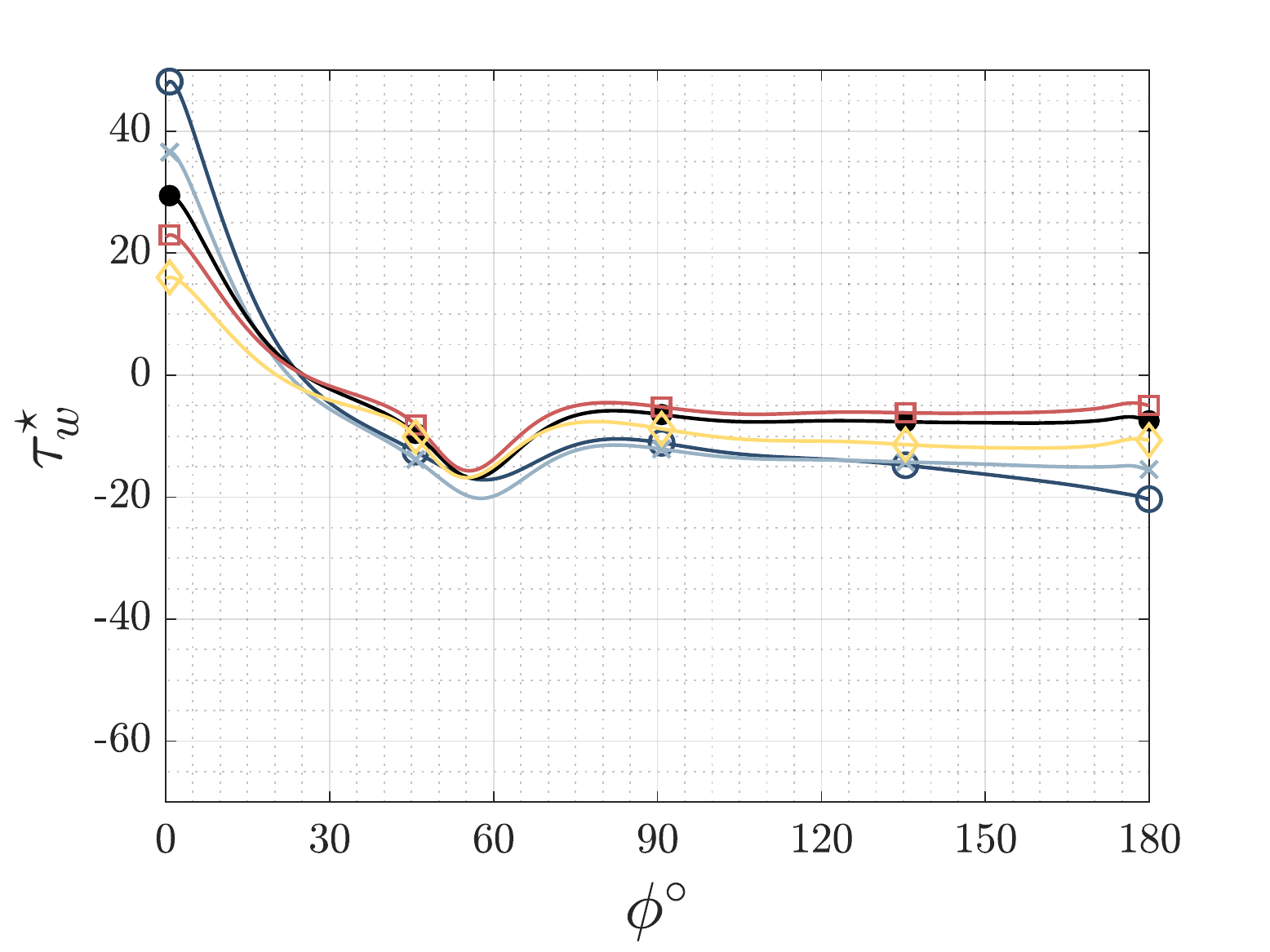}
    }
    
    \caption{Instantaneous wall shear stress $\bm{\tau}^\star_w$ at the inner wall ($\theta=180^\circ$) as a function of toroidal position $\phi$ under (a) WEC and (b) UEC for $t^\star=0.23$ ($\circ$), $t^\star=0.24$ ($\times$), $t^\star=0.25$ ($\bullet$), $t^\star=0.26$ ($\square$), $t^\star=0.27$ ({\large$\diamond$}). Under WEC, locally increased wall shear stresses occur in the range $30^\circ < \phi < 45^\circ$, with maximum $|\bm{\tau}^\star_w|$ approximately four times larger than UEC.}
    \label{f:tauw}
\end{figure*}
\begin{figure}[t]
    \centering\setcounter{subfigure}{0}
    \includegraphics[width=0.48\textwidth,keepaspectratio]
    {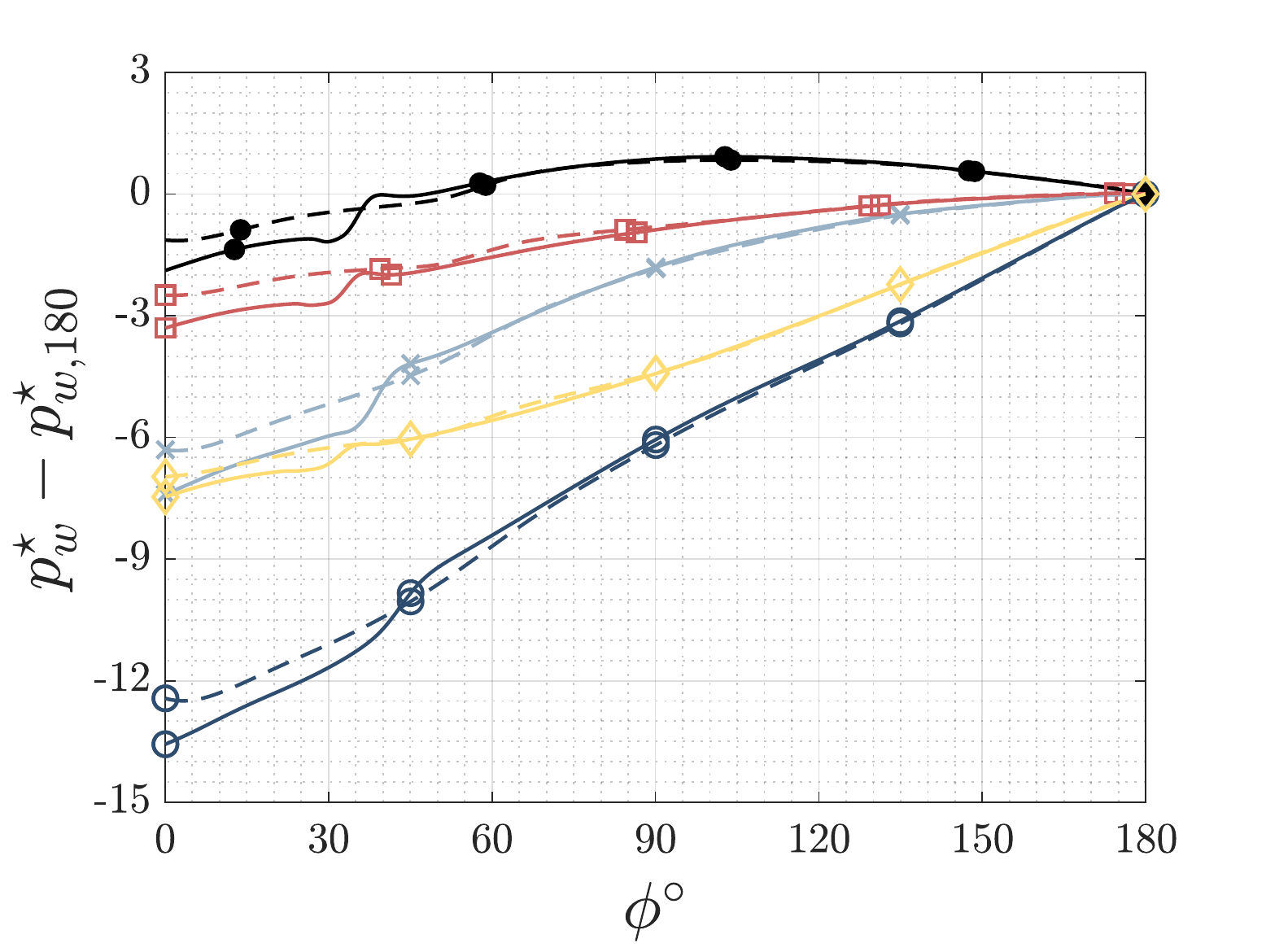}
    
    \caption{Instantaneous wall pressure $p^\star_w$ at the inner wall ($\theta=180^\circ$) as a function of toroidal position $\phi$ under WEC (\textit{solid}) and UEC (\textit{dashed}) for $t^\star=0.23$ ($\circ$), $t^\star=0.24$ ($\times$), $t^\star=0.25$ ($\bullet$), $t^\star=0.26$ ($\square$), $t^\star=0.27$ ({\large$\diamond$}). Values are adjusted by wall pressure at the curvature exit. Under WEC, a sudden drop in local wall pressure occurs in the range $30^\circ < \phi < 45^\circ$, corresponding to the increased velocities and subsequent wall shear stresses in this region when compared to UEC results. The phase $t^\star=0.25$ displays an overall increase in wall pressure with respect to the exit pressure compared to other instances in time.}
    \label{f:pw}
\end{figure}

In this section, we describe the method used to properly compute the wall shear stress. To define a formulation, we use the second-order stress tensor $\bmm{T}$ for a Newtonian fluid
\begin{align}
    \bmm{T} &= 2 \mu \bmm{S}.
\end{align}
Written explicitly, the stress tensor takes the form
\begin{align}
    \bmm{T} &= \mu
    \begin{bmatrix}
        2 \dfrac{\partial u}{\partial x}
        &    \dfrac{\partial v}{\partial x}+\dfrac{\partial u}{\partial y}
        &    \dfrac{\partial w}{\partial x}+\dfrac{\partial u}{\partial z} \\
        \dfrac{\partial v}{\partial x}+\dfrac{\partial u}{\partial y}
        &  2 \dfrac{\partial v}{\partial y}
        &    \dfrac{\partial w}{\partial y}+\dfrac{\partial v}{\partial z} \\
        \dfrac{\partial w}{\partial x}+\dfrac{\partial u}{\partial z}
        &    \dfrac{\partial w}{\partial y}+\dfrac{\partial v}{\partial z}
        &  2 \dfrac{\partial w}{\partial z}
    \end{bmatrix},
\end{align}
where the components of the velocity gradient tensor $\nabla \bmm{u}$ are directly computed using the flux reconstruction methodology discussed in Sec.~\ref{s:numerical_scheme}. We use the following full decomposition approach to accurately compute the wall shear stress vector. For each element face along the wall, we compute the wall traction $\bmm{T}\bmm{n}$ vector and the amount of traction in the wall normal direction as $\left[ \left( \bmm{T}\bmm{n} \right) \cdot \bmm{n} \right] \bmm{n}$, where the vector $\bmm{n}$ is normal to the wall surface. We then subtract these two terms to obtain the shear stress vector that lies in the wall as
\begin{align}
    \bmm{\tau}_w = \bmm{T}\bmm{n}
                 - \big[ \big( \bmm{T}\bmm{n} \big) \cdot \bmm{n} \big] \bmm{n}.
\label{e:instantaneous_wss}
\end{align}
From Eq.~\ref{e:instantaneous_wss}, we can compute the direction and magnitude of the wall shear stress vector over the entire curved surface throughout the pulse cycle. Since the flow is symmetric about the $xy$ plane, we plot results from the upper surface only. For visualization purposes, we map the surface of the curved geometry onto a plane in Fig.~\ref{f:wss_axis} and provide orientation of the mapping along with flow direction. In general, knowledge of the wall shear stress is valuable since it is used to compute important haemodynamic metrics commonly used to assess localized atherosclerotic disease in curved and branched sections of the human arterial network.

Results shown in Figs.~\ref{f:wss_wec_uec}\subref{f:wss_wec_uec:a} and \ref{f:wss_wec_uec}\subref{f:wss_wec_uec:g}
demonstrate the inner to outer wall shift in maximum wall shear stress discussed in Sec.~\ref{s:pulsatile_primary_flow} under both WEC and UEC. The effect of secondary velocity near the entrance to the curve is small and the skewness of the velocity profile towards the inner wall comes from the fact that the boundary layer at the outer wall is thicker because of its longer wall length and the boundary layer at the inner wall is thinner because of its shorter wall length, with the fluid in the core accelerating due to the displacement effect caused by the growing boundary layer. This figure also shows that UEC exhibits a larger maximum wall shear stress at the inner wall near the entrance and is caused by heavier skewness of the velocity profile towards the inner wall. This heavier skewness is due to the lack of flow development inherent to the uniform velocity profile prescribed at the inlet to the computational domain.

Overall, Fig.~\ref{f:wss_wec_uec} demonstrates that the wall shear stress vectors reflect the amount of secondary flow under WEC and UEC throughout fluid deceleration from peak flow rate. At peak flow rate, the wall shear stress vectors are mostly angled toward the inner wall over a large portion of the curvature due to the secondary flow. At the inner and outer wall, the vectors are aligned with the positive streamwise direction. In general, curved pipe flows exhibit larger values of WSS along the outer wall due to larger velocity gradients resulting from the outwardly shifted velocity profile. Under WEC and UEC, results in Fig.~\ref{f:wss_wec_uec} show that as the fluid moves inward along the upper/lower wall, the WSS increases then decreases sharply to a minimum near $\theta=150^\circ$, eventually increasing slightly at the inner wall. This sharp inflection in $|\bm{\tau}^{\star}_w|$ is caused by flow reversal occurring along the inner wall. Large values of WSS at $\phi \approx 22^\circ$ and $\theta \approx 98^\circ$ are coincident with deformed Dean vortex pair formations, which are more intense under WEC than UEC by a factor of 3.5 (see Figs.~\ref{f:womersley_lambda2} and \ref{f:uniform_lambda2} at $(t^\star,\theta)=(0.23,22^\circ)$.


As flow deceleration begins, shear stresses decrease globally while severely altering direction under WEC near the $\phi=30^\circ$ plane such that vectors become less aligned with the streamwise direction and more aligned with the radius of curvature, particularly near the entrance at $\phi=15^\circ$. At $t^\star=0.21$, the fluid has slowed near the inner wall, causing the shear stress vector to decrease in magnitude before reversing direction. After this phase the flow decelerates further, increasing the amount of reverse flow along the inner wall and causing the shear vectors to point upstream. Over the entire deceleration phase of the waveform, we calculate a reduction of the spatially averaged shear stress by 86\% (WEC) and 90\% (UEC) as the flow rate drops by a factor of seven.

Under WEC at $t^\star=0.23$, a small circular pocket of locally increased WSS appears at $\phi=40^\circ$. As the flow decelerates, this pocket of large shear stress moves upstream, reaches a local maximum at $t^\star=0.25$, and decreases thereafter. Under UEC, a similar but smaller pocket of WSS can be observed. From Figs.~\ref{f:tauw} and \ref{f:pw} we see the locally increased WSS under WEC at the inner wall in the range $30^\circ < \phi < 45^\circ$ and the corresponding sudden drop in wall pressure in comparison to the results obtained under UEC. The oscillatory and multidirectional distribution of the wall shear stress vector indicates the importance in understanding secondary flow morphologies and spatially/temporally varying Dean-type and Lyne-type vortical structures. With this analysis, we can conclude that a lack of flow development near the inlet to the curvature corresponds to a lower degree of multidirectionality and reduced magnitude of WSS, particularly along the inner wall and within the first half of the curved artery model.

These wall shear stress results indicate it may be pathologically favorable for the flow to be less developed entering a curved artery. For example, flow emanating from the heart and entering nearby curved arteries (e.g. arch of aorta) is rather undeveloped due to the lack of straight sections upstream of said arteries; if the flow were more developed, the prevalence of cardiovascular disease---especially along the inner wall---might be higher. Therefore, formation of vortical structures and their relation to wall shear stress patterns under varying degrees of flow development is potentially physiologically significant since cardiovascular disease is highly correlated to altering shear stresses at the wall. Consequently, it behooves medical professionals to assess the level of blood flow development and its influence on disease in curved arteries especially if surgical intervention, such as an arterial graft, is necessary.

\section{CONCLUSIONS}
\label{s:conclusions}

We performed numerical simulations using a physiological (pulsatile) inflow of a Newtonian blood-analog fluid in a $180^\circ$ curved rigid pipe without taper or torsion---a simple model for a human artery, with circular cross-section and constant curvature. In particular, we investigated the effect of flow development at the entrance to the curve on the formation of vortical structures and subsequent wall shear stress patterns using two conditions. The first pulsatile entrance flow condition was fully developed while the second condition was undeveloped. We observed large differences in secondary flow patterns particularly during the deceleration phase of the physiological waveform, where multiple vortical structures of both Dean-type and Lyne-type coexist. We find that peak axial velocity under an undeveloped entrance condition is less skewed towards the outer wall and smaller than that under a fully developed condition. This decreased axial velocity produces smaller centrifugal forcing and pressure gradients in-plane, thereby driving the secondary motion of the fluid at a slower velocity. This reduced secondary flow ultimately inhibits growth of any interior flow vortices. Furthermore, we conclude from our analysis that a lack of flow development near the inlet to the curvature corresponds to a lower degree of multidirectionality and a reduced magnitude in wall shear stresses, particularly along the inner wall and within the first half of the model. These wall shear stress results indicate it may be pathologically favorable for the flow to be less developed entering a curved artery, such as that which occurs in arteries and branches nearby/downstream of the heart. If the flow were more developed, these findings suggest that the prevalence of cardiovascular disease---especially along the inner wall of curved segments---might be higher. Therefore, formation of vortical structures and their relation to wall shear stress patterns under varying degrees of flow development is potentially physiologically significant, and it is advantageous to understand the influence of blood flow development on disease in curved arteries, especially if surgical intervention is intended.


\section*{ACKNOWLEDGMENTS}
\label{s:acknowledgments}

This study was conducted under the support of the Presidential Merit Fellowship and the Center for Biomimetics and Bioinspired Engineering at The George Washington University.

%

\section*{REFERENCES}
\bibliography{./cox_arxiv_2021}

\end{document}